\documentclass[10pt,aps,floatfix,prd,onecolumn,merge,nofootinbib,
        superscriptaddress,notitlepage]{revtex4} 
\usepackage{graphicx} 
 \graphicspath{{./figures/}}
\usepackage{dcolumn} 
\usepackage{amsmath,amsfonts} 
\usepackage{physics}
\usepackage{slashed}
\usepackage{cancel}
\usepackage[caption=false]{subfig}
\usepackage[usenames,dvipsnames,svgnames]{xcolor}
\usepackage{braket}
\usepackage{multirow}
\usepackage[shortlabels]{enumitem}

\usepackage{mathabx}

\usepackage{hyperref} 

\usepackage[capitalise]{cleveref}

\hypersetup{%
  colorlinks=true, linktocpage=true, pdfstartpage=1, pdfstartview=FitV,%
  breaklinks=true, pdfpagemode=UseNone, pageanchor=true, pdfpagemode=UseOutlines,%
  plainpages=false, bookmarksnumbered, bookmarksopen=true, bookmarksopenlevel=1,%
  hypertexnames=true, pdfhighlight=/O,
  urlcolor=Cerulean, linkcolor=Cerulean, citecolor=Cerulean, 
}

\def\d{\partial}
\def\nl{\nonumber\\}

\usepackage[normalem]{ulem}

\newcommand{\comment}[1]{\ignorespaces}

\begin{document}
\title{Multilepton signatures for scalar dark matter searches in coannihilation scenario}
\author{Sreemanti Chakraborti}
\email{sreemanti@iitg.ac.in}
\affiliation{Department of Physics,
               Indian Institute of Technology Guwahati,
               Assam 781039, India}
\author{Rashidul Islam}
\email{rislam@iitg.ac.in}
\affiliation{Department of Physics,
               Indian Institute of Technology Guwahati,
               Assam 781039, India}
%

%
\begin{abstract}
We revisit the scalar singlet dark matter (DM) scenario with a pair of dark lepton partners which form a vectorlike Dirac fermionic doublet. The extra  doublet couples with the Standard Model (SM) leptonic doublet and the scalar singlet via a non-SM-like Yukawa structure. As a result, (i) since the extra fermionic states interact with other dark sector particles as well as the SM via gauge and Yukawa interactions, it gives rise to new DM annihilation processes including pair annihilation as well as coannihilation channels, and (ii) such a Yukawa structure opens up new production channels for leptonic final states giving much enhancement in cross sections to search for dark matter in the LHC. Using suitable kinematic observables, we train a {\em boosted decision tree} (BDT) classifier to separate enhanced but still feeble light leptonic signals from the background in an effective manner. On the other hand, the same technique is applied to study $\tau$-tagged jets in the search for DM signals.
\end{abstract}

\maketitle
%
\section{Introduction}
\label{sec:intro}
Cosmological considerations and astrophysical observations have established beyond any reasonable doubt the existence of the dark matter (DM). The satellite-borne experiments such as WMAP~\cite{Hinshaw:2012aka} and Planck~\cite{Aghanim:2018eyx} measured extremely precisely the {\em cosmological relic abundance}, and it is given by $\Omega_{\rm DM} h^2 = 0.1199 \pm 0.0027$, $h$ being the reduced Hubble constant. Though DM constitutes about 27\% of the energy budget of the Universe, the particle nature of it remains an enigma. The search for a suitable candidate for particle dark matter is a longstanding problem~\cite{Bergstrom:2000pn,Bertone:2004pz,Feng:2010gw}. The so-called {\em weakly interacting massive particle} (WIMP) is the most widely explored sector to resolve the discrepancy. Within the WIMP paradigm, the scalar singlet dark matter or scalar ``Higgs-portal'' scenario is perhaps the most studied of all the relevant scenarios of dark matter to explain the relic density~\cite{Silveira:1985rk,McDonald:1993ex,Burgess:2000yq}. Consequently, it went through immense scrutiny theoretically as well as experimentally (see, for example, Refs.~\cite{Athron:2017kgt,Arcadi:2019lka} for recent reviews of the current status of the Higgs-portal scenario). We now know that the {\em direct detection}~\cite{Agnes:2015ftt,Akerib:2016vxi,Aprile:2017iyp,Cui:2017nnn}, {\em indirect detection}~\cite{Fermi-LAT:2016uux,Ackermann:2015lka,Abramowski:2013ax}, and {\em invisible Higgs decay}~\cite{Belanger:2013xza,Aad:2015pla,Khachatryan:2016whc} searches put a strong bound on the coupling of the Standard Model (SM) Higgs boson, $h$ with the said scalar singlet, say, $S$. Let us call this coupling $\lambda_{hS}$. These experiments constrain $\lambda_{hS}$ to be very small. As a result, it gives an overabundance of relic density except around a small window around the resonance region, \(m_S \sim m_h/2\).

However, one can improve the situation with scalar singlet DM using various alternatives, such as considering other symmetries within the dark sector~\cite{Belanger:2012zr,Bhattacharya:2013hva,Bian:2013wna,Bhattacharya:2017fid} or adding new particles in the particle spectrum so as to arrange other portals~\cite{Batell:2016ove,Bandyopadhyay:2017tlq,Han:2018bni} for DM annihilation without worsening the existing constraints. An interesting possibility in this context, called {\em coannihilation}~\cite{Griest:1990kh}, is a widely studied feature in DM dynamics where the DM annihilates with another dark sector particle and the chemical equilibrium between the annihilating particles ensures the substantial depletion of DM number density. This feature is a very useful handle to revive the scenarios where direct detection bounds push relic density to overabundance. In such scenarios, coannihilation works efficiently as a DM number changing process without affecting the direct search measurements, because the direct detection channels are relevant only for DM DM $\to q \bar{q}$ interactions.

In the present work, we will revisit the scenario of the scalar singlet dark matter with a pair of accompanying dark leptons which form a vectorlike Dirac fermionic doublet. Gauge invariance requires that the two components be degenerate at tree level and only a small mass splitting of the order of 300~MeV can be generated by the radiative corrections~\cite{Buckley:2009kv}. One can generate finite mass splitting at tree level in a gauge-invariant way by increasing the particle content in the model. In this article, we introduce an additional scalar triplet for this purpose whose vacuum expectation value (VEV) will be responsible for generating a small but sufficient mass splitting between the dark leptons. This mechanism is used to generate finite mass of neutrinos in {\em type-II seesaw} models~\cite{Hambye:2003ka}. However, the triplet may or may not play any role in the phenomenology of the dark matter depending on the values of their masses. We assume them to be very heavy so that their effect is negligible apart from generating sufficient mass splitting between the dark leptons which will play an important role in our study. Now this dark sector doublet couples with the SM leptonic doublet and the scalar singlet via a novel Yukawa interaction which is less explored in the literature. There are two distinct interesting features of this model: (i) Since the new dark sector fermions form a doublet, they will interact with the SM via gauge interaction as well as the new Yukawa coupling, which, in turn, will give rise to new annihilation channels, and (ii) such a Yukawa structure will open up new production channels for leptonic final states giving much enhancement in cross sections to search for dark matter in collider environments like the LHC through the said channel. Depending on the choice of parameters, here the DM annihilation can have three distinguishable stages, namely, pair annihilation, coannihilation, and mediator annihilation. Here, it is to be noted that coannihilation scenarios in WIMPs are mostly studied in the literature in the context of supersymmetric (SUSY)~\cite{Ellis:1998kh,Ellis:2015vaa,Baker:2018uox,Arganda:2018hdn} and coloured coannihilating particles~\cite{Baker:2015qna,Biondini:2019int,Ibarra:2015nca}. Our model discusses a leptophilic context, and the coannihilation channels play an important role here due to the gauge interaction in the dark sector in addition to the new Yukawa coupling. This feature is significantly different from the cases explored in the literature where the leptophilic Yukawa structure involves singlet dark sector partners with the DM candidate~\cite{Khoze:2017ixx,Borah:2017dqx}.

As mentioned above, since the coannihilating partner\footnote{Nonobservance of any new fermionic partner state in LEP2 which interacts with the SM leptons via Yukawa-type interaction puts a bound of $m_{\psi} > 104$~GeV on their masses. However, please note that here the new fermions interact strongly only with the $\tau$ lepton. Since LEP2 $\tau$ detection was not very precise, we must take this limit with a pinch of salt.} couples to the SM with gauge as well as Yukawa coupling, the leptonic search channels get a boost in cross section from it, and it is only logical to probe the said channel for collider signatures. Moreover, the leptonic channel gives cleaner signals than the other channels. Still, the collider searches of dark matter are a very challenging prospect. Note that any leptophilic DM model like ours contributes to the calculation of muon $g-2$. Very good agreement between the theoretical calculation and experimental measurements of muon $g-2$, $\Delta a_\mu = a^{\rm Exp}_\mu - a^{\rm SM}_\mu = 268(63)(43) \times 10^{-11}$~\cite{Tanabashi:2018oca}, put a strong constraint on the new Yukawa couplings of the light SM leptons. However, there is no such bound for the production of $\tau$ leptons. Hence, it would be a good prospect to probe that channel for dark matter signatures in colliders. Our case is similar to the SUSY theories where stau is the coannihilating partner~\cite{Godbole:2008it,Ellis:1998kh,Jittoh:2005pq,Kaneko:2008re,Jittoh:2010wh,Citron:2012fg,Konishi:2013gda,Desai:2014uha}. In the SUSY scenario, the particle content is much larger than our minimalistic model, leading to more involved phenomenology. On the other hand, in a minimalistic model like ours, we have more handle to pinpoint the effects of coannihilation, and it is less probable to be lost in the midst of other effects.

Despite the leptonic channel getting a boost, the cross section can still be smaller. So, to probe light leptonic channels effectively, one must follow sophisticated techniques to separate signals from the backgrounds. The {\em multivariate analysis} is one such prospect. We perform {\em boosted decision tree} (BDT) response to separate feeble light leptonic signals from the background in an effective manner. On the other hand, despite further enhancement in cross section, the $\tau$ leptons mostly decay into hadronic jets, resulting in difficulty in their reconstruction. We used $\tau$-tagged jets from the detector simulation with 60\% $\tau$-tagging efficiency to perform the BDT response.

We organised the paper as follows. In \cref{sec:model}, we describe the contents of our model. The dark matter phenomenology, its formalism, and the observations from the relic density, direct, and indirect detection calculation are discussed in \cref{sec:dmpheno}. \cref{sec:collider} contains the study of collider signatures at the LHC through multivariate analysis of light dilepton as well as di-$\tau$ lepton channels. Finally, we conclude our results in \cref{sec:conc}.

%
\section{Model description}
\label{sec:model}
As we described briefly in the introduction, we want a model where the dark sector will consist of one or more coannihilating partners in addition to the scalar singlet dark matter. So, we consider a vectorlike Dirac fermionic doublet \(\varPsi^T = (\psi^0, \psi^-)\) and a real scalar singlet \(\phi\) in addition to the SM particles. To achieve the stability of the dark sector, both these new fields are odd under \(\mathbb{Z}_2\) symmetry, whereas the SM fields are \(\mathbb{Z}_2\)-even. Gauge invariance makes the masses of each field in the fermionic doublet degenerate. The difference between the masses can come only from radiative corrections which are of the order of 300~MeV. However, one can introduce extra fields in the model to increase the mass splitting. That is the reason why we introduce a \(\mathbb{Z}_2\)-even scalar triplet in addition to the SM doublet scalar. However, this scalar triplet does not affect the phenomenology of dark matter in any way and serves the only purpose of tuning the mass splitting of new fermions. In \cref{tab:quant}, the quantum number assignments of the particles relevant to new interactions are shown.
\begin{table}[!ht]
  \centering
  {\setlength{\tabcolsep}{1em}
  \begin{tabular}{c|c c c c|c c}
    \hline
    & $\ell_L$     & $e_R$  & $H$  &  $\Delta$   & $\varPsi$
                                                               & $\phi$
    \\ \hline\hline
    $SU(2)_L$      & $\mathbf{2}$
                            & $\mathbf{1}$
                                   & $\mathbf{2}$
                                           & $\mathbf{3}$
                                                      & $\mathbf{2}$
                                                               & $\mathbf{1}$
    \\ \hline
    $U(1)_Y$       & $-1/2$ & $-1$ & $1/2$ & $1$ & $-1/2$ & $0$
    \\ \hline
    $\mathbb{Z}_2$ &   $+$  &  $+$ &  $+$  & $+$ &   $-$  & $-$
    \\ \hline
  \end{tabular}
  }
  \caption{Quantum number assignment of the relevant fields in our model. Electromagnetic charges are given by \(Q = t^3 + Y\).}
  \label{tab:quant}
\end{table}

\noindent The real scalar singlet \(\phi\) which is our DM candidate interacts with the SM via the Higgs-portal. As the other dark sector particles $(\psi^0, \psi^\pm)$ form an $SU(2)_L$ doublet, it interacts with the SM through gauge bosons. The coannihilating doublet $\varPsi$ couples with the SM leptonic doublet $\ell_L$ and the scalar singlet $\phi$ via a Yukawa interaction. This is novel in the sense that the widely used Yukawa structure in any new physics model consists of a scalar doublet which is the replica of the SM Yukawa interaction. Although this particular Yukawa structure is less explored in the literature, it fits the bill for all our requirements for this study. Although, as mentioned previously, the scalar triplet does not play any role in the DM phenomenology, we write the relevant terms in the Lagrangian nonetheless for the sake of completeness.

Hence, the resulting Lagrangian takes the form
\begin{align}
 {\cal L} =& {\cal L}_{\rm SM}
 + \widebar\varPsi i\slashed D \varPsi - \widebar\varPsi M_\varPsi \varPsi
 \nl
 +& \frac{1}{2} (\d_\mu \phi)^2 - \frac{\mu^2_\phi}{2} \phi^2 - \frac{\lambda_\phi}{4} \phi^4
 - \frac{\lambda_{h\phi}}{2} (H^\dag H) \phi^2
 \nl
 +& {\rm tr}[(D^\mu\Delta)^\dag (D_\mu\Delta)] - \mu^2_\Delta {\rm tr}[\Delta^\dag\Delta]
 \nl
 -& [\mu H^T i\tau^2\Delta^\dag H + {\rm h.c.}] - \lambda_\Delta ([{\rm tr}(\Delta^\dag\Delta)]^2 + {\rm tr}[(\Delta^\dag\Delta)^2])
 \nl
 -& \lambda_{H\Delta} [H^\dag H {\rm tr}(\Delta^\dag\Delta) + H^\dag \Delta\Delta^\dag H]
 - \lambda_{\phi\Delta} \phi^2 {\rm tr}(\Delta^\dag\Delta)
 \nl
 -& \frac{1}{\sqrt{2}} \left[ y_\Delta \widebar\varPsi^c\ i \tau_2 \Delta \varPsi + {\rm h.c.} \right]
 - \left[ y_\alpha (\widebar\ell_{\alpha L} \varPsi) \phi + {\rm h.c.} \right]  ,
 \label{Lag_tot}
\end{align}
where ${\cal L}_{\rm SM}$ is the SM Lagrangian, \(M_\varPsi = m_{\psi^0} = m_{\psi^-}\) is the bare mass term of the new fermionic doublet, and \(D_\mu = \d_\mu + ig_W\,t^a\,W^a + ig^\prime\,Y\,B_\mu\) is the covariant derivative. The mass of the scalar singlet $\phi$ is given by \(m^2_\phi = \mu^2_\phi + \lambda_{h\phi}\,v^2/2\), where the VEV $v = \sqrt{v_H^2 + v_\Delta^2} \approx 246$~GeV, $v_H$ and $v_\Delta$ being the VEVs of the doublet and triplet scalar fields, respectively. The mass splitting between the dark lepton fields in the doublet comes out to be
\begin{align}
  \delta = |m_{\psi^0} - m_{\psi^+}| = y_\Delta v_\Delta
  \label{eq:deltam}
\end{align}

A discussion is in order here on the existing bounds that constrain the model parameters. The couplings which will play a significant role in the DM dynamics are the Higgs-portal coupling $\lambda_{h\phi}$ and the Yukawa couplings $y_\ell, \ell = e, \mu, \tau$. To put the bounds from the direct detection searches at bay, we have considered $\lambda_{h\phi} \lesssim 10^{-4}$, which also takes care of the invisible decay measurement. On the other hand, the muon $g-2$ measurement puts a bound on the value of the Yukawa couplings of light leptons. Although it need not be so stringent, we still take a conservative choice of values at $y_e \sim y_{\mu} \lesssim 10^{-9}$. This leaves the third-generation Yukawa coupling $y_\tau$ to be the only one free from experimental constraints. However, one must note that, to keep our model in the perturbative regime, we must have $y_\tau \leq 4\pi$. The measurement of the $\rho$ parameter~\cite{Tanabashi:2018oca} puts a bound on the value of the triplet VEV: $v_\Delta \sim 3$~GeV. So for a safe choice we have taken the value of the mass splitting $\delta \leq 10$~GeV.

%
\section{Dark matter phenomenology}
\label{sec:dmpheno}

\subsection{Formalism}
\label{sec:formalism}
In the proposed model, DM number changing processes are (i) pair annihilation ($\phi \phi \to \rm SM\,SM$), (ii) coannihilation ($\phi \psi^{\pm0} \to \rm SM\,SM$) and (iii) mediator annihilation ($\psi^{\pm0} \psi^{\mp0} \to \rm SM\,SM$). The choice of parameters will determine the relative contribution of these processes towards the relic density as we discuss in the following sections. In agreement with the common assumption of thermal freeze-out, the dark sector particles are in equilibrium with the thermal bath in the early Universe. At the same time, they are also in chemical equilibrium with each other, due to substantial interaction strength between themselves. Keeping all these in mind, one can write the Boltzmann equation as follows~\cite{Griest:1990kh}
\begin{gather}
  \frac{dn}{dt}
  =
  - 3\,H\,n - \ev{\sigma_{\rm eff}\,v}\,(n^2-n^2_{\rm eq}) \,,
\end{gather}
where $n$ and $n_{\rm eq}$ are the DM number density and the equilibrium number density respectively. Now, the effective velocity averaged annihilation cross section $\langle \sigma_{\rm eff}\,v\rangle$ specific to this model can be written as
\begin{align}
  \langle\sigma_{\rm eff}\,v\rangle
  =&
  \frac{1}{[g_\phi + \bar g_{\psi^0} +\bar g_{\psi^\pm}]^2}\,
  \big[ g_\phi^2\,\langle\sigma_{\phi \phi \to \rm SM\,SM}v\rangle
  \nl
  +& g_\phi\,\bar g_{\psi^0}\,\langle\sigma_{\phi \psi^0 \to \rm SM\,SM}v\rangle
  + g_\phi\,\bar g_{\psi^\pm}\,\langle\sigma_{\phi \psi^\pm \to \rm SM\,SM}v\rangle
  \nl
  +& \bar g_{\psi^0}^2\,\langle \sigma_{\psi^0 \psi^0 \to \rm SM\,SM}v\rangle
  + \bar g_{\psi^\pm}^2 \langle\sigma_{\psi^\pm \psi^\mp \to \rm SM\,SM}v\rangle
  \nl
  +& \bar g_{\psi^\pm}\bar g_{\psi^0} \langle\sigma_{\psi^\pm \psi^0 \to \rm SM\,SM}v\rangle \big]
\label{BEQ}
\end{align}
where
\begin{gather}\label{g_BEQ}
\begin{gathered}
  \bar g_{\psi^0} = g_{\psi^0}\,(1 + \Delta m_0)^{3/2}\,\exp[-x\,\Delta m_0],
  \\
  \bar g_{\psi^\pm} = g_{\psi^\pm}\,(1 + \Delta m_{\rm ch})^{3/2}\,\exp[-x\,\Delta m_{\rm ch}] \,.
\end{gathered}
\end{gather}
In the expressions above, $g_\phi = 1, g_{\psi^0} = g_{\psi^\pm} = 2$ are the internal degrees of freedom and $x=m_\phi/T$. $\Delta m$'s are dimensionless mass splitting parameters defined as
\begin{gather}
  \begin{gathered}
    \Delta m_0 = (m_{\psi^0} - m_\phi)/m_\phi\,,
    \\
    \Delta m_{\rm ch} = (m_{\psi^\pm} - m_\phi)/m_\phi\,.
  \end{gathered}
  \label{eq:mass_ratio}
\end{gather}

As previously mentioned, the pair annihilation and coannihilation channels predominantly control the DM freeze-out. The mass splitting between $\phi$ and other dark sector particles and the Yukawa couplings mainly determines the contribution of these processes towards the total DM annihilation cross section. These mass splittings play a very important role, especially for the coannihilation and mediator-annihilation processes, as the Boltzmann factor in \cref{BEQ} gives rise to a significantly increased annihilation cross section for small values of $\Delta m$'s.

Before we go into the details of the freeze-out mechanisms, let us discuss the parameters used in the analysis. Since the mass splitting parameters defined in \cref{eq:mass_ratio} between DM and other dark sector particles play an important role in freeze-out of $\phi$, we will use them as independent parameters along with DM mass. As discussed in the previous section, the only important Yukawa coupling here will be $y_\tau$, which couples $\phi$ to the third-generation $SU(2)_L$ lepton doublet and the new fermionic doublet $\varPsi$. In summary, we take the following values of the parameters throughout our analysis:
\begin{align}
\boxed{
\begin{aligned}
  \text{Free parameters: }&\quad m_\phi,\ \Delta m_0,\ \Delta m_{\rm ch},\ y_\tau\,,
  \\
  \text{Fixed parameters: }&\ \lambda_{h\phi}=10^{-4}, y_e \sim y_{\mu}\sim 10^{-9}\,.
\end{aligned}
}
\label{eq:param}
\end{align}

\subsection{Analysis and observations}

\subsubsection{Relic density}
\label{sec:relic}
In addition to the Higgs-portal annihilation channels of scalar singlet DM, the present model introduces a Yukawa interaction between the dark sector particles and SM. Unlike the Higgs-DM quartic coupling ($\lambda_{h\phi}$), the new Yukawa coupling ($y_{\tau}$) is unconstrained except for the perturbative limits. This provides an excellent tool to explain the relic density for a wide parameter space even with negligible $\lambda_{h\phi}$, which, in turn, alleviates the direct search bounds. We have performed DM analysis using {\texttt micrOMEGAs}~\cite{Belanger:2014vza}.

\begin{figure}[!ht]
  \centering
  \includegraphics[width=0.22\linewidth]{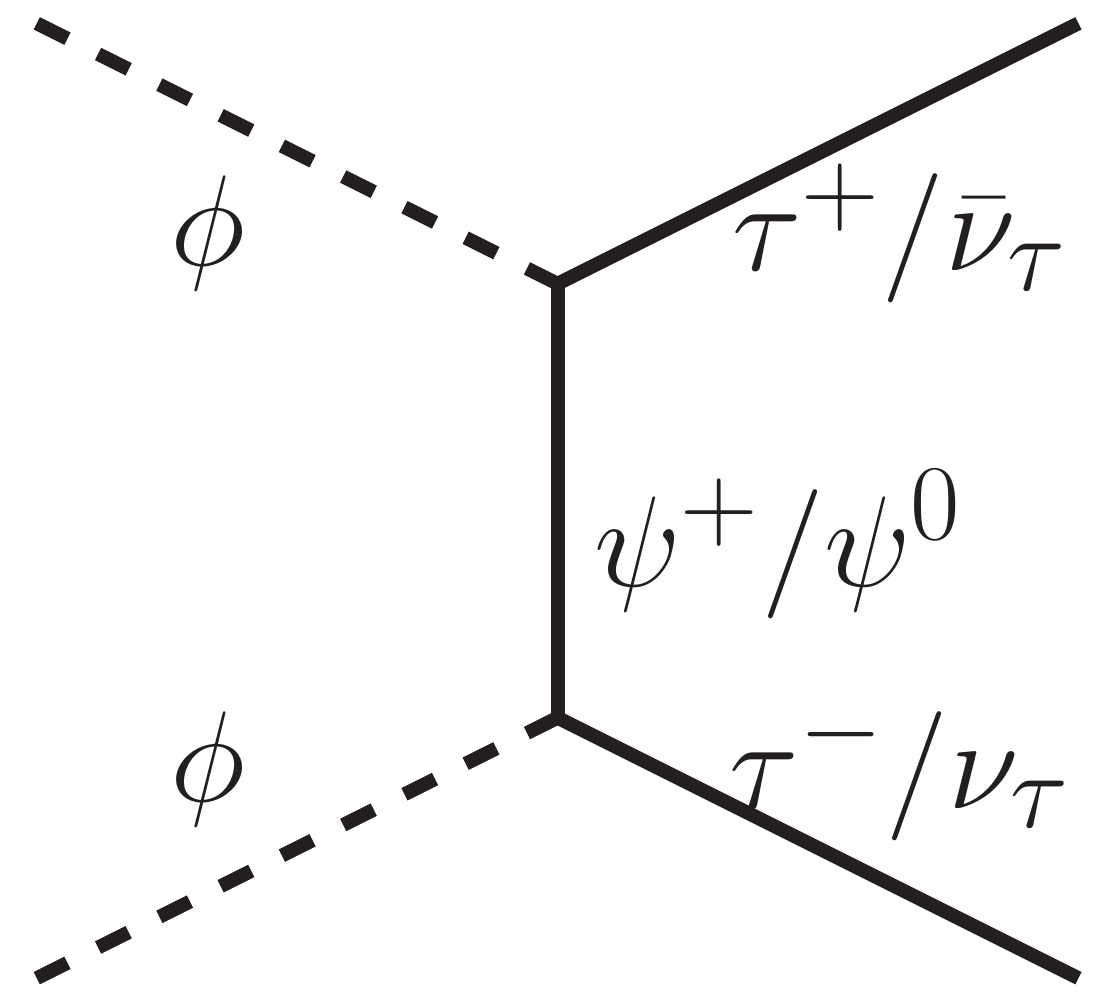}
  \caption{Feynman diagrams depicting the pair annihilation channels of $\phi$ }
  \label{ann}
\end{figure}

\begin{figure}[!ht]
  \centering
  \includegraphics[height=0.18\linewidth]{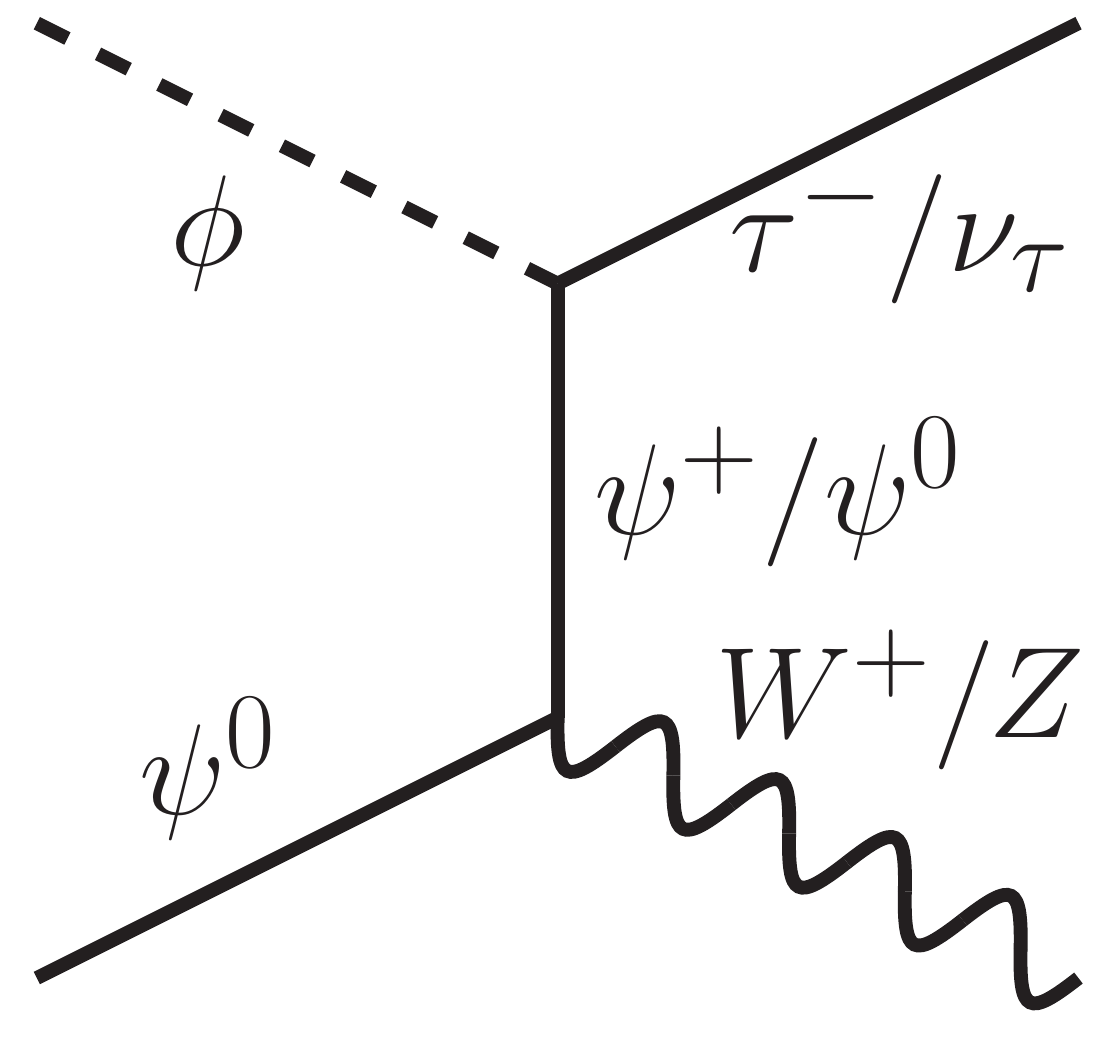}
  \includegraphics[height=0.18\linewidth]{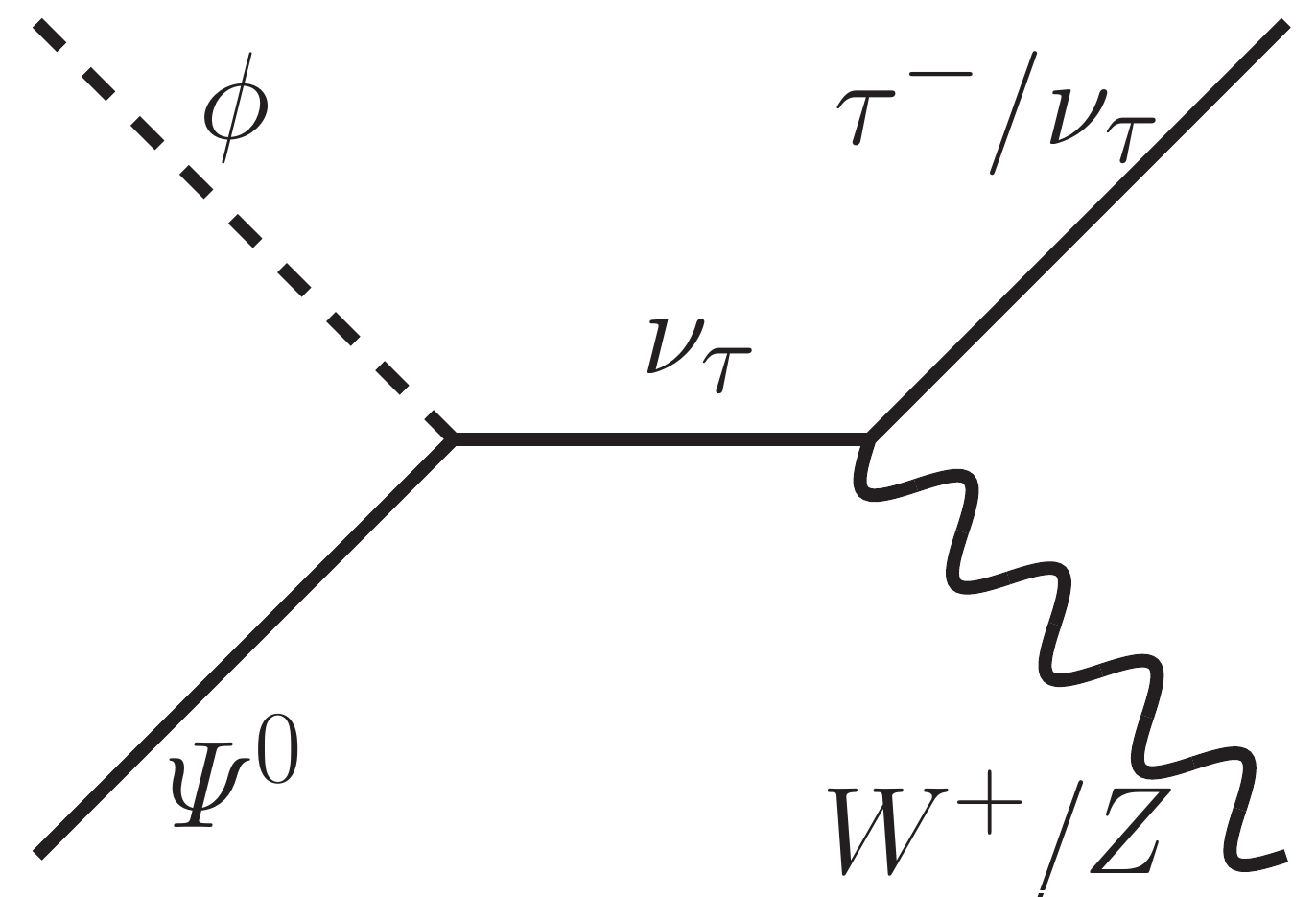}
  \includegraphics[height=0.18\linewidth]{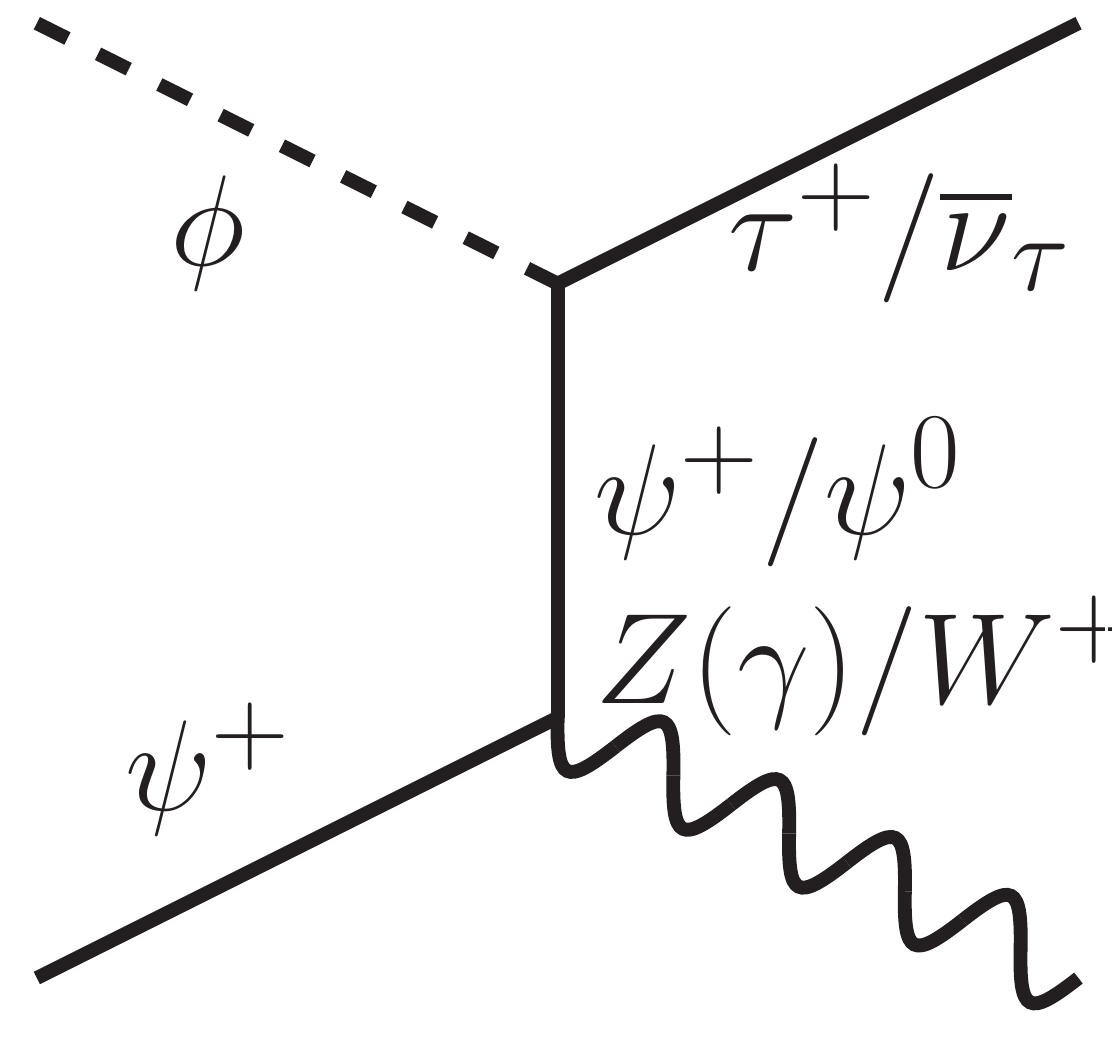}
  \includegraphics[height=0.18\linewidth]{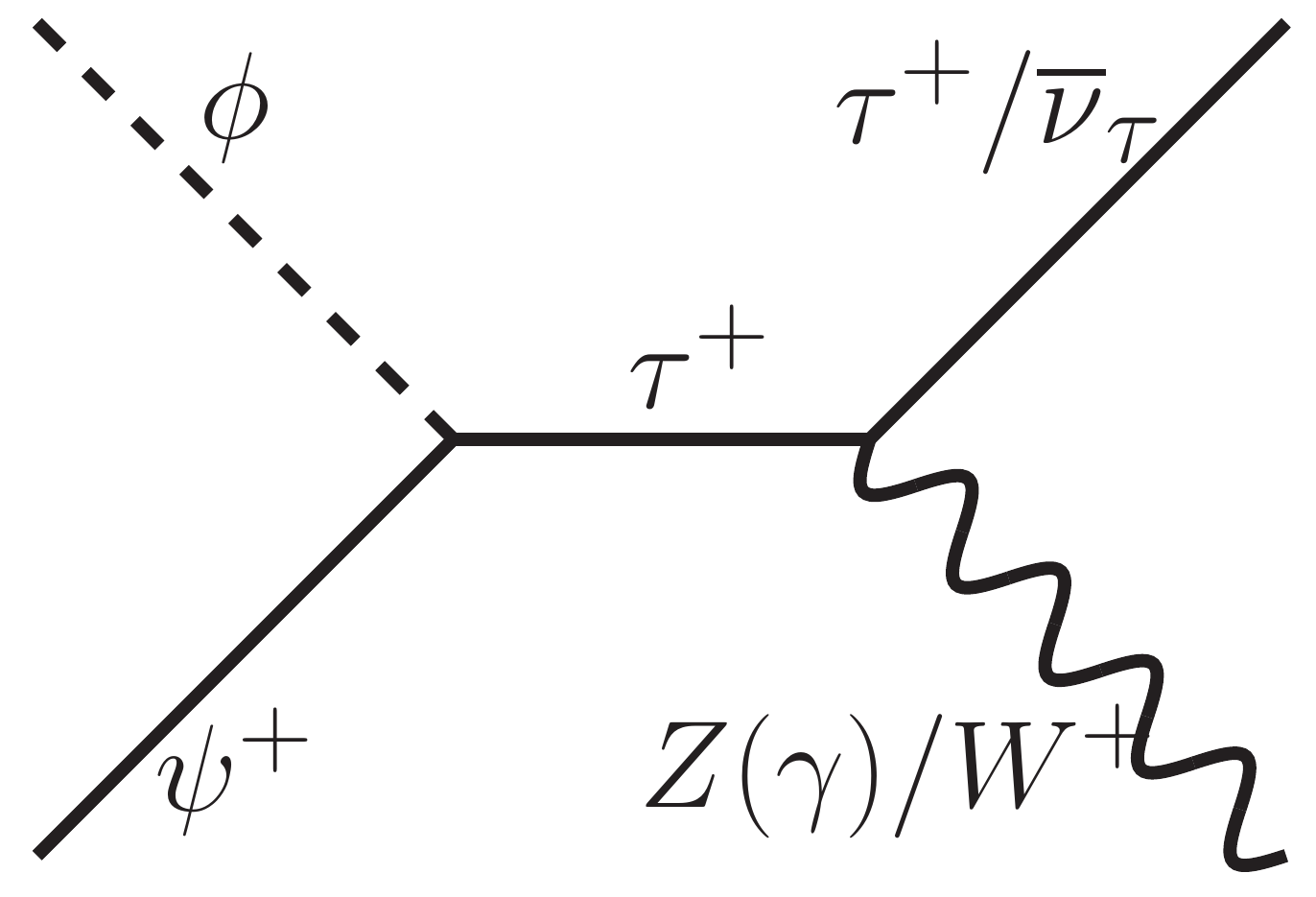}
  \caption{Feynman diagrams depicting the coannihilation channels of $\phi$ }
  \label{coann}
\end{figure}

\begin{figure}[!ht]
  \centering
  \includegraphics[height=0.18\linewidth]{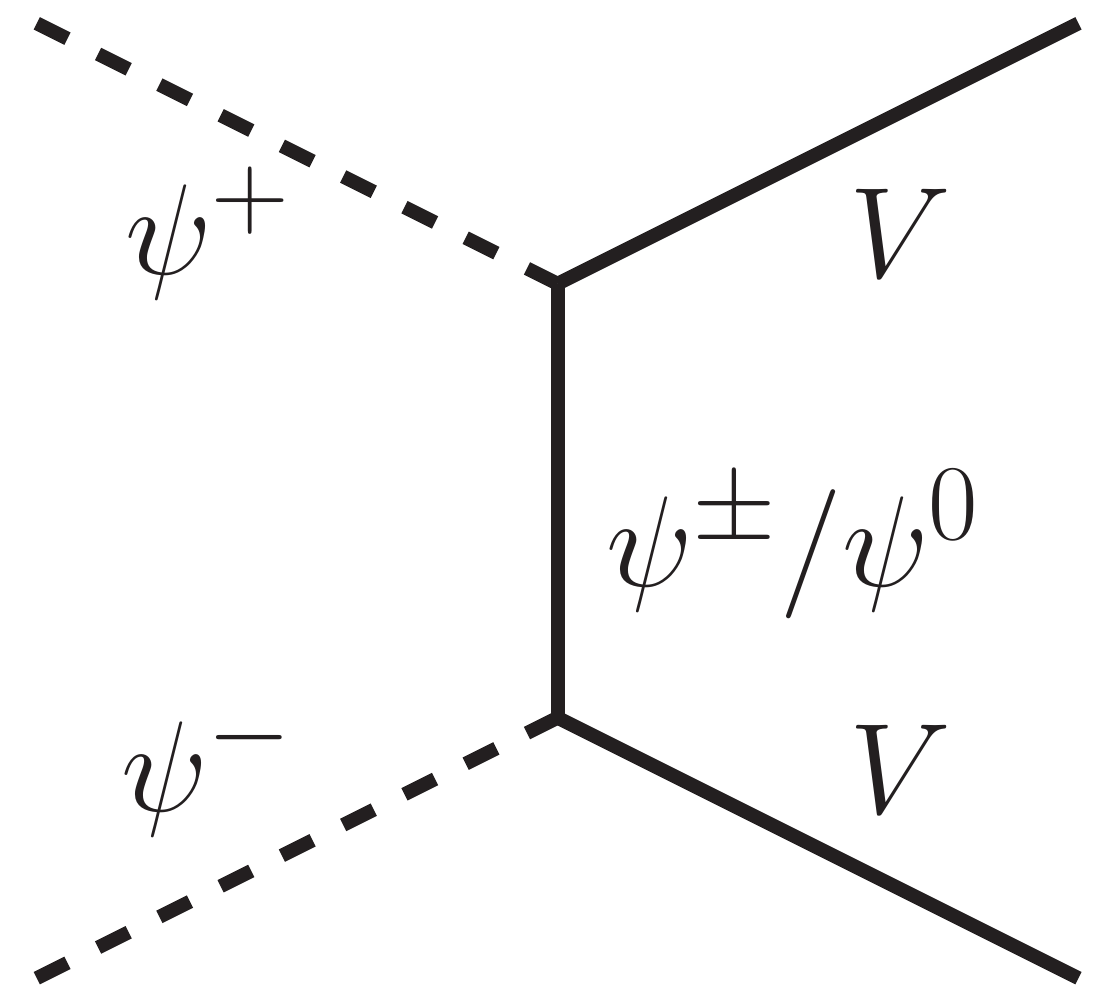}
  \hskip1cm
  \includegraphics[height=0.18\linewidth]{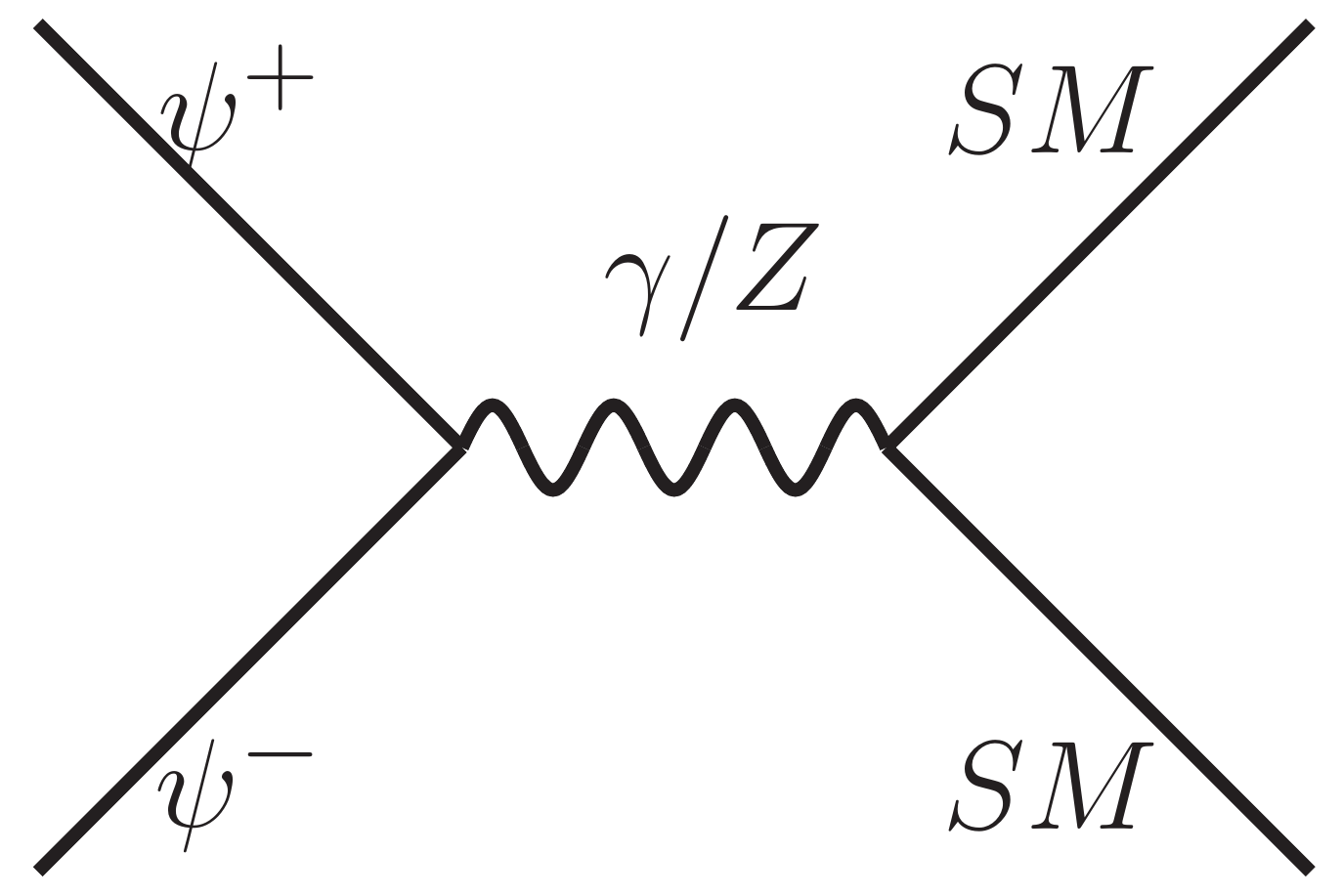}
  \hskip1cm
  \includegraphics[height=0.18\linewidth]{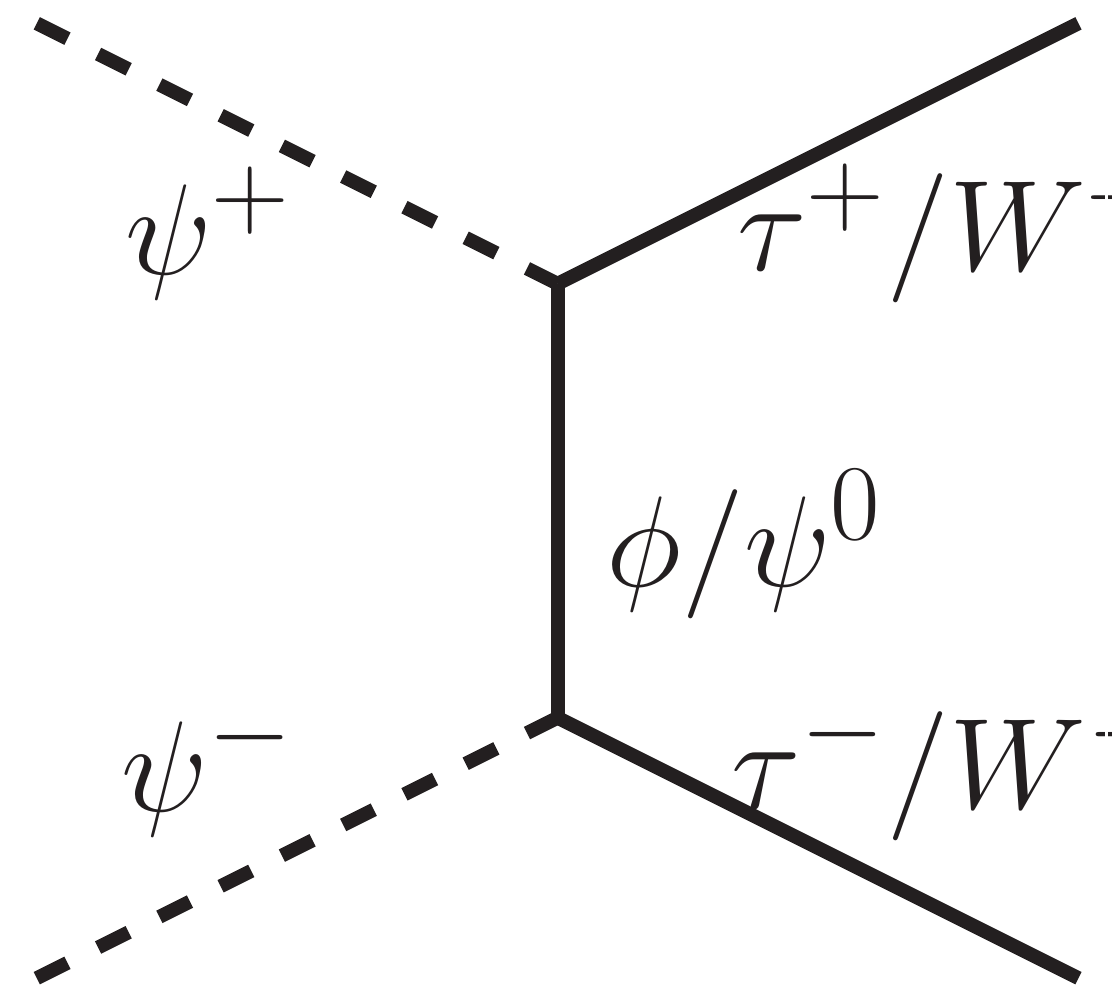}
  \\
  \includegraphics[height=0.18\linewidth]{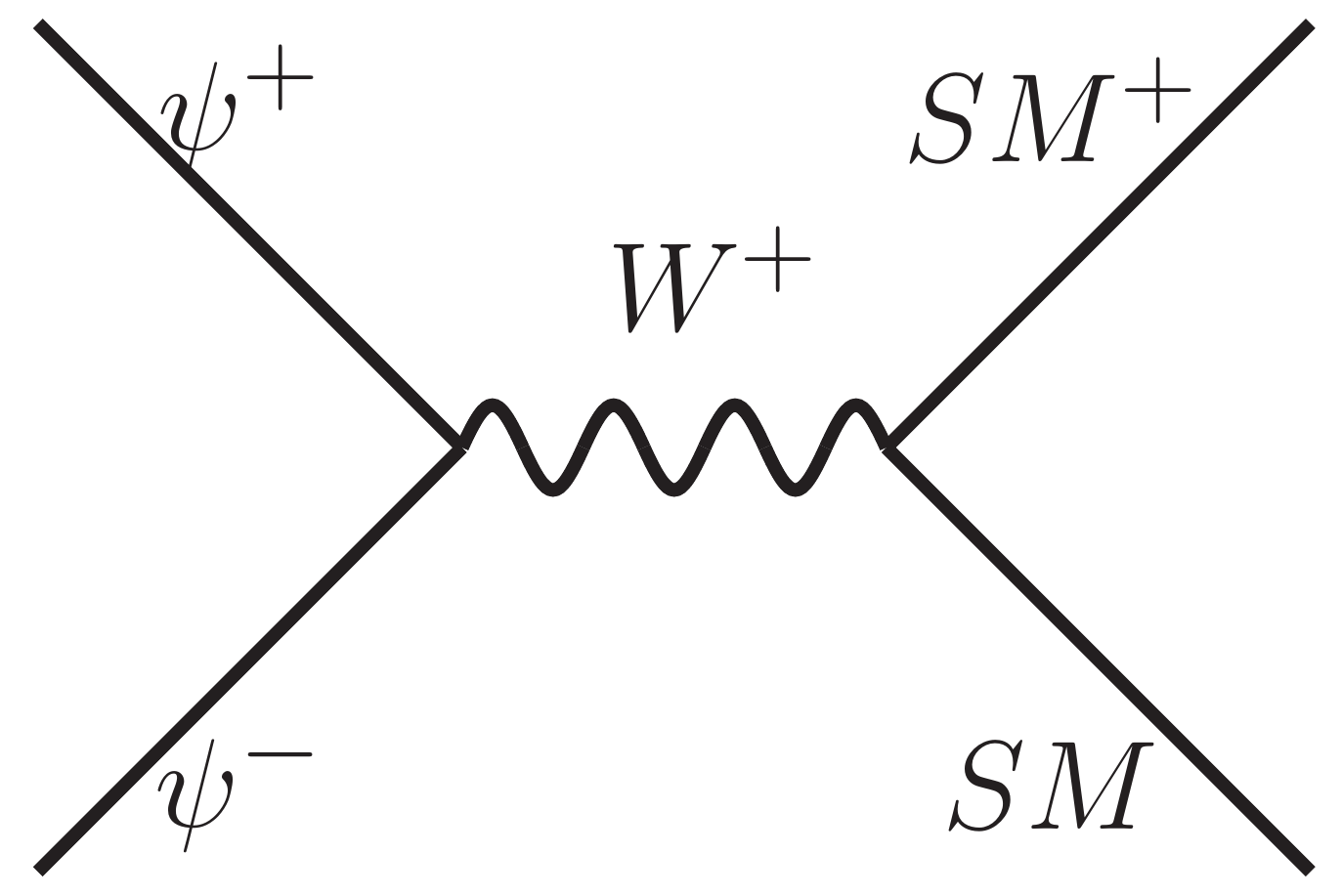}
  \hskip1cm
  \includegraphics[height=0.18\linewidth]{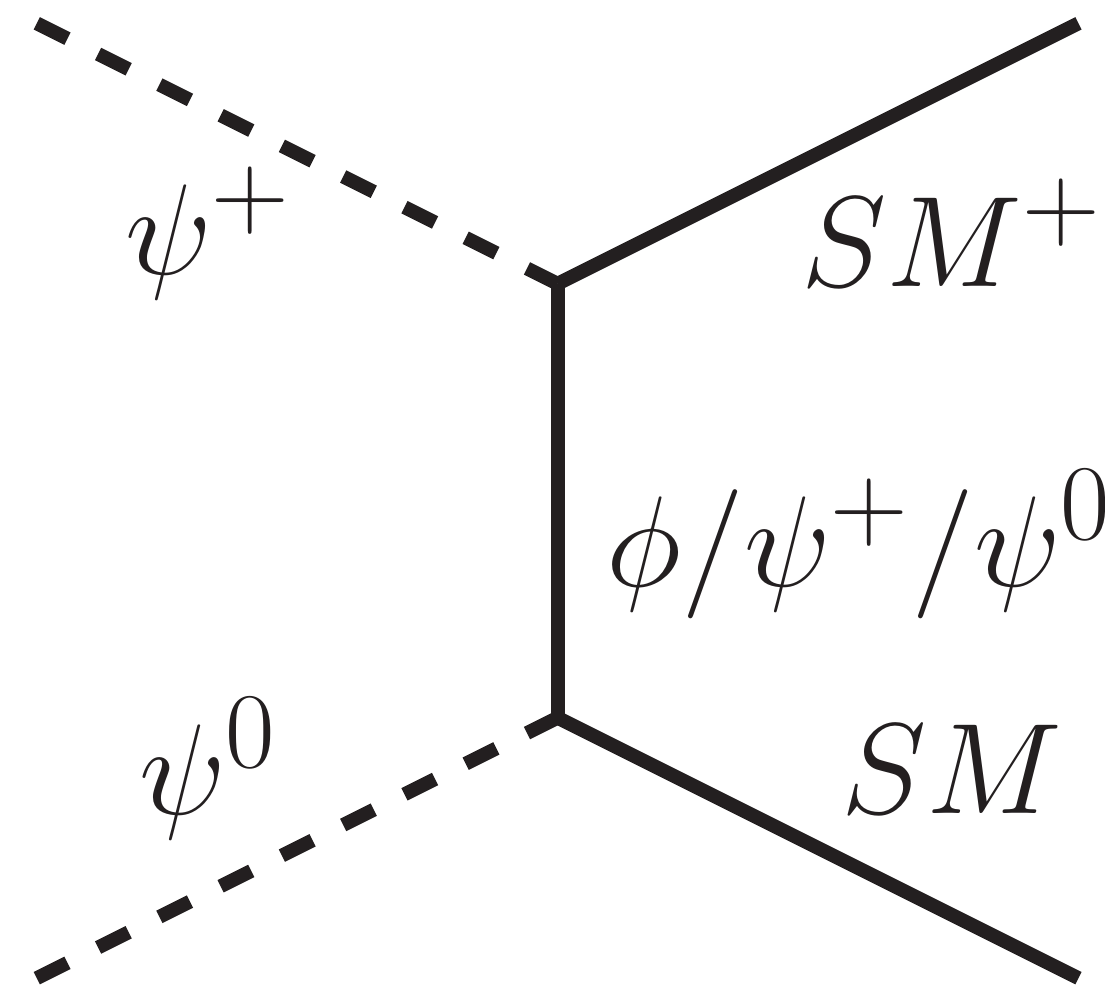}
  \hskip1cm
  \includegraphics[height=0.18\linewidth]{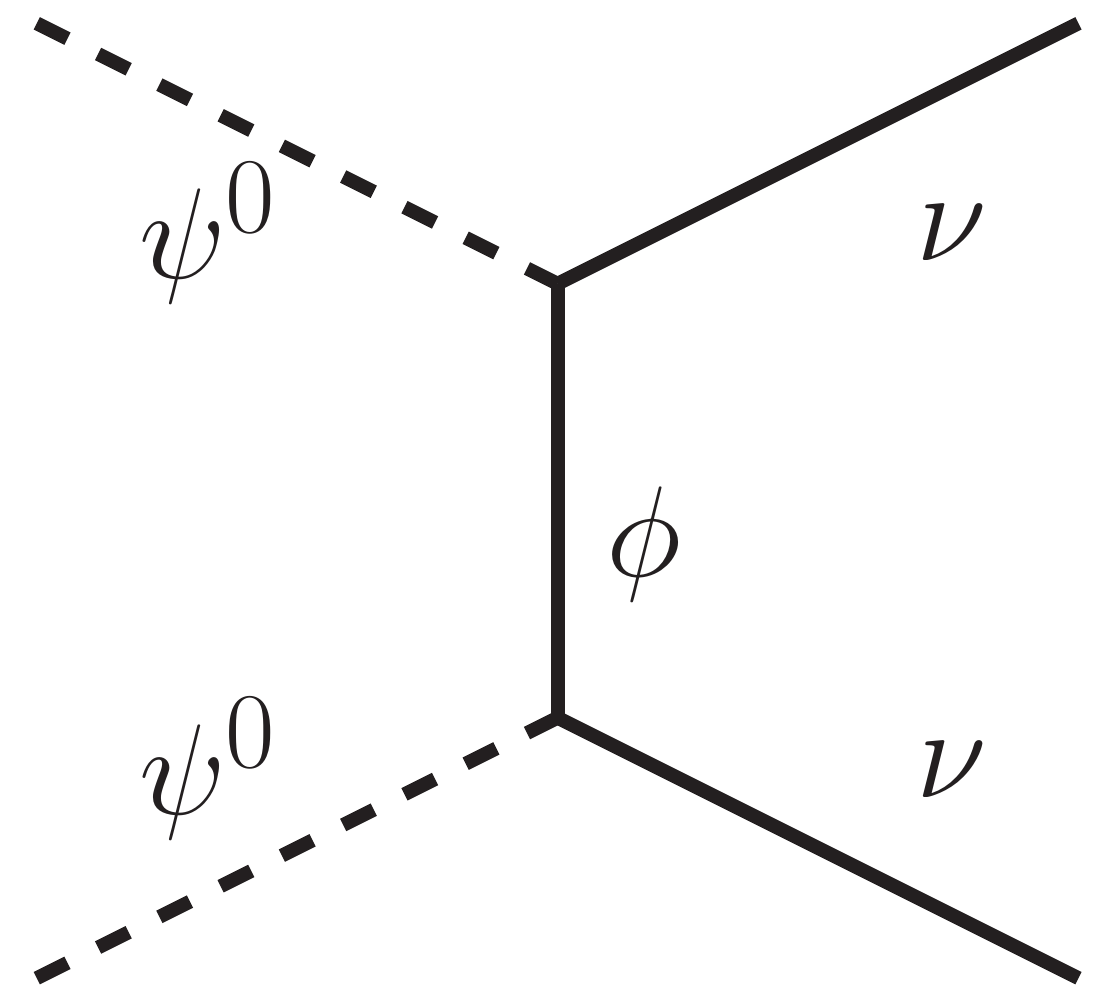}
  \caption{Feynman diagrams depicting the mediator annihilation channels }
  \label{conv}
\end{figure}

Because of minuscule $\lambda_{h\phi}$, the Higgs-portal annihilation channels have a negligible contribution towards DM relic density, and, hence, we will focus on the newly introduced channels only. All these annihilation channels can be broadly classified into three categories:
\begin{itemize}
\item pair annihilation ($\phi \phi \to \rm SM\,SM$) (\cref{ann}),
\item Coannihilation ($\phi \psi^{\pm0} \to \rm SM\,SM$) (\cref{coann}), and
\item Mediator annihilation ($\psi^{\pm0} \psi^{\mp0} \to \rm SM\,SM$) (\cref{conv}).
\end{itemize}
All these categories can coexist or supersede each other, depending on the choice of parameters. The coannihilation and mediator annihilation processes become efficient only for small mass splittings between the DM and the dark sector particles. This is due to the exponential factor sitting in the expression for the respective $\langle \sigma_{\rm eff}\,v \rangle$ [see \cref{BEQ}]. Moreover, in this model, the cross sections of these processes are larger than the DM pair annihilation cross section, which is the precise reason why these processes significantly reduce DM relic density~\cite{Boehm:1999bj}. In the following analysis, we will see that, over the entire parameter space, these DM number changing processes supersede the pair annihilation contribution to the relic density by 1 or 2 orders whenever $\delta m\ =\ m_{\psi^{\pm}}(m_{\psi^0})-m_{\phi}$ are small.    

Similar to $\delta m$'s, the two $\Delta m$'s also account for the strength of the coannihilation and mediator annihilation channels for two heavier dark sector particles $\psi^0$ and $\psi^{\pm}$. Apart from this, the Yukawa coupling $y_{\tau}$ also plays a significant role. In this context, it is worth noting that, for the pair annihilation channels in \cref{ann}, the cross section depends on $y_{\tau}^4$, while, for coannihilation channels, it is only a $y_{\tau}^2$ dependence, and the mediator annihilation channels, being mostly gauge mediated, have very little dependence on $y_{\tau}$. One can easily verify this from the analytical expressions in \cref{eq:eq1,eq:eq2}.

Since all the above three categories of DM annihilation can coexist in the parameter space, it would be interesting to identify the limiting cases where the transition from one category to another is perceivable. It is worth mentioning here that the interaction channels between $\phi$ and two other dark sector particles are exactly similar (see \cref{ann,coann}), so in the degenerate mass limit $m_{\psi^0} \sim m_{\psi^{\pm}}$ their contribution will be the same. However, here the extra scalar triplet allows a finite mass difference between $\psi^{\pm}$ and $\psi^0$. This mass splitting helps to identify the dominant dark lepton in the number-changing DM coannihilation and mediator annihilation channels.


\begin{figure}[!ht]
    \centering
    \begin{minipage}{0.48\linewidth}
        \centering
        \includegraphics[trim=0 0 0 -25,clip,width=\linewidth]{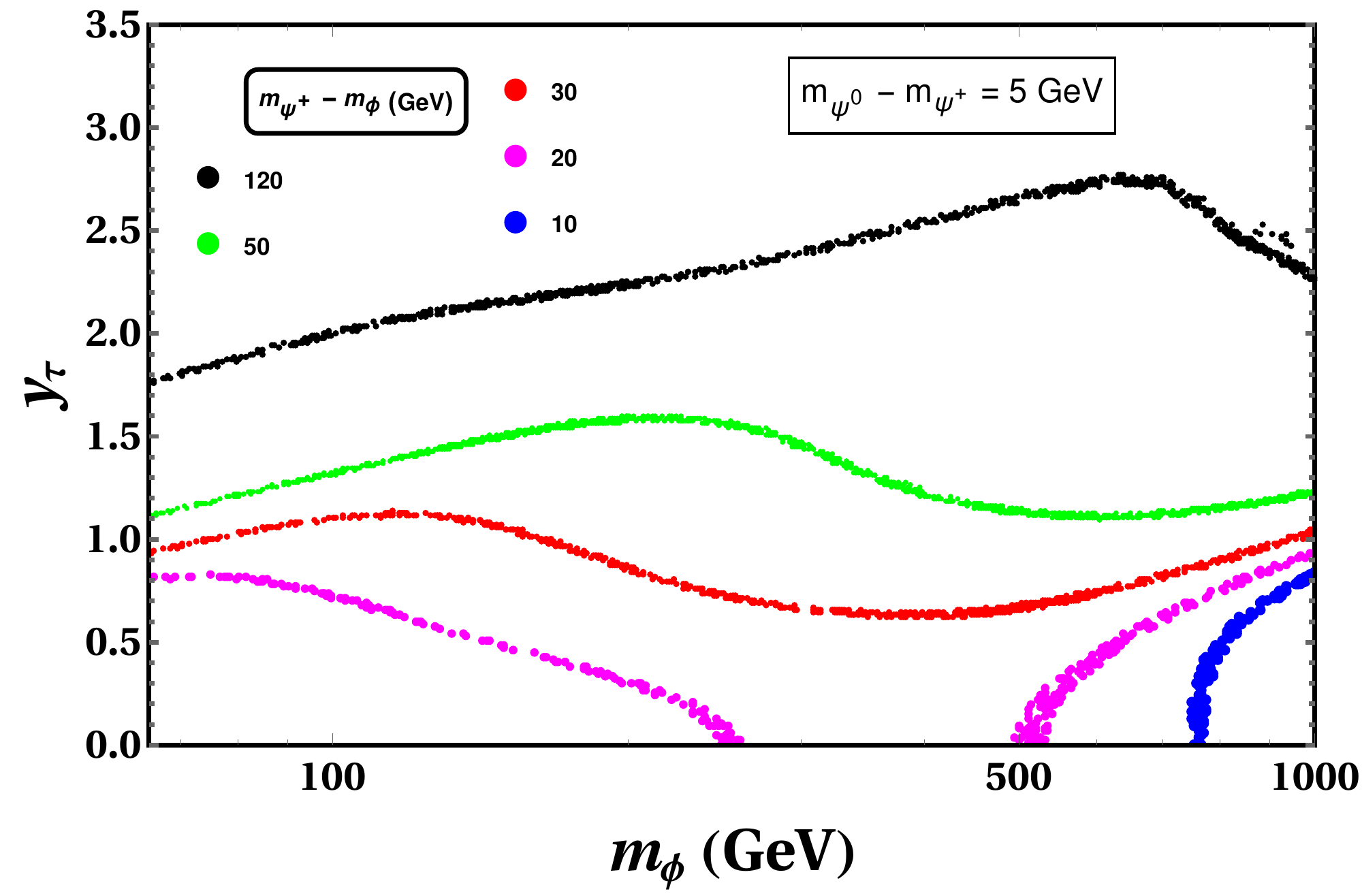}
        \caption{$y_\tau$ vs $m_\phi$ variation showing the transition from mediator annihilation to coannihilation and pair annihilation regime for different values of $m_{\psi^{\pm}}(m_{\psi^0})-m_{\phi}$ in GeV.}
        \label{ann_plots}
    \end{minipage}%
    \hfill
    \begin{minipage}{0.48\linewidth}
        \centering
        \includegraphics[width=\linewidth]{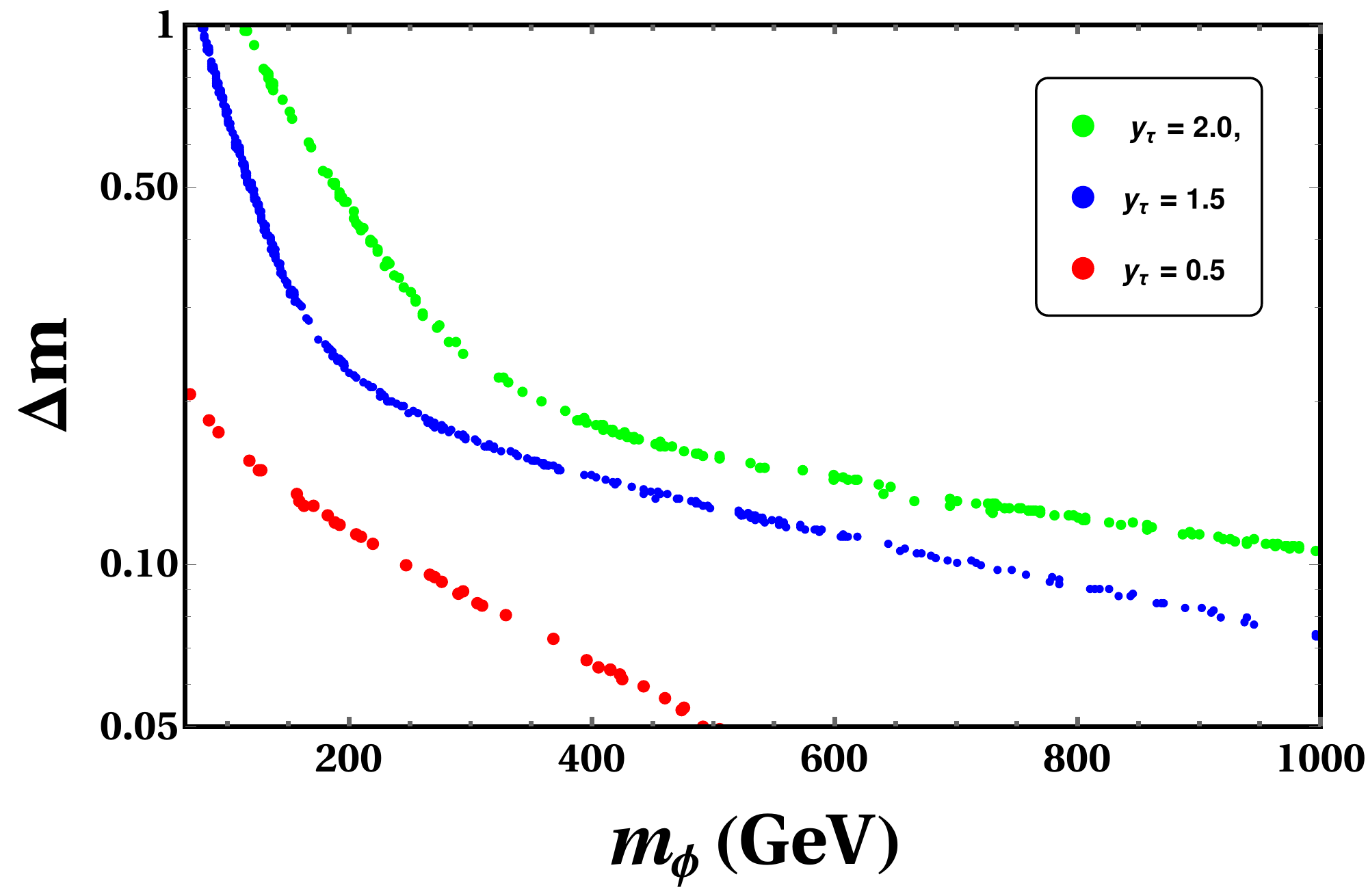}
        \caption{$\Delta m~(=\Delta m_{\rm ch}=\Delta m_0)$ vs $m_\phi$ variation for fixed values of $y_{\tau}$. All points satisfy relic density.}
        \label{coann_plot_1}
    \end{minipage}
\end{figure}

In \cref{ann_plots}, the transition between the above-mentioned categories is depicted in the $y_{\tau}$ vs $m_{\phi}$ plane for some fixed values of $\delta m=m_{\psi^{\pm}}(m_{\psi^0})-m_{\phi}$. For pair annihilation, $\langle \sigma_{\rm eff}\,v\rangle$ has $m_{\psi^{\pm}}$ dependence only in the $t$-channel propagator, but, for coannihilation, $m_{\psi^{\pm}}$ appears in the propagator and the initial state along with the Boltzmann factor [\cref{g_BEQ}]. The functional dependence on the Boltzmann factor is even stronger for the mediator annihilation channels, which implies that for small mass splitting between the dark leptons and $\phi$, mediator annihilation channels will deplete the DM number density most efficiently. For slightly larger splittings, the coannihilation channels take over. A large enough value of $\delta m$, however, makes the coannihilation processes negligible due to substantially large Boltzmann suppression, and so the pair annihilation predominantly dictates DM annihilation. This is clear from \cref{ann_plots}, where the blue line depicts the relic density allowed $y_{\tau}$ vs $m_{\phi}$ correlation for the mediator annihilation dominated channels. As required for this feature, $m_{\psi^{\pm}}-m_{\phi}$ is small ($\lesssim$ 10 GeV). In magenta and red lines, however, coannihilation channels dominate due to comparatively larger $\delta m$. It is to be noted here that $m_{\psi^{0}}-m_{\psi^{\pm}}$ is fixed at 5 GeV in this plot, which implies that $\phi-\psi^{\pm}$ coannihilation is stronger than the neutral dark lepton counterpart. As the DM-dark lepton splittings increase, for a fixed value of $m_{\phi}$,  $\langle \sigma_{\rm eff}\,v \rangle$ becomes further Boltzmann suppressed, and, hence, larger $y_{\tau}$ is required to keep it within observed limits. For even larger $\delta m$, e.g., green and black lines, coannihilation contribution becomes more suppressed and pair annihilation becomes dominant. In order to obtain a sufficient annihilation cross section for the right relic, this implies a fairly large $y_{\tau}$.

In \cref{coann_plot_1}, the variation of $\Delta m$ ($\Delta m_{\rm ch}=\Delta m_0=\Delta m$) vs $m_\phi$ is plotted for all points satisfying the right relic. In the smaller $m_{\phi}$ region, the major share in relic density comes from $\phi \psi^{{\pm}(0)}$ coannihilation channels. Larger values of $\Delta m$ cause Boltzmann suppression in $\ev{\sigma_{\rm eff}\,v}$, which, in turn, is compensated by larger values of the coupling, as clearly seen in the plot. However, this feature is more prominent for smaller values of $m_\phi$. This is due to the fact that, for a fixed $\Delta m$, larger values of $m_\phi$ imply large $m_{\psi^{{\pm}(0)}}-m_\phi$, which rules out any substantial effect from the coannihilation channels. In fact, for larger $m_{\phi}$, pair annihilation channels begin to dominate the total annihilation. Propagator suppression causes $y_{\tau}$ to decrease towards higher $m_{\phi}$, but here the decrease is at a much slower rate than the smaller $m_{\phi}$ regime.

\begin{figure}[!ht]
    \centering
    \begin{minipage}{0.48\linewidth}
        \centering
        \includegraphics[trim=0 0 0 -15,clip,width=\linewidth]{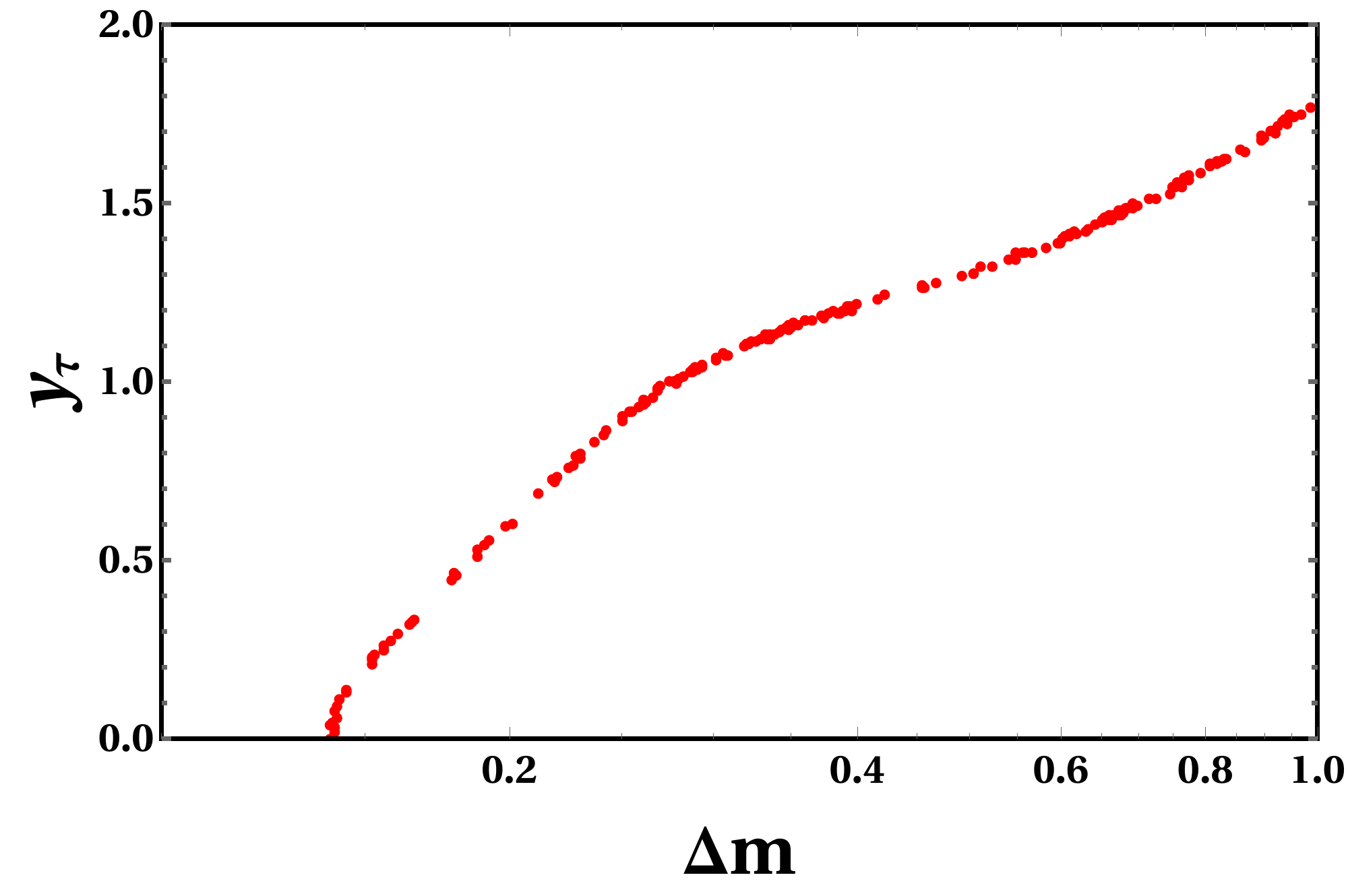}
        \caption{Variation of relic density allowed points in $y_\tau$ vs $\Delta m~(=\Delta m_{\rm ch}=\Delta m_0)$ plane for $m_\phi$=100~GeV. The coupling gradually increases for larger $\Delta m$, compensating for larger Boltzmann suppression in $\langle \sigma_{\rm eff}\,v \rangle$.}
        \label{coann_plot_3}
    \end{minipage}%
    \hfill
    \begin{minipage}{0.48\linewidth}
        \centering
        \includegraphics[width=\linewidth]{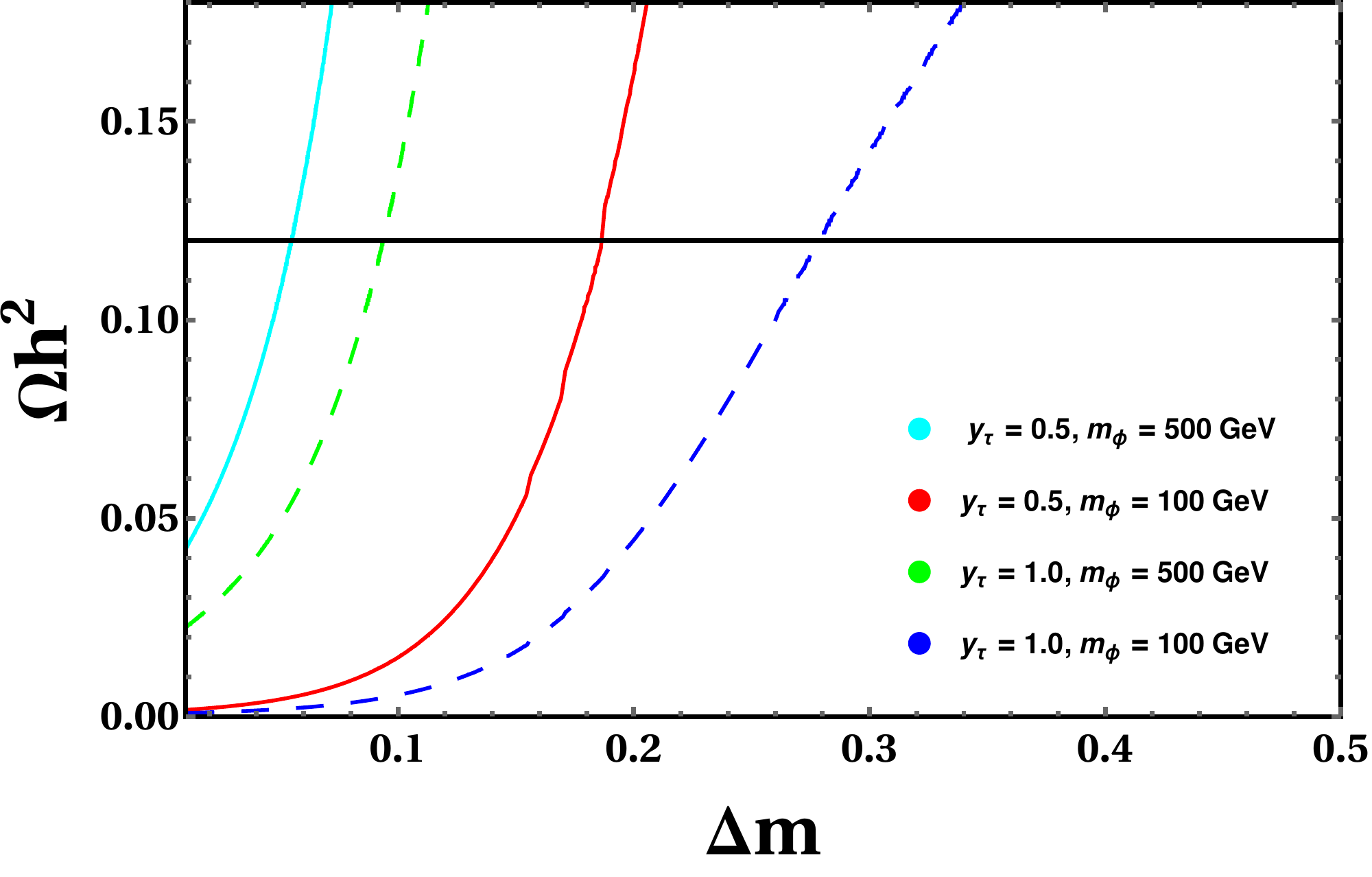}
        \caption{Variation of relic density with $\Delta m~(=\Delta m_{\rm ch}=\Delta m_0)$ in the coannihilation regime for fixed values $y_{\tau}$ and $m_{\phi}$. The black line represents the right relic density at $\Omega h^2=0.1215$.}
        \label{coann_plot_4}
    \end{minipage}
\end{figure}

As mentioned previously, for very small values of $\Delta m$'s, mediator annihilation or the freeze-out of $\psi^0$ and $\psi^{\pm}$ contributes to the relic density of $\phi$. As observed from \cref{BEQ}, the dependence on $\Delta m$'s is stronger in the Boltzmann factor of $\langle \sigma_{\rm eff}\,v \rangle$ than coannihilation, and this leads to the fact that, for very small values of $\Delta m$'s, the mediator-driven annihilations almost entirely dominate the total DM annihilation. It is also worth noting that, being mostly gauge mediated, these channels substantially contribute to the DM annihilation even for very small values of $y_{\tau}$. For our choice of parameters, we have observed that mediator annihilation is effective for $(m_{\psi^{\pm(0)}}-m_\phi) \lesssim 10$~GeV, and then the coannihilation processes take over. This feature is clear from \cref{coann_plot_3}, where we can see that there is no relic density allowed $y_{\tau}$ for $\Delta m \lesssim$ 0.1 GeV. Beyond this range, as $\Delta m$ increases, the required coupling also increases gradually to compensate for the Boltzmann suppression. $m_\phi$ is fixed at 100~GeV. Contributions from both the dark leptons are equal in the total annihilation cross section of DM, since they are considered degenerate.

In \cref{coann_plot_4}, the variation of the relic density is plotted with $\Delta m$ for some fixed Yukawa couplings and DM mass. As already argued, for a fixed $m_{\phi}$, larger coupling corresponds to larger $\Delta m$ due to Boltzmann suppression in $\ev{\sigma_{\rm eff}\,v}$ as well as larger mass suppression of $\psi^{{\pm}(0)}$ in the $t$-channel propagator of the coannihilation channels. This explains the shift along the $X$ axis from the red to the blue line where $m_\phi$ is 100~GeV and the Yukawa coupling $y_\tau$ varies from 0.5 to 1.0. We see the same trend for the cyan and green lines, but the amount of shift is relatively less, because, in this case, $m_\phi$ is larger (500~GeV), which automatically implies a fairly large splitting between $m_{\psi}^0/m_{\psi^{\pm}}$ and $m_\phi$, and, consequently, the coannihilation effect is not so prominent. We can argue that, for a fixed value of $y_\tau$, larger DM mass obtains the correct relic density with a relatively smaller $\Delta m$; hence, the red line with $m_\phi$=100~GeV shifts left towards the cyan line with the same $y_\tau$ but larger $m_\phi$ = 500~GeV. The same logic applies to the shift between the blue and the green line. This trend also agrees with \cref{coann_plot_1}. As expected, very small values of $\Delta m$ give an underabundance for the choice of parameters due to a fairly large increase in the Boltzmann factor of \cref{g_BEQ}.

\begin{figure}[!ht]
  \centering
  \includegraphics[width=0.48\linewidth]{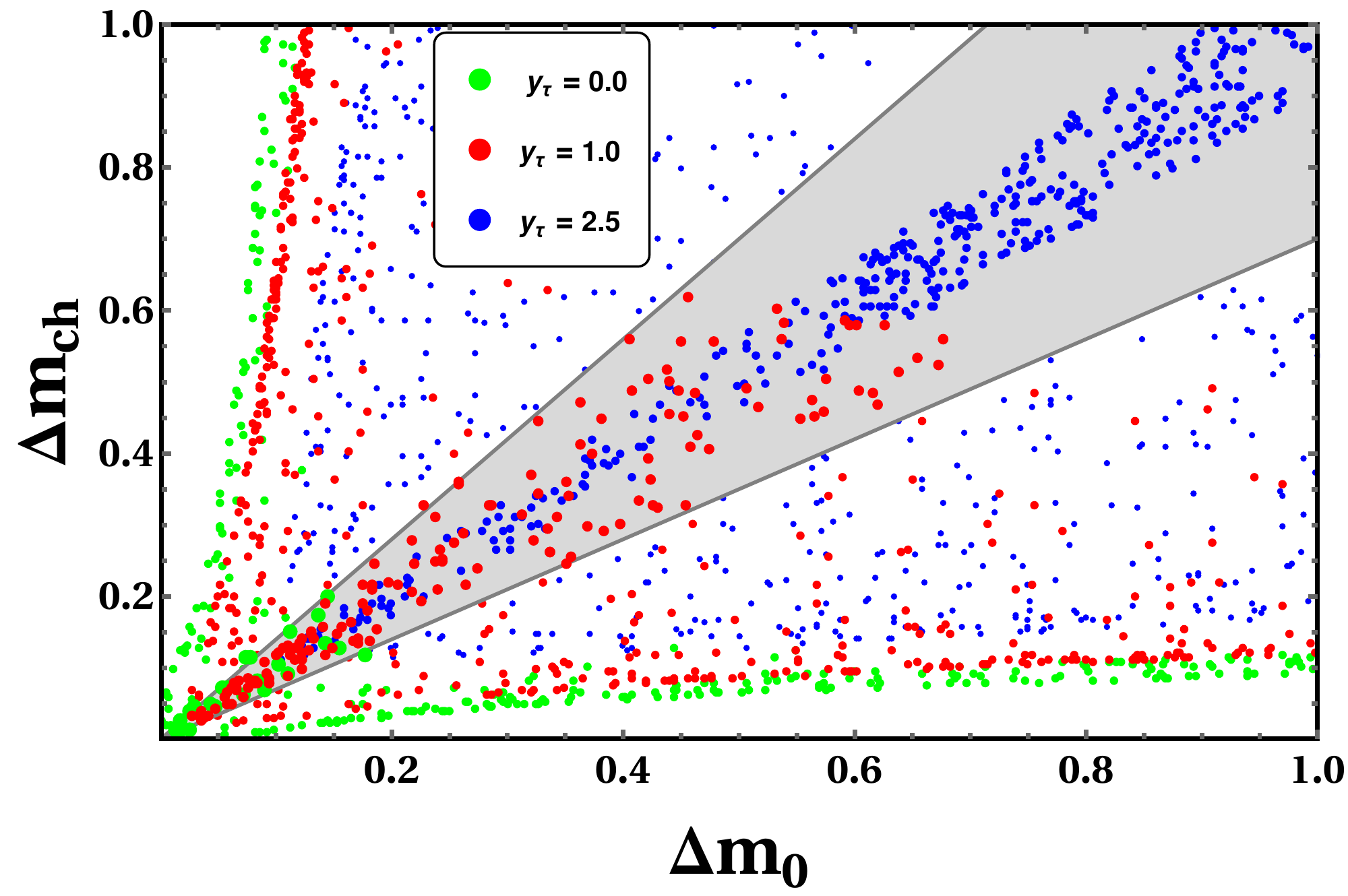}
  \caption{Correlation between $\Delta m_0$ and $\Delta m_{\rm ch}$ for different values of $y_\tau$. $m_\phi$ varies from 65~GeV to 1 TeV. The three colours indicate three different couplings and distinguish the regions of parameter space where different DM number-changing processes dictate the relic density. The shaded region is allowed by the constraint on mass splitting between the dark leptons.}
  \label{coann_plot_5}
\end{figure}

Now we observe the possibility of coexistence of all the possible annihilation regimes by varying both $\Delta m_0$ and $\Delta m_{\rm ch}$ along with the DM mass. \cref{coann_plot_5} gives a correlation plot between $\Delta m_0$ and $\Delta m_{\rm ch}$ at different values of $y_{\tau}$. For very small values of the coupling, mediator annihilation and coannihilation are dominant over pair annihilation depending on $\Delta m$, because, as seen from \cref{eq:eq1,eq:eq2}, annihilation cross sections are proportional to $y_\tau^4$ while for coannihilation it is only $y_\tau^2$ and mediator annihilations are mostly gauge mediated. The green points, corresponding to $y_\tau=0.0$, show that one can obtain the correct relic density only if one of the $\Delta m$'s is fairly small. Mediator annihilation is the only possibility here, because some of the channels in \cref{conv} are $y_{\tau}$ independent. The red shading, on the other hand, corresponding to a larger coupling ($y_\tau$=1.0), shows some scattered points in the allowed region. This implies that, with the increase in the coupling, the coannihilation and pair annihilation channels become stronger, and, to maintain the right relic, the mediator annihilation channels are automatically suppressed. This is an artifact of the larger $\Delta m$ values in the allowed region. The spread even increases for blue points, which corresponds to even larger $y_\tau=2.5$. In this region, due to such large coupling, pair annihilation is the most dominant, and other number-changing processes are supressed. Therefore, it becomes obvious that all three possible modes of annihilation can coexist in the present model for a wide parameter space where $m_\phi$ varies from a few GeV up to the TeV scale, and the coupling ranges from 0 to 3. However, the constraint on mass splitting between the dark leptons excludes the white region in the correlation plot. The introduction of a scalar triplet to generate the finite mass splitting plays the key role in relaxing the parameter space here , because in the degenerate limit the allowed grey shaded region narrows down to the line along $\Delta m_0 = \Delta m_{\rm ch}$. Nevertheless, even with the imposed constraint, the essential feature remains unchanged, where large Yukawa coupling and larger values of $\Delta m$ favour pair annihilation while the smaller values facilitate coannihilation and mediator annihilation.

\subsubsection{Direct and indirect detection}
\begin{description}[leftmargin=0pt,labelindent=0pt]
\item{{\bf Direct search prospect ---}}
As known from the direct detection of scalar DM models, DM undergoes elastic scattering with detector nuclei through Higgs mediation. The spin-independent scattering cross section in our model is~\cite{Barbieri:2006dq}
\begin{equation}
\sigma_{SI}=\frac{\lambda_{\phi h}^2}{16 \pi m_h^4}f^2\frac{m_N^4}{(m_{\phi}+m_N)^2}
\end{equation}
where the form factor ($f \sim$ 0.3) contains all the contributions from the nuclear matrix elements. Throughout the study, we have fixed the DM-Higgs coupling $\lambda_{\phi h}$ at $10^{-4}$. This keeps $\sigma_{SI}$ 2-4 orders below the experimental bounds~\cite{Aprile:2018dbl}. The new physics Yukawa coupling $y_{\tau}$ being leptophilic plays no role in direct searches.

\item{\bf{Indirect search prospect ---}}
The indirect detection experiments further constrain the DM velocity averaged cross section for relevant channels contributing to high-energy $\gamma$-ray flux in the Universe. In the context of our model, as far as these possibilities are concerned, due to DM-Higgs coupling $\lambda_{\phi h}=10^{-4}$, $\langle\sigma v\rangle_{\gamma\gamma}$ and $\langle\sigma v\rangle_{b\bar{b}}$ contributions will be minuscule. However, the annihilation channels in \cref{ann} give rise to $\langle\sigma v\rangle_{\tau^+\tau^-}$ possibility. 

In \cref{fig:DDID}, $\langle \sigma v \rangle_{\tau\tau}$ is plotted against $m_{\phi}$ for two different values of $\Delta m$'s, whereas $y_{\tau}$ is varied in the colour bar. Both $m_{\phi}$ and $y_{\tau}$ are varied over the full range. To be specific, some values of $y_{\tau}$ are taken above our conservative choice for the perturbative limits to demonstrate the entire parameter space. The allowed limit ($y_{\tau} \lesssim 3.0$) is up to the green shade, whereas the purple region above is not allowed by perturbativity. It is visible from \cref{ID_2} that the large $\Delta m$'s considered here suggest larger propagator suppression for the relevant channels and, consequently, shift the parameter space downwards along the $Y$ axis compared to \cref{ID_1}.

\begin{figure}[!ht]
  \centering
  \subfloat[]{\includegraphics[width=0.48\linewidth]{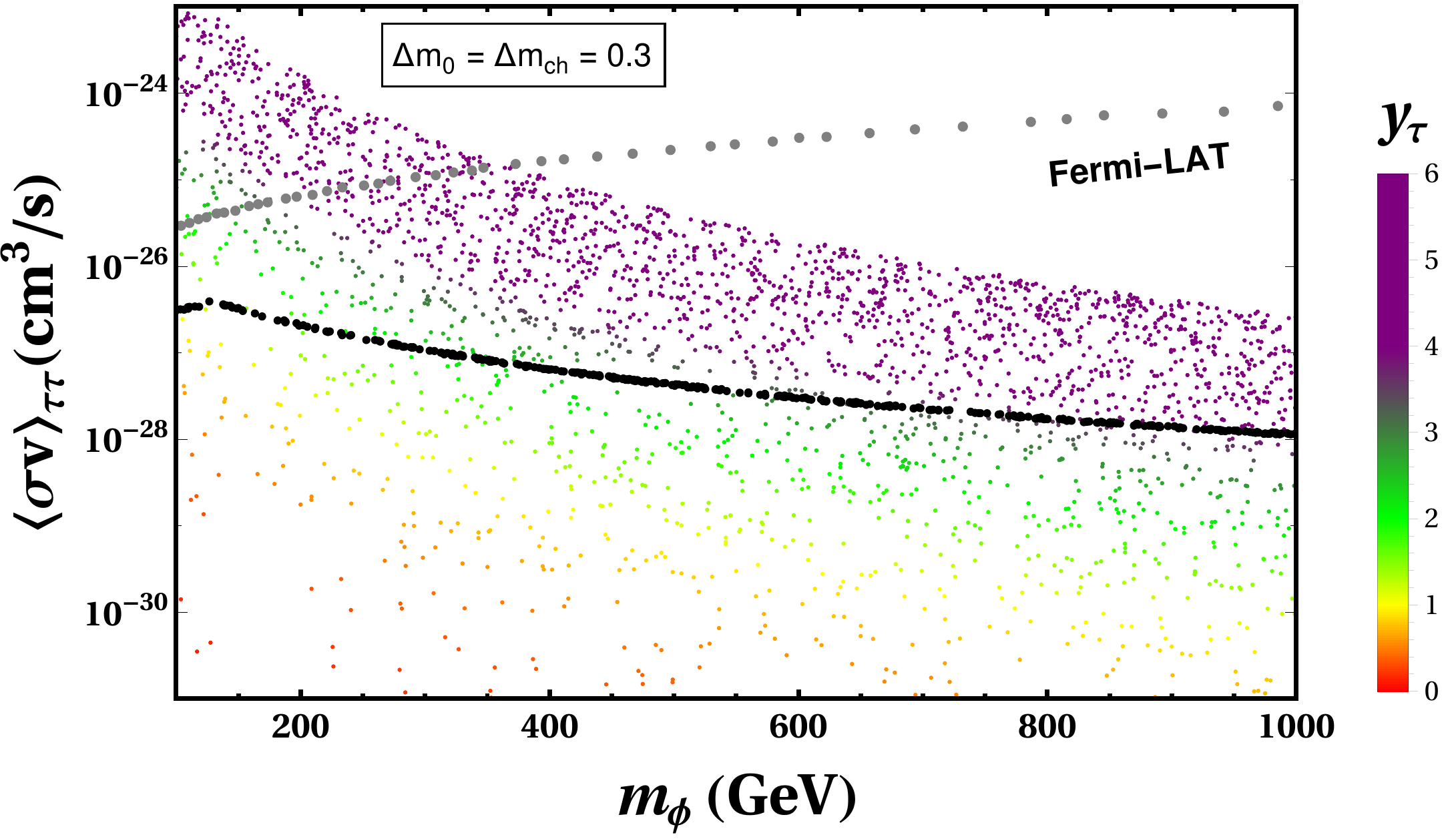}\label{ID_1}}
  \hfill
  \subfloat[]{\includegraphics[width=0.48\linewidth]{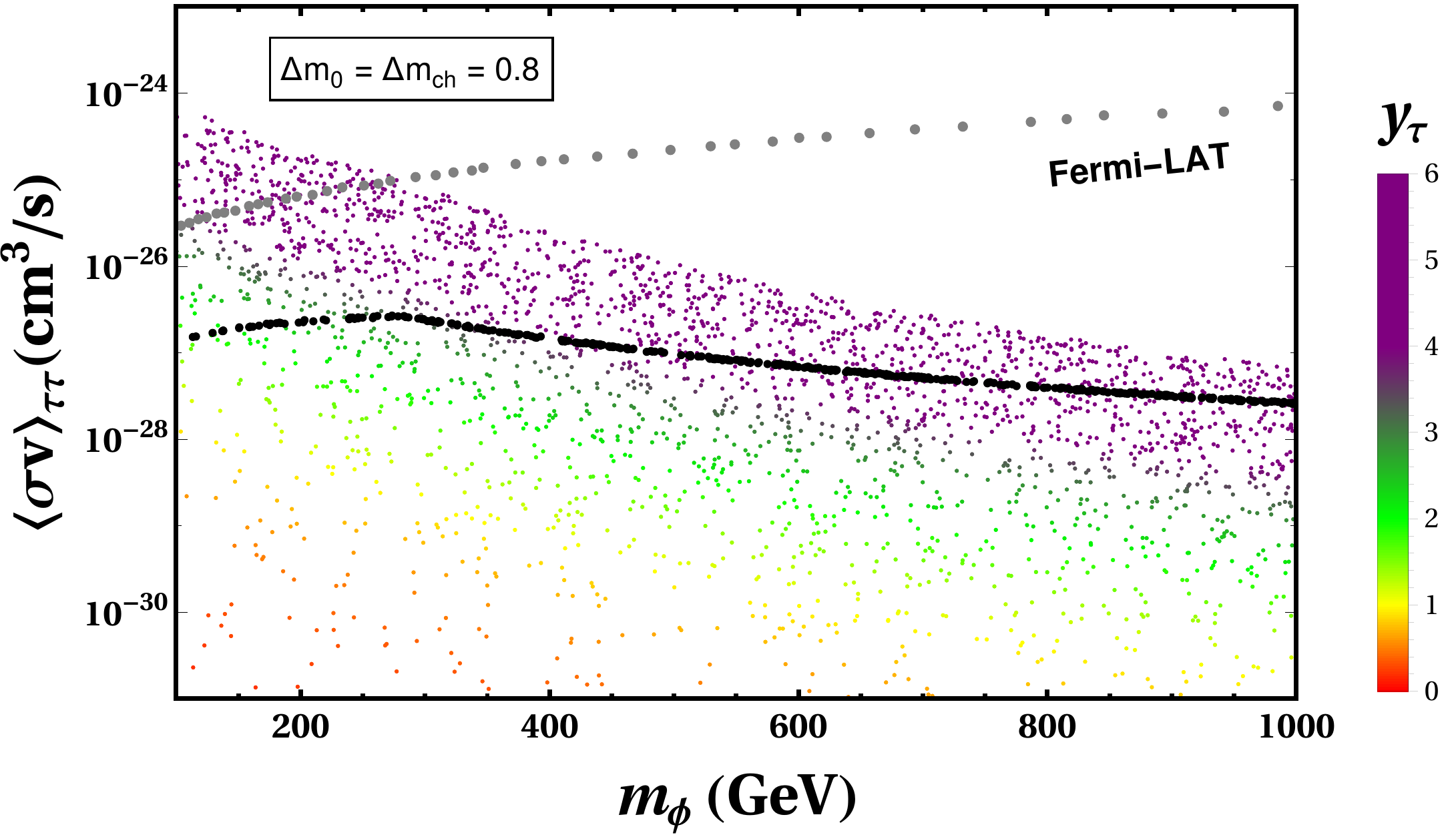}\label{ID_2}}
  \caption{Plots indicating the parameter space and the constraints relevant to indirect detection. The top plot is for $\Delta m_0=\Delta m_{ch}=0.3$, whereas for the bottom plot $\Delta m_0=\Delta m_{ch}=0.8$. $y_{\tau}$ is varied in the colour bar, and our conservative choice for the perturbative limit is depicted by the region between red and green. The gray dots represent the latest bounds from $\tau^{+}\tau^{-}$ measurements observed in Fermi-LAT. The black dots represent the relic density allowed region.}
  \label{fig:DDID}
\end{figure}

The parameter space below the indirect search limits is depicted by the region below the gray dots, which is the latest experimental bound from Fermi-LAT data\cite{Ackermann:2015lka}. Because of the shift of the parameter region downwards along the $Y$ axis for larger $\Delta m$'s as explained above, a larger region remains below the experimental limits for \cref{ID_2}, except a small portion towards smaller values of $m_{\phi}$. However, it is clearly seen that the relic density allowed region, depicted by the black dots, remains safely below the experimental limits in both the plots. For \cref{ID_1}, since the $\Delta m$'s are smaller, the relic density allowed region remains below the perturbative limits for $y_{\tau}$ for a large range of $m_{\phi}$, whereas for \cref{ID_2}, a substantial portion remains above. This feature is also seen in \cref{coann_plot_3}. There, for a fixed value of DM mass ($m_{\phi}$ = 100 GeV), the relic density allowed points correspond to $y_{\tau} \simeq 1.0$ for $\Delta m$ = 0.3, whereas  $y_{\tau} \simeq 1.7$ for $\Delta m$ = 0.8.

\end{description}

%
\section{Collider signatures}
\label{sec:collider}
The challenges of discovering dark matter in colliders are manifold. They manifest themselves as \emph{missing energy} ($E^{miss}_T$). Hence, the focus shifts entirely on the characteristics and precise measurements of associated production of visible particles. The charged multilepton channels are the most suitable to probe dark matter because of its clean signal, whereas QCD backgrounds overshadow the multijet channel, and it is very difficult to separate signals from the background. Since ours is a leptophilic model, these channels bear more significance than the others for our case. Having said so, please note that muon $g-2$ constrains the new light leptonic couplings severely, whereas the coupling with the $\tau$ lepton remains unbounded, as we already mentioned previously. Here, we are going to study the collider signatures of DM though charged multilepton $+~E^{miss}_T$ channels. Our analysis will include both light charged leptons, since they give by far the cleanest signals, as well as $\tau$ leptons, as the unbounded couplings give a greater cross section than light leptons, paving the way to better analysis to see the effects of the new couplings.

Although $\tau$ lepton analysis poses more challenges, at the same time it unravels more unique features that can come in handy in the analysis for any collider like the LHC. The $\tau$ lepton is the charged lepton of the third generation and the heaviest among them. It is even heavier than most of the light quark mesons. As a result, $\tau$ leptons decay hadronically, which sets them apart from all other leptons. Because of the lepton number conserving weak interactions, the $\tau$ final states are always accompanied by one neutrino in the hadronic final states and two neutrinos in the leptonic final states. Since the neutrinos add to the missing energy, the full $\tau$ energy cannot be measured. The leptonic decays of the $\tau$ are difficult to distinguish from prompt leptons in a $\ell~+~E^{miss}_T$ final state. Therefore, only the hadronically decaying $\tau$'s are suitable for the collider signatures.

We have used {\texttt FeynRules}~\cite{Alloul:2013bka} to generate model files for our model. Events have been generated using {\texttt MadGraph5}~\cite{Alwall:2014hca} and showered with {\texttt Pythia 8}~\cite{Sjostrand:2014zea}. Finally, the detector simulation has been performed using {\texttt Delphes}~\cite{deFavereau:2013fsa}. We use the $\tau$-tagged jets from {\texttt Delphes} and reconstruct them with the help of {\texttt FastJet}~\cite{Cacciari:2011ma} using the anti-$k_T$ algorithm. The separation $\Delta R$ of two adjacent $\tau$ jets is taken to be 0.4, and the $\tau$-tagging efficiency is taken to be 60\%. We carried out our analysis for the LHC at the CM energy $\sqrt{S} = 13$~TeV. We used the dynamic factorisation and renormalisation scale for the signal as well as the background events.

For the generation of parton-level events, we apply minimum or maximum cuts on the transverse momenta $p_T$ and rapidities $\eta$ of light jets, $b$ jets, leptons, photons, and missing transverse momentum. Also, distance cuts between all possible final objects in the rapidity-azimuthal plane are applied, with the distance between two objects $i$ and $j$ defined as $\Delta R_{ij} = \sqrt{(\phi_i - \phi_j)^2 + (\eta_i - \eta_j)^2}$, where $\phi_i$ and $\eta_i$ are the azimuthal angle and rapidity of the object $i$, respectively.

The preliminary selection cuts used in the analysis are
\begin{itemize}
  \item $p_T > 10$ GeV and $|\eta| < 2.5$ for all charged light leptons,
  \item $p_T > 20$ GeV and $|\eta| < 5$ for all $non$-$b$ jets, and
  \item $\Delta R_{ij} > 0.4$ between all possible jets or leptons.
\end{itemize}
After this, the {\scshape .lhe} files obtained through parton level events are showered with {\em final state radiation} (FSR) with {\texttt Pythia 8} where {\em initial state radiation} (ISR) and multiple interactions are switched off and fragmentation or hadronisation is allowed.

\begin{figure}[!ht]
  \centering
  \subfloat[]{\includegraphics[height=0.2\linewidth]{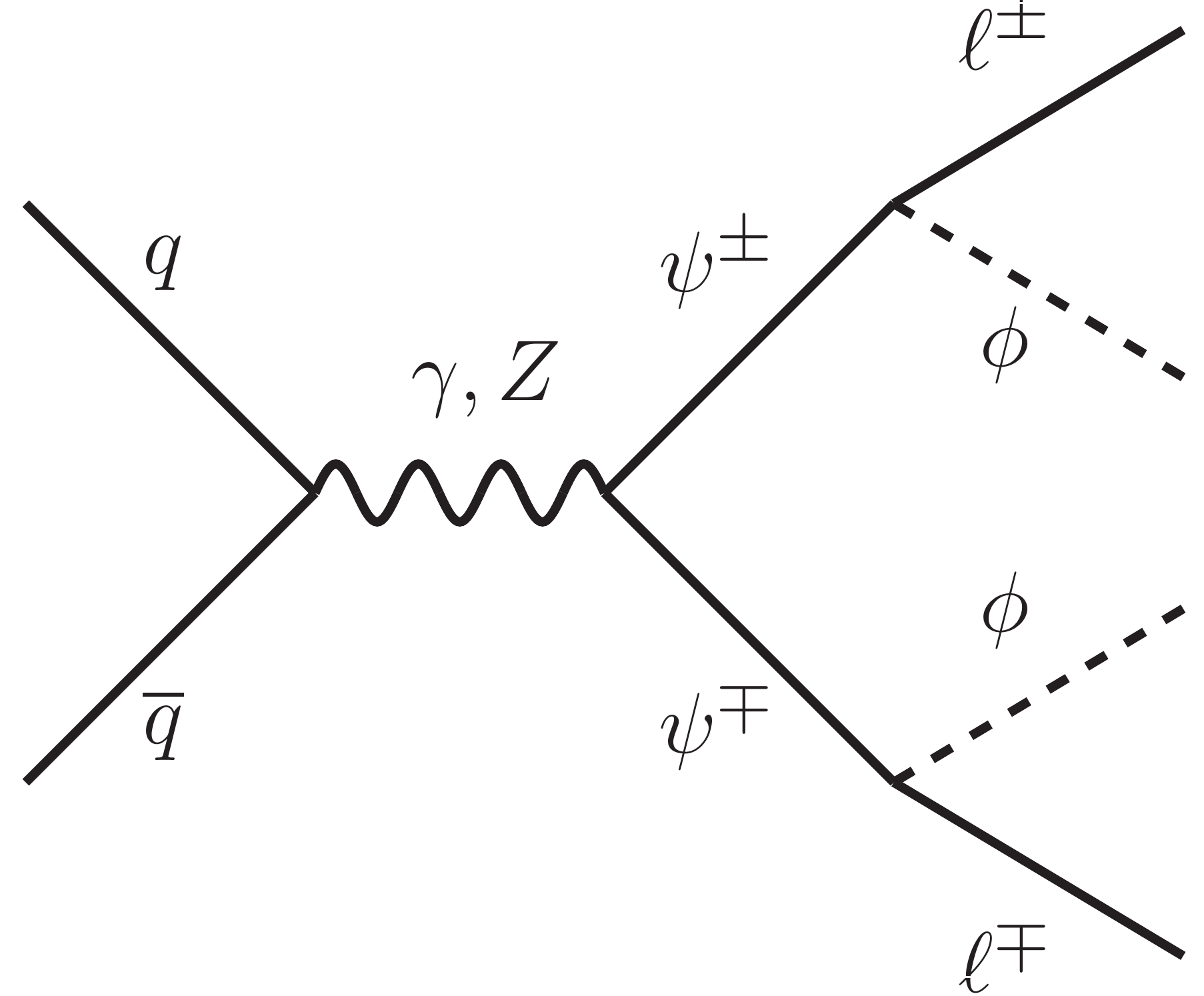}\label{coll_sig1_2lep}}
  \hskip1cm
  \subfloat[]{\includegraphics[height=0.2\linewidth]{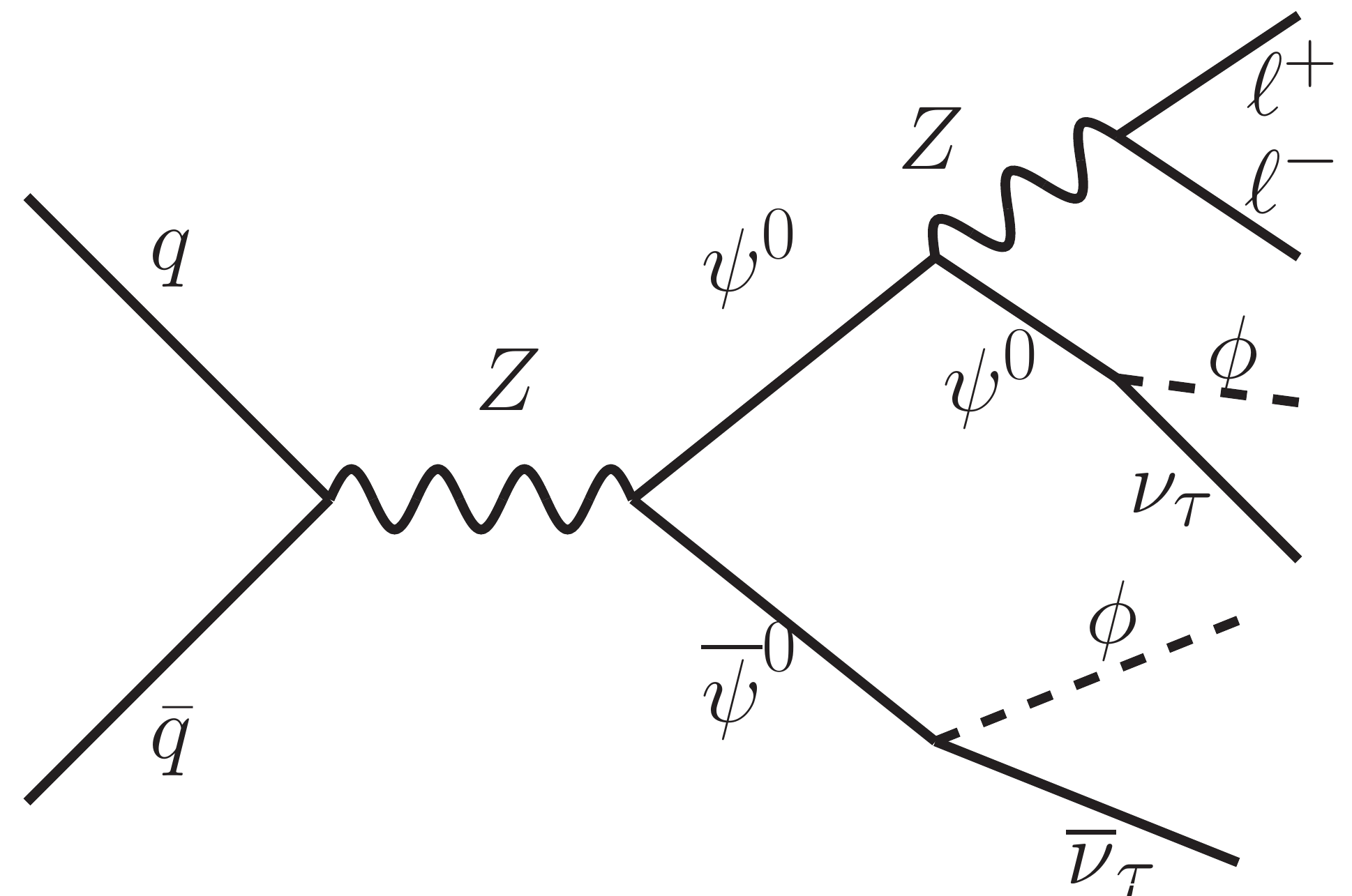}\label{coll_sig2_2lep1}}
  \hskip1cm
  \subfloat[]{\includegraphics[height=0.2\linewidth]{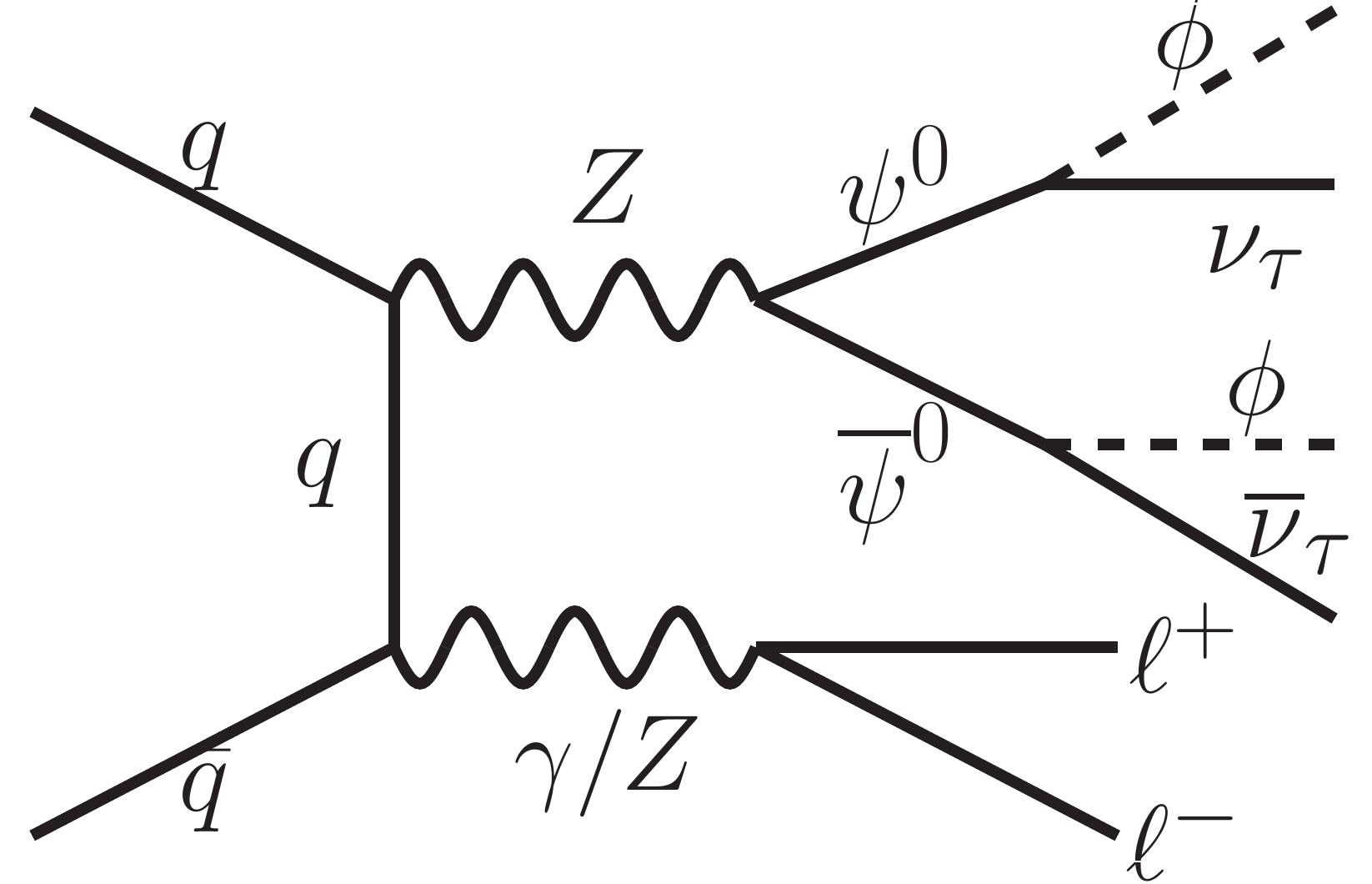}\label{coll_sig2_2lep2}}
  \caption{Feynman diagrams contributing to the dilepton channels.}
  \label{fig:coll_sig_2lep}
\end{figure}

The most important modes of production in dilepton channels are
\begin{enumerate}[(1)]
  \item $pp \to \ell^+ \ell^-\,2\phi$\,;
  \item $pp \to$ $\ell^+ \ell^- \nu \bar\nu\,2\phi$\,,
\end{enumerate}
where $\ell$ stands for all three generations of charged leptons, namely, $e, \mu,$ and $\tau$. Among the two classes of signal processes, the set (1) proceeds as follows:
\begin{enumerate}[(i)]
  \item $pp \to \psi^+ \psi^-$, followed by the decay of both $\psi$'s as $\psi^\pm \to \ell^\pm\,\phi$ [See \cref{coll_sig1_2lep}] \,,
\end{enumerate}
The couplings which play a role in the above processes are only the ones involving light leptons and, hence, are very suppressed for the light dilepton channel as shown in \cref{eq:param}. In the same vein, this channel will be dominant for the $\tau$ lepton analysis as a large value of $y_\tau$ is in effect.

On the other hand, the set (2) of processes mentioned above will proceed as
\begin{enumerate}[(i)]
  \item $pp \to \psi^0 \bar\psi^0$, followed by $\psi^0 \to \phi\,\nu$ and $\bar\psi^0 \to \ell^+\ell^-\,\bar\nu\phi$ [See \cref{coll_sig2_2lep1}]\,.
  \item $pp \to Z Z(\gamma^*)$, followed by $Z \to \nu\bar\nu\,2\phi$ and $Z(\gamma^*) \to \ell^+\ell^-$ [See \cref{coll_sig2_2lep2}]\,.
\end{enumerate}
As the couplings involved can also be either gauge couplings or that involving the $\tau$ lepton, both being considerably large, the set (2) of processes give us sufficient cross sections even for light dilepton channel to proceed with our analysis.

To highlight the features of our model clearly, we have selected the following benchmark points (see \cref{tab:bp}). The significance of the choice in benchmark points will be clear as we elaborate on our analysis in the following discussions.
\begin{table}[!ht]
 \centering
  {\setlength{\tabcolsep}{1em}
 \begin{tabular}{c|c|c|c|c}
  \hline
      & $m_\phi$
              & $m_{\psi^0}$
                      & $m_{\psi^\pm}$
                              & \multirow{2}*{$y_\tau$}
                                                  \\
      & (GeV) & (GeV) & (GeV) &           \\ \hline\hline
  BP1 & $100$ & $125$ & $120$ & $0.7$  \\ \hline
  BP2 & $ 80$ & $100$ & $ 90$ & $0.1$  \\ \hline
 \end{tabular}
 }
 \caption{Benchmark points used for the collider analysis.}
 \label{tab:bp}
\end{table}

For the benchmark points given in \cref{tab:bp} we get the cross sections for the light dilepton channel as shown in \cref{tab:cs_lep}.
\begin{table}[!ht]
  \centering
  {\setlength{\tabcolsep}{1em}
  \begin{tabular}{ l | c | c }
    \hline
    \multicolumn{1}{c|}{\multirow{2}*{Processes}}
      & \multicolumn{2}{c}{Cross section (pb)} \\ \cline{2-3}
        & BP1      
          & BP2      
                   \\ \hline\hline
      $(2)\,pp \to \ell^+ \ell^- \nu \bar\nu\,2\phi$
          & $232.17\times10^{-6}$
            & $409.30\times10^{-6}$
                   \\ \hline
  \end{tabular}
  }
 \caption{Cross sections of the light dilepton signal processes. The set (1) of subprocesses is highly suppressed for the coupling choices of \cref{eq:param}, whereas the set (2) further involves gauge and $\tau$ lepton couplings and, hence, is dominant in this scenario.}
 \label{tab:cs_lep}
\end{table}

\cref{tab:cs_2tau} shows the signal cross sections for di-$\tau$-jet channel. As mentioned previously, we can see the distinction between the cross sections of process (1), which is substantially greater than that of process (2) for this case. This is because, although the large value of $y_\tau$ and gauge couplings dictate both the processes, process (2) is suppressed by branchings and phase space.

\begin{table}[!ht]
  \centering
  {\setlength{\tabcolsep}{1em}
  \begin{tabular}{ l | r@{}l | r@{}l }
    \hline
    \multicolumn{1}{c|}{\multirow{2}*{Processes}}
      & \multicolumn{4}{c}{Cross section (pb)} \\ \cline{2-5}
        & BP1     &
          & BP2     &
            \\ \hline\hline
      $(1)\,pp \to \tau^+ \tau^-\,2\phi$
          & $805.09$&$\times10^{-3}$
            & $786.33$&$\times10^{-3}$
              \\ \cline{1-5}
      $(2)\,pp \to \tau^+ \tau^- \nu \bar\nu\,2\phi$
          & $2.72$&$\times10^{-3}$
            & $3.33$&$\times10^{-3}$
              \\ \hline
  \end{tabular}
  }
 \caption{Cross sections of the signal processes for the di-$\tau$-jet channel. Despite the large $y_\tau$ and gauge couplings, process (2) is suppressed due to more branchings and phase space.}
 \label{tab:cs_2tau}
\end{table}

The major backgrounds at the LHC for the light dilepton channel are as follows
\begin{enumerate}[align=left]
  \item[{\bf Bkg1 ---}] $pp \to t \bar t$, followed by the top (anti)quark decaying into the leptonic channel, $t (\bar t) \to \ell^\pm \nu (\bar \nu) b (\bar b)$.
  \item[{\bf Bkg2 ---}] $pp \to W^+ W^-$. $W^\pm$ further decays via leptonic channel as $W^\pm \to \ell^\pm \nu (\bar \nu)$.
  \item[{\bf Bkg3 ---}] $pp \to W^\pm Z (\gamma^*)$, followed by $W^\pm \to \ell^\pm \nu (\bar \nu)$, and $Z /\gamma^*$ decays into leptonic channel, $Z (\gamma^*) \to \ell^+ \ell^-$.
  \item[{\bf Bkg4 ---}] $pp \to Z Z (\gamma^*)$, followed by leptonic decays $Z \to \nu \bar \nu$ and $Z (\gamma^*) \to \ell^+ \ell^-$.
\end{enumerate}

The major backgrounds at the LHC for the $\tau$-jet channel will be similar as above with light lepton replaced by $\tau$'s. Since $\tau$'s can decay into hadronic channels we also have to consider light jets as backgrounds. In the following we show all the backgrounds for this particular channel.
\begin{enumerate}[align=left]
  \item[{\bf Bkg1 ---}] $pp \to t \bar t$, followed by the top (anti)quark decaying into the $\tau$-jet channel, $t (\bar t) \to \tau^\pm \nu (\bar \nu) b (\bar b)$.
  \item[{\bf Bkg2 ---}] $pp \to W^+ W^-$. $W^\pm$ further decays via $\tau$-jet channel as $W^\pm \to \tau^\pm \nu (\bar \nu)$.
  \item[{\bf Bkg3 ---}] $pp \to W^\pm Z (\gamma^*)$, followed by $W^\pm \to \tau^\pm \nu (\bar \nu)$, and $Z /\gamma^*$ decays into $\tau$-jet channel, $Z (\gamma^*) \to \tau^+ \tau^-/ 2j$.
  \item[{\bf Bkg4 ---}] $pp \to Z Z (\gamma^*)$, followed by $Z \to \nu \bar \nu$ and jet decays $Z (\gamma^*) \to \tau^+ \tau^-/ 2j$.
\end{enumerate}
\cref{tab:cs_bkg} shows the cross sections for the above backgrounds.
\begin{table}[!ht]
  \centering
  {\setlength{\tabcolsep}{1em}
  \begin{tabular}{ l | r@{}l | r@{}l }
    \hline
    \multirow{2}*{Processes}
               & \multicolumn{4}{c}{Cross section (pb)}               \\ \cline{2-5}
               & Leptonic&                & Jet     &                 \\ \hline\hline
    {\bf Bkg1} & $ 21.27$&                & $  5.31$&                 \\ \hline
    {\bf Bkg2} & $  3.13$&                & $781.43$&$\times10^{-3}$  \\ \hline
    {\bf Bkg3} & $402.68$&$\times10^{-3}$ & $ 14.75$&$\times10^{3}$   \\ \hline
    {\bf Bkg4} & $272.43$&$\times10^{-3}$ & $  2.78$&$\times10^{3}$   \\ \hline
  \end{tabular}
  }
 \caption{Cross sections of the backgrounds.}
 \label{tab:cs_bkg}
\end{table}

\begin{figure*}[!ht]
  \centering
  \includegraphics[width=0.3\linewidth]{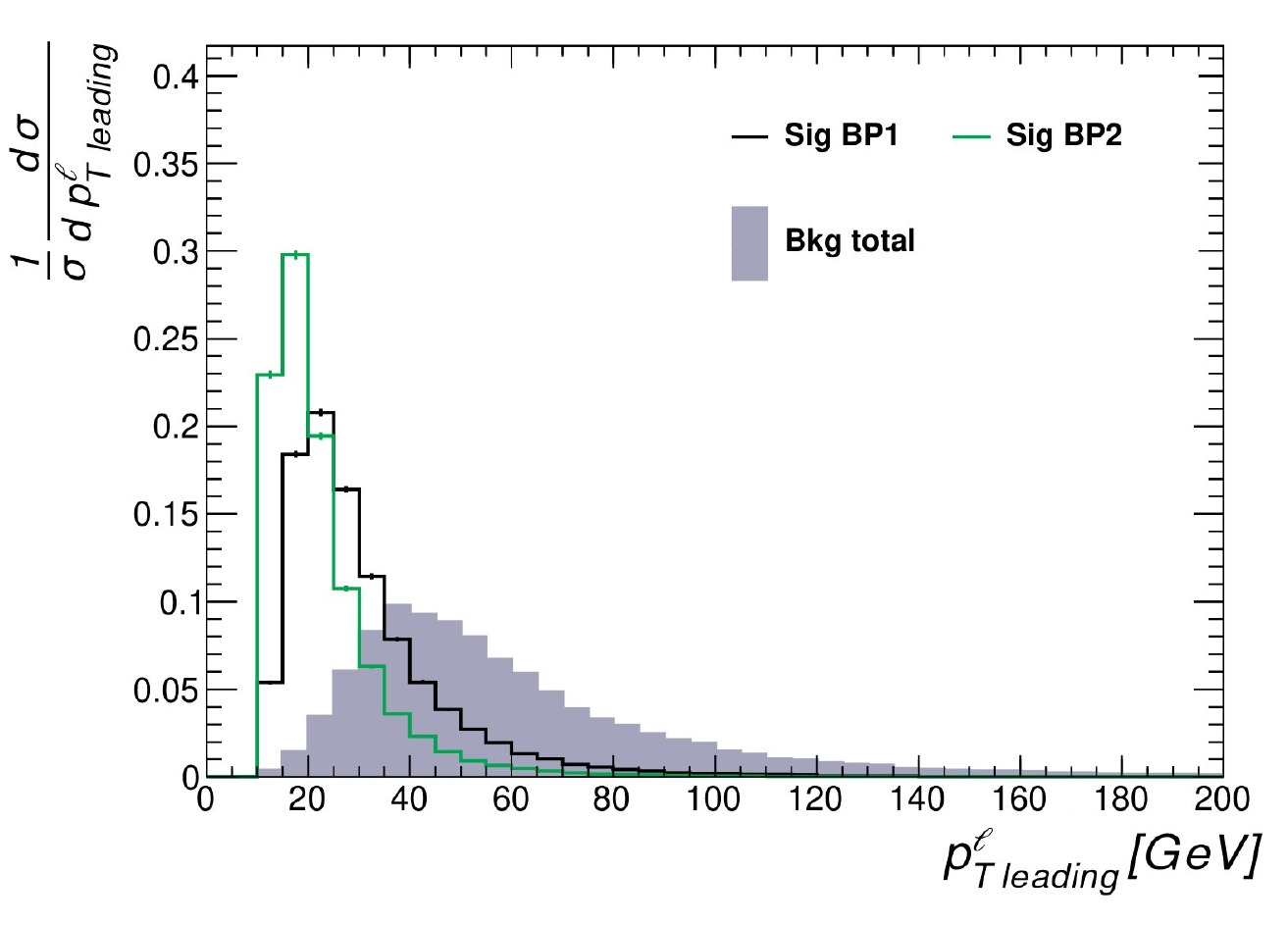}
  \includegraphics[width=0.3\linewidth]{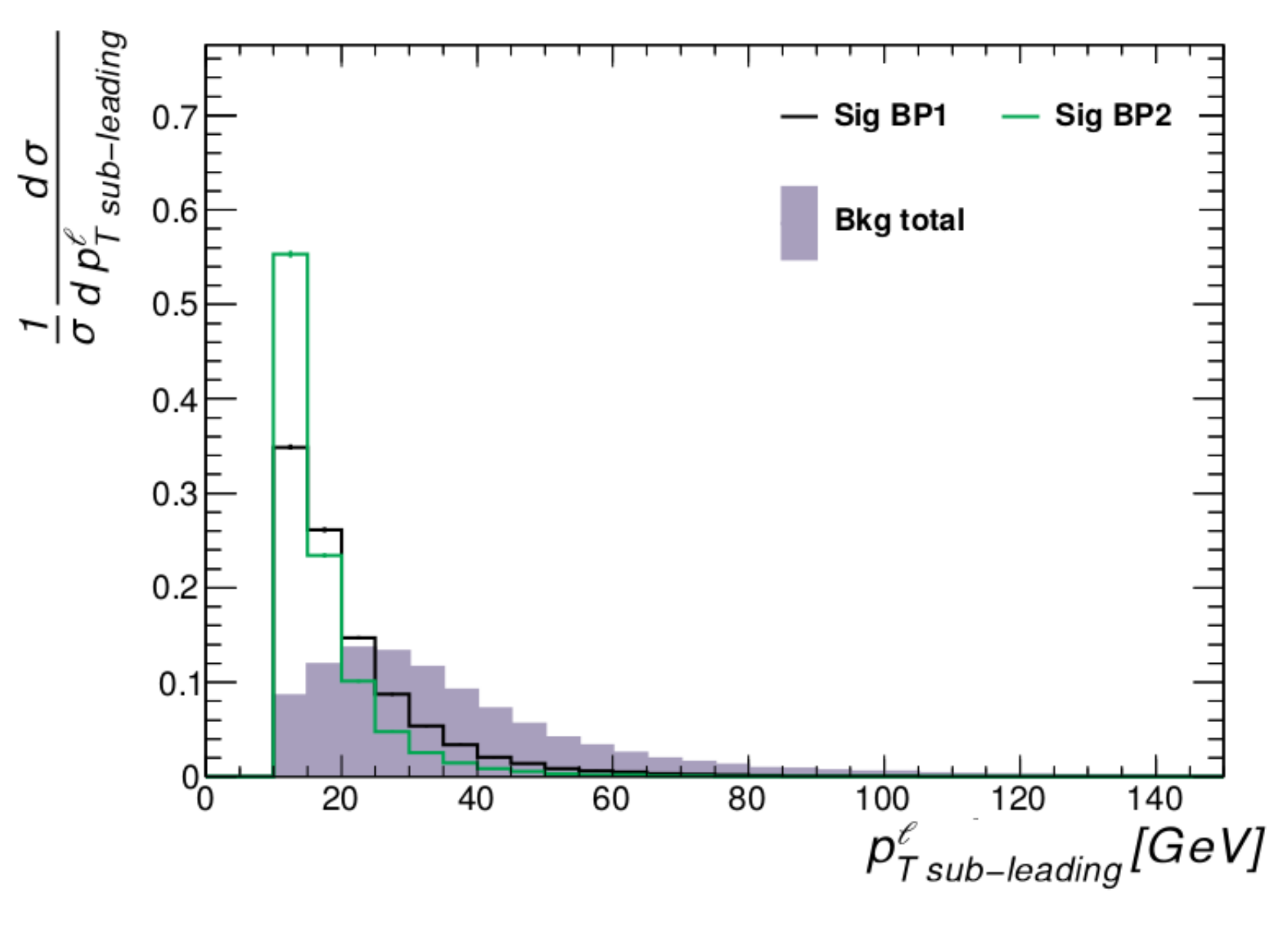}
  \includegraphics[width=0.3\linewidth]{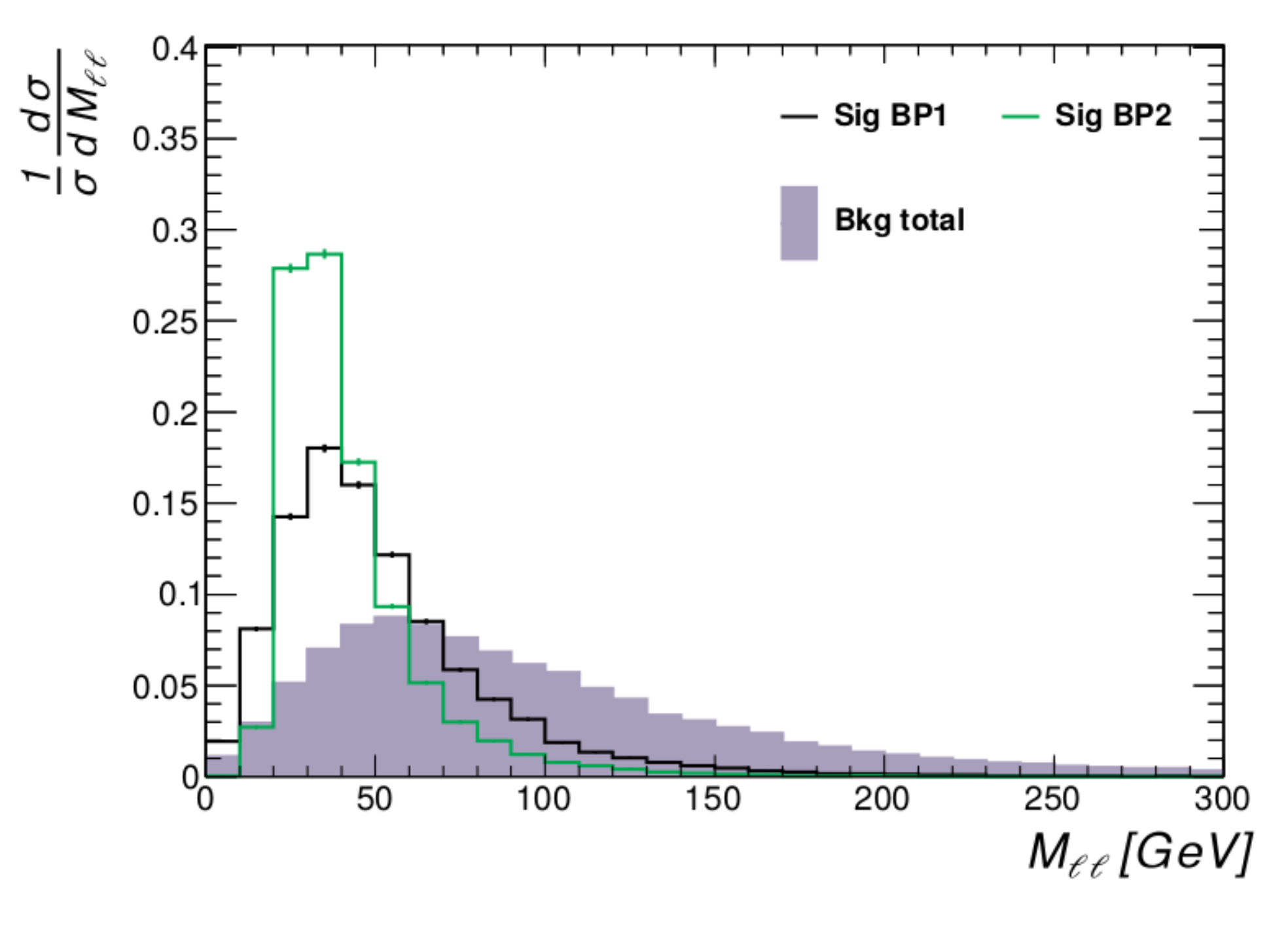}
  \\
  \includegraphics[width=0.3\linewidth]{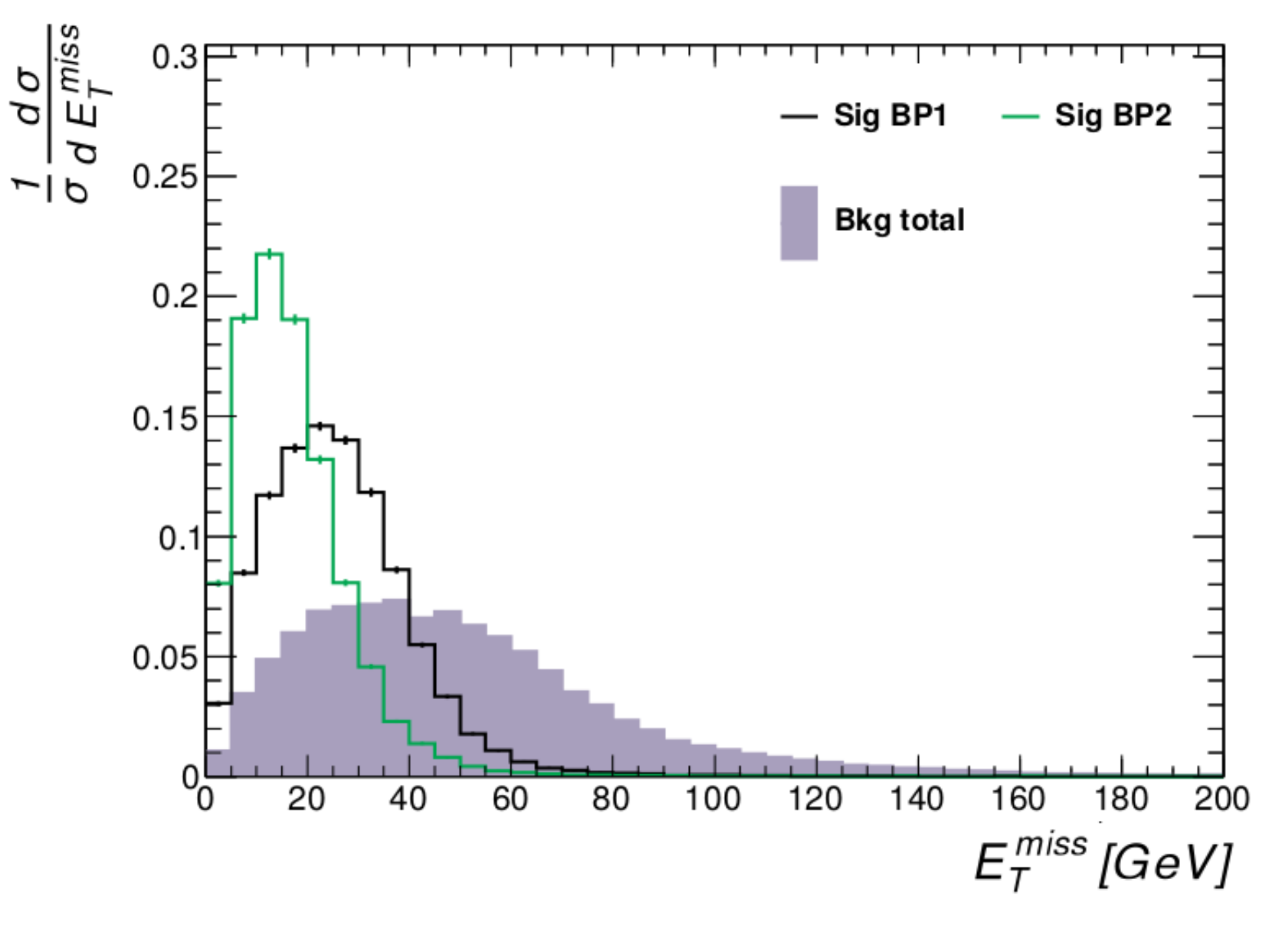}
  \includegraphics[width=0.3\linewidth]{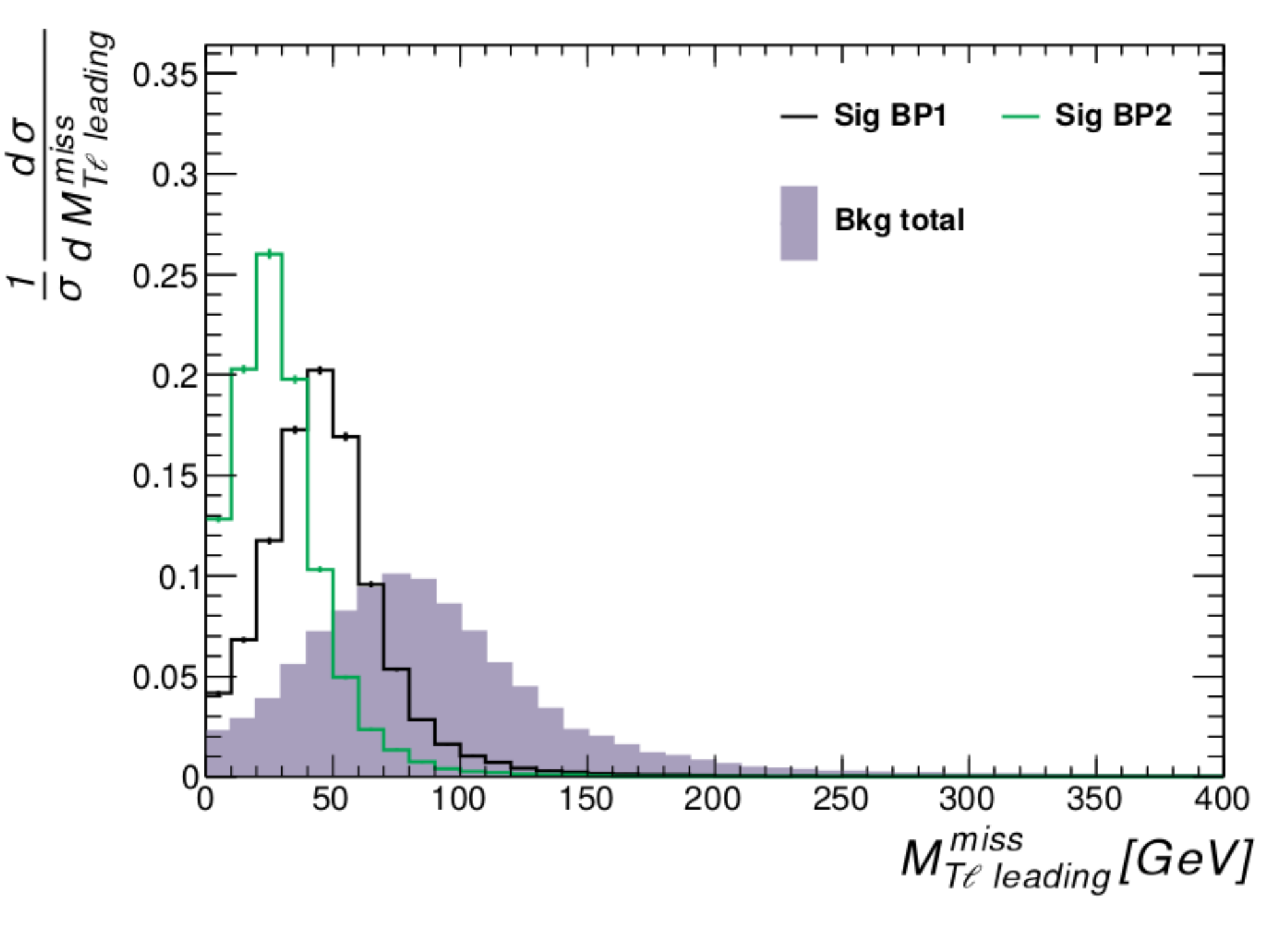}
  \includegraphics[width=0.3\linewidth]{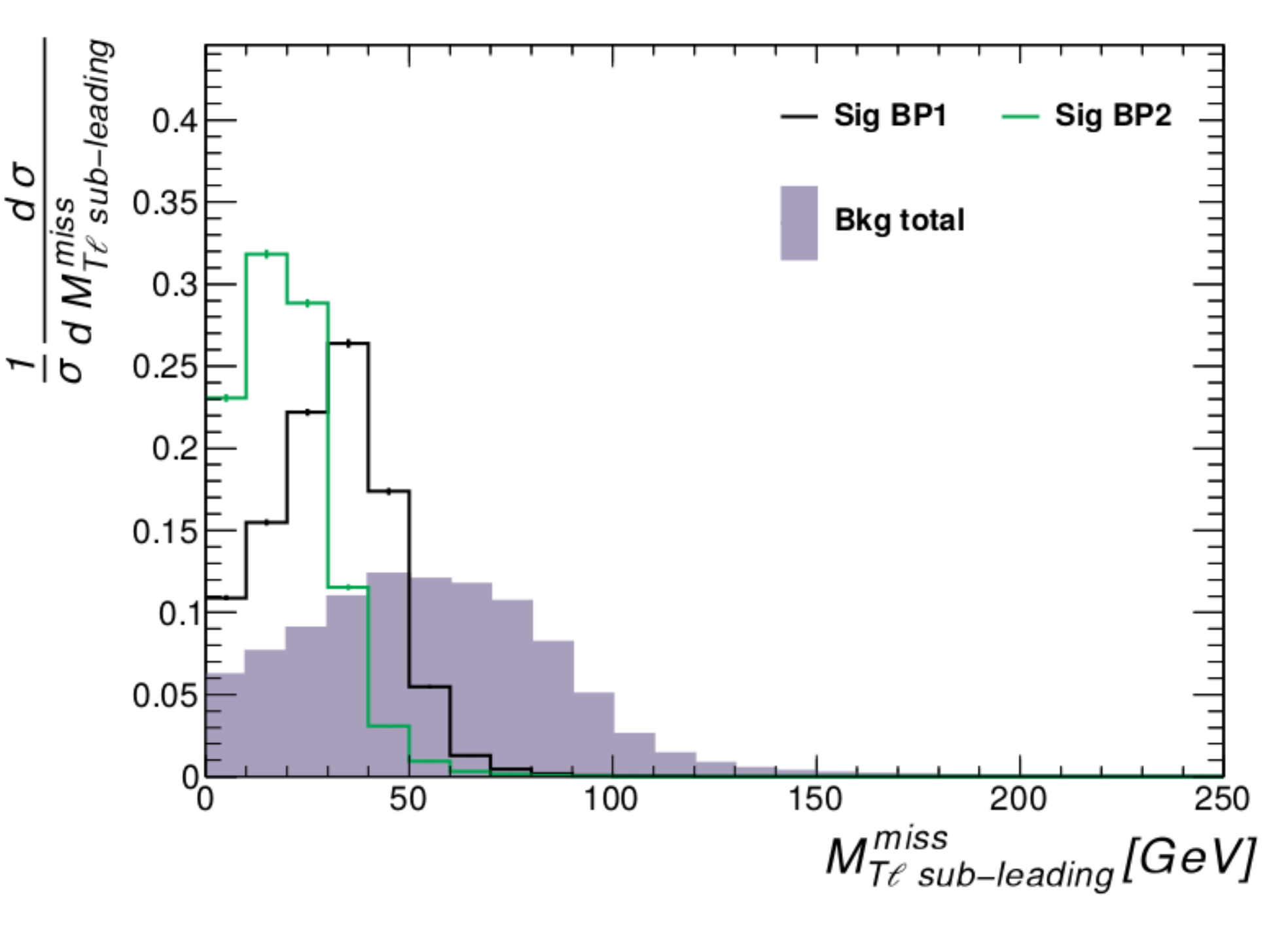}
  \caption{Distribution plots for the dilepton channel. The top panel shows $p_T$ distributions of the leading and subleading $p_T$-ordered leptons and the lepton invariant mass. The bottom panel shows the distributions of the missing transverse energy and missing transverse masses with leading and subleading leptons.}
  \label{fig:2lep}
\end{figure*}

\begin{figure*}[!ht]
  \centering
  \includegraphics[width=0.3\linewidth]{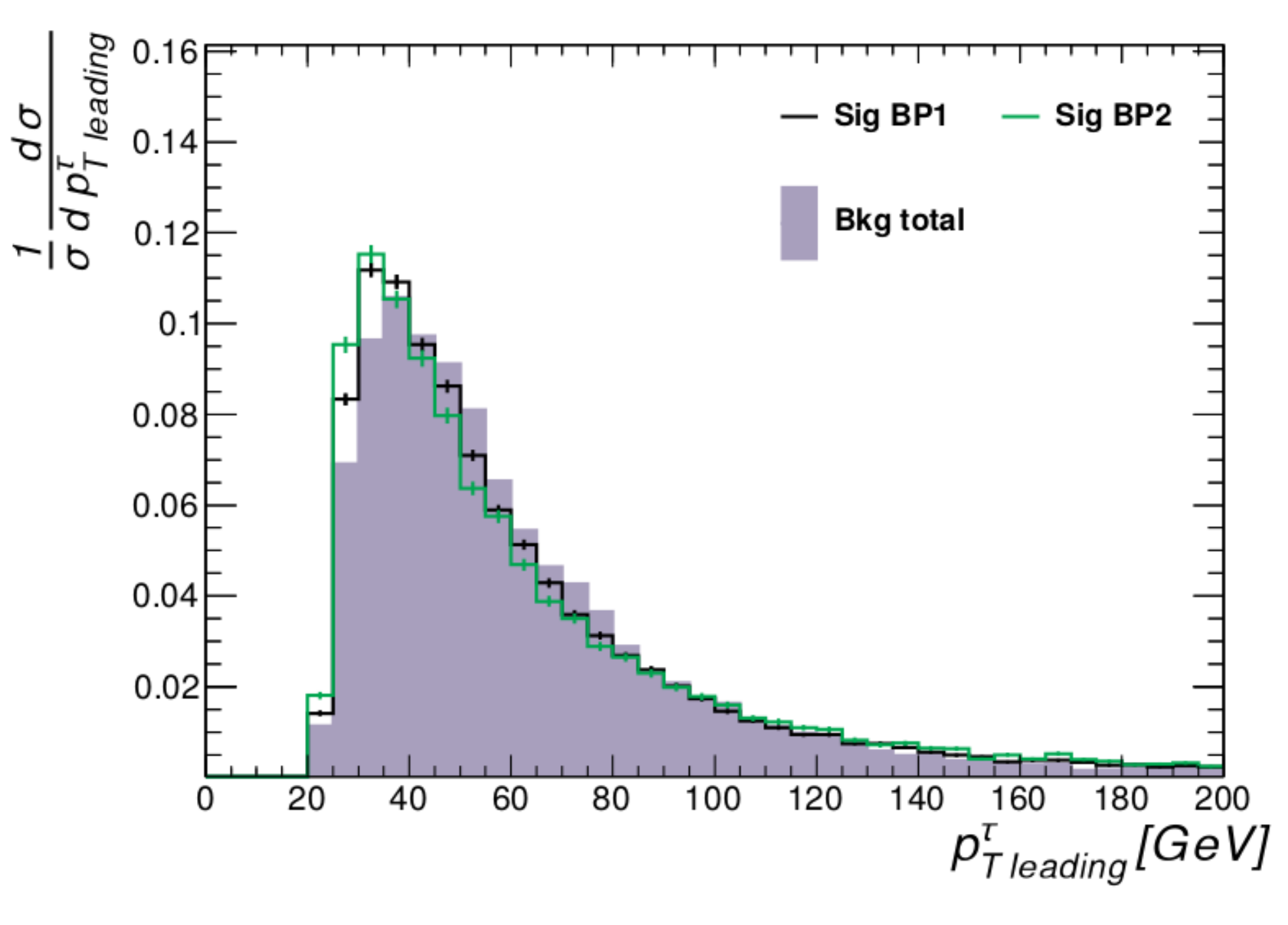}
  \includegraphics[width=0.3\linewidth]{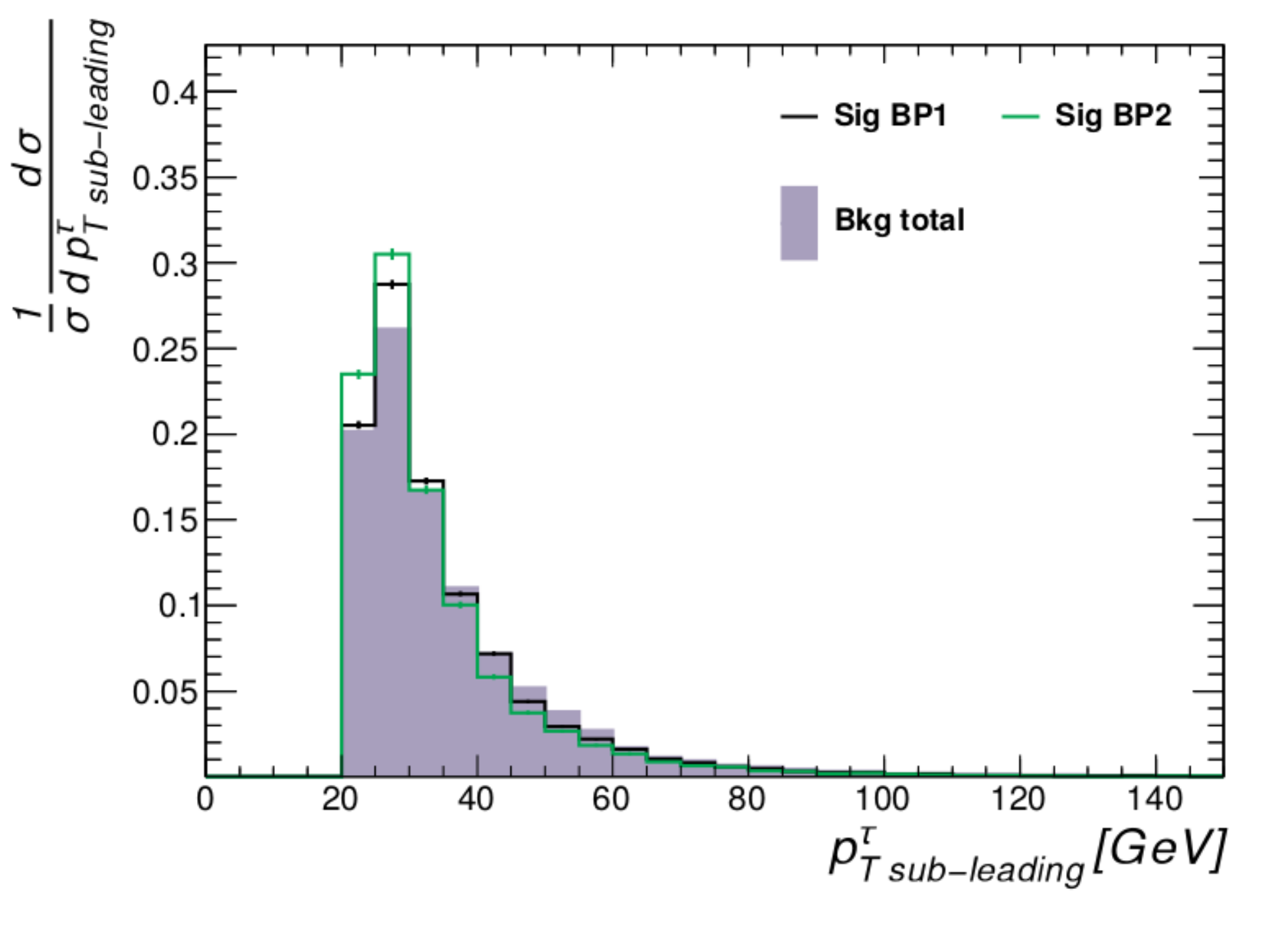}
  \includegraphics[width=0.3\linewidth]{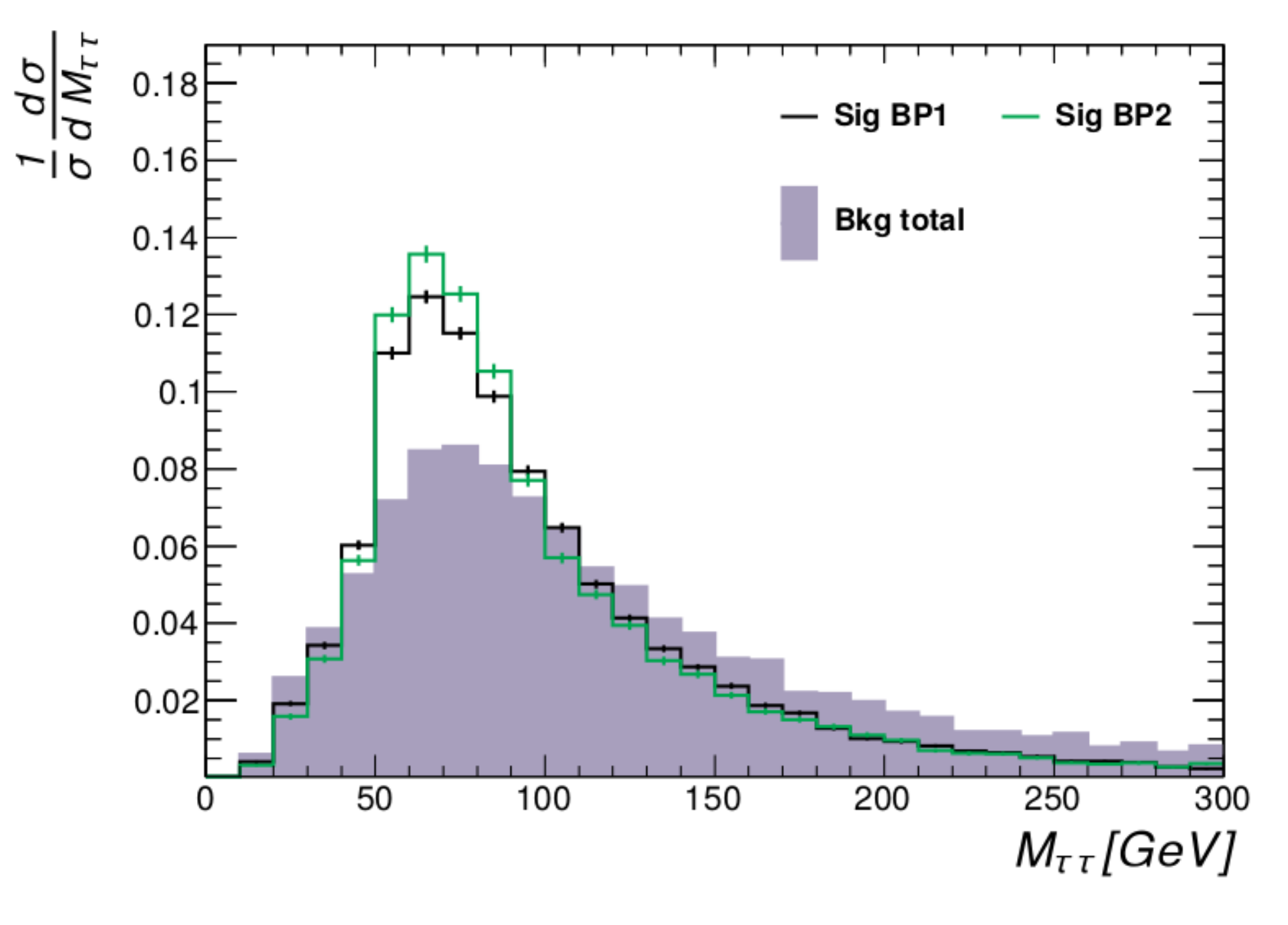}
  \\
  \includegraphics[width=0.3\linewidth]{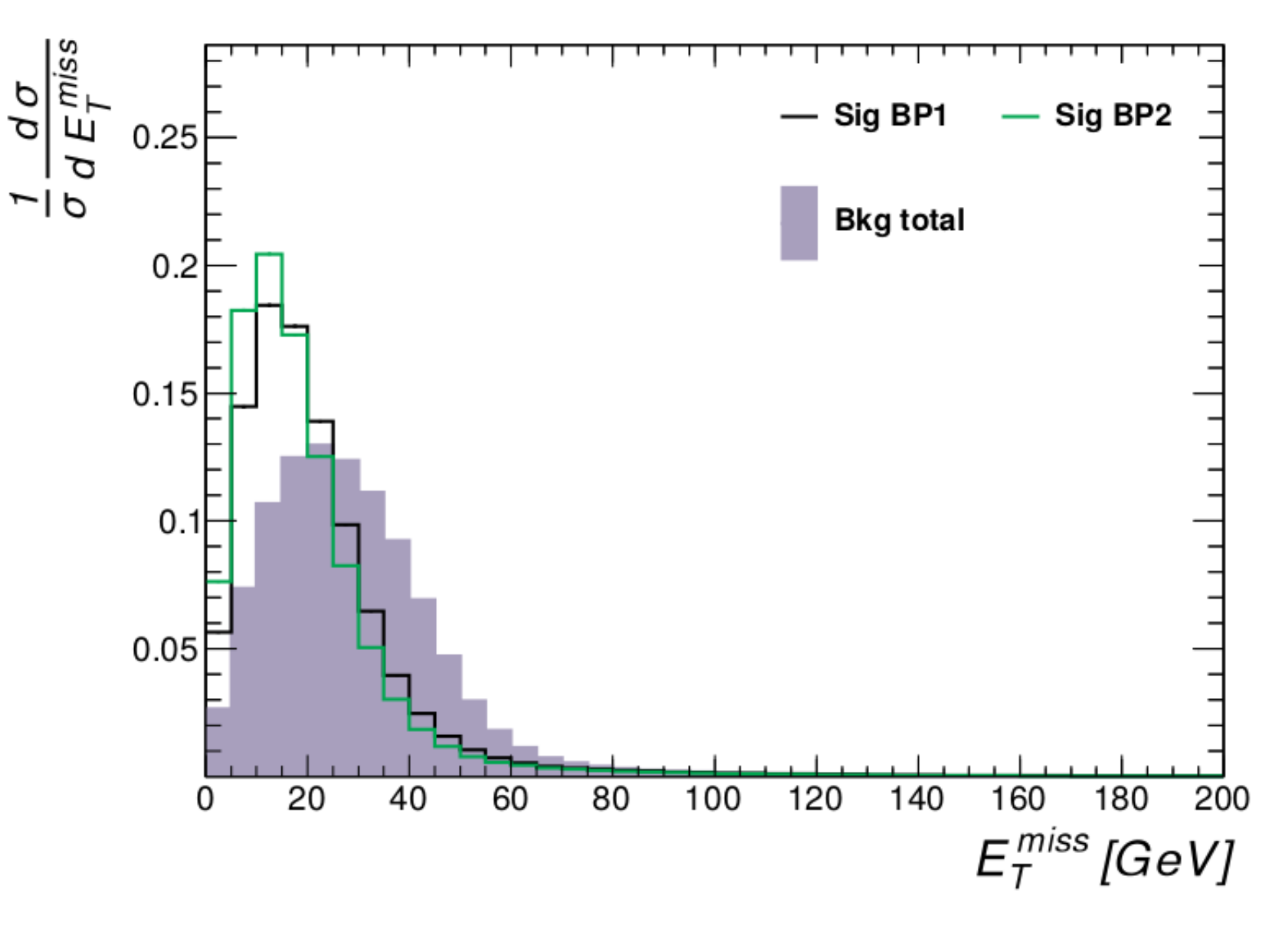}
  \includegraphics[width=0.3\linewidth]{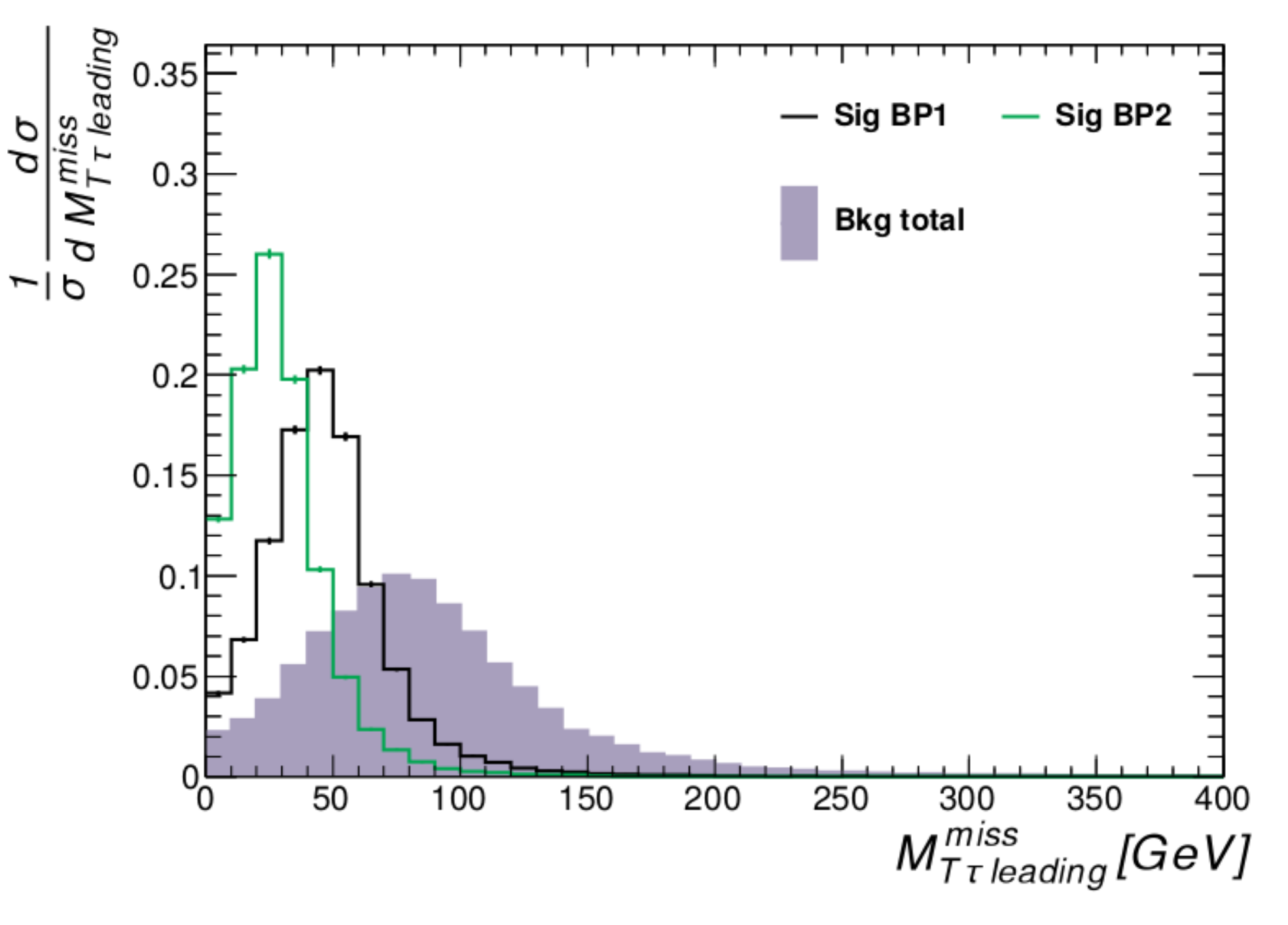}
  \includegraphics[width=0.3\linewidth]{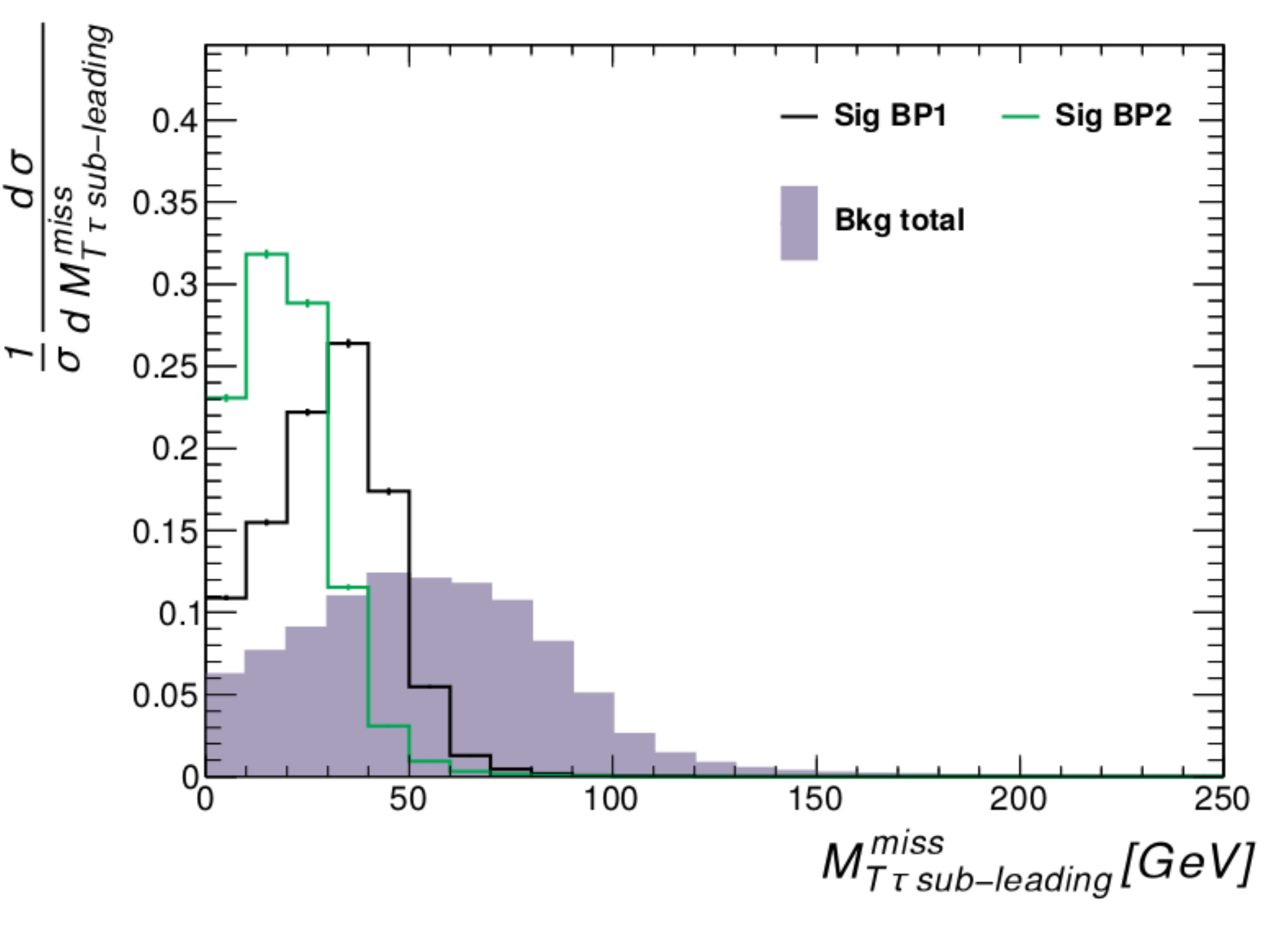}
  \caption{Distribution plots for the di-$\tau$-jet channel. The top panel shows $p_T$ distributions of the leading and subleading $p_T$-ordered $\tau$ jets and the $\tau$ jet invariant mass. The bottom panel shows the distributions of the missing transverse energy and missing transverse masses with leading and subleading $\tau$ jets.}
  \label{fig:2tau}
\end{figure*}

Before getting involved in a more intricate analysis, we shall first discuss the kinematic distributions for this channel. The kinematic observables at our disposal are only the 4-momenta of the leptons and the missing energy $E^{miss}_T$. We order them according to the magnitude of their transverse momentum $p_T$. As a result, a {\em leading lepton} would always mean the leading-$p_T$ lepton. We can also construct other observables from them, such as the invariant mass of the lepton pair. In a similar vein, we would construct the so-called transverse mass of the lepton-$E^{miss}_T$ system. We shall call this quantity the missing transverse mass and define it for each lepton-$E^{miss}_T$ system as
\begin{gather}
 M^{miss}_{T\ell} = \sqrt{2\,E_{T}^\ell\,E^{miss}_T\,(1 - \cos\Delta\phi_{\ell E^{miss}_T})} \,,
\end{gather}
where $E_{T}^\ell = p_{T}^\ell$, which is the magnitude of the transverse momentum of a given lepton, and $\Delta\phi_{\ell E^{miss}_T}$ is the difference between the azimuthal angles of the lepton and missing transverse momentum. One can also construct similar observable for $\tau$'s. The missing transverse mass plays an important role in distinguishing the massless invisible particles (such as neutrinos) from the massive ones (as is the case for our dark matter candidates), and, hence is very crucial for our analysis. With all these observables at our disposal, we show some of the distributions in \cref{fig:2lep} for the light dilepton channel and \cref{fig:2tau} for the di-$\tau$-jet channel.

Before going further into the analysis, let us discuss the features of the benchmark points which we mentioned previously. It is clear from the distributions that the benchmark points 1 and 2 have very similar patterns and that the distributions are more populated in the lower region of each observable. It is also clear from these distributions that is very difficult to separate the signal from the background by simple cut-flow analysis. We will lose too many signal events with respect to the background leading to very low signal-to-background efficiency. That is the reason we apply the multivariate analysis which we will elaborate next.

In the next level of our study, we use {\em Toolkit for Multivariate Data Analysis} (TMVA)~\cite{Hocker:2007ht} in ROOT, to distinguish the signal events from the backgrounds efficiently. For this, we use the distributions of \cref{fig:2lep} (\cref{fig:2tau}) and some other kinematic observables to train a BDT for the light dilepton (di-$\tau$-jet) channel. The complete list of observables used to train BDT are as follows:
\begin{itemize}
 \item $p_T$ and $\eta$ of the leading and subleading light leptons ($\tau$-jets) and the invariant mass of the pair,
 \item missing transverse momentum $E^{miss}_{T}$,
 \item missing transverse mass $M^{miss}_{T}$ of the leading and subleading light leptons ($\tau$ jets),
 \item the difference of the azimuthal angles $\Delta\phi_{\ell E^{miss}_{T}}$ of the leading and subleading light leptons ($\tau$-jets) with the missing transverse energy.
\end{itemize}
We use these distributions as discriminators to the BDT analysis. The discrimination of the signal and background can be improved further using proper cuts in addition to the preliminary selection cuts to the signal and/or background events. The resulting BDT response functions give us an estimate of the signal efficiency vs the rejection of the background.

\begin{figure}[!ht]
  \centering
  \includegraphics[width=0.8\linewidth]{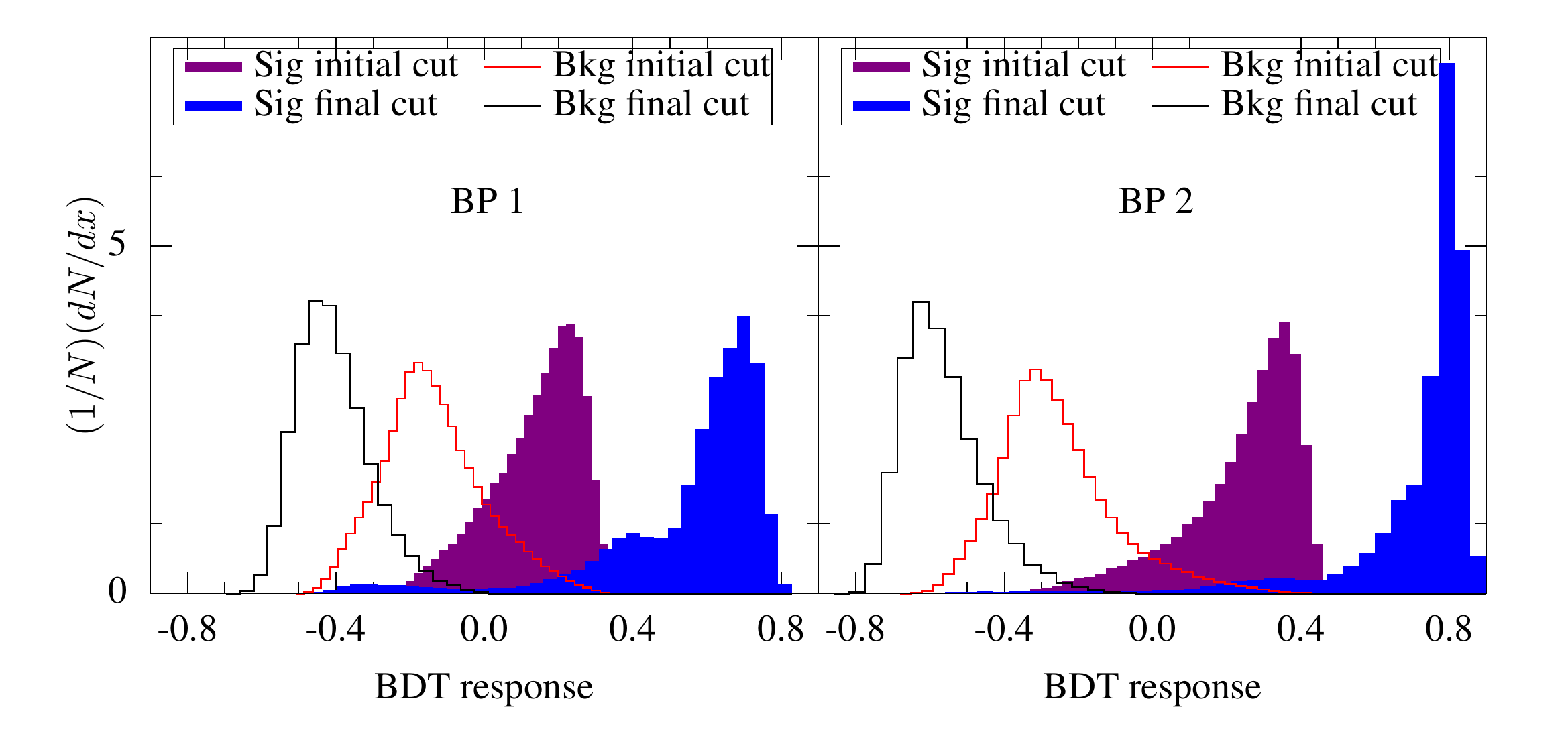}
  \caption{BDT response curves for the dilepton channel. The solid purple and the hollow red ones are before any additional cuts with only taking into account the preliminary selection cuts, whereas the solid blue and the hollow black ones are after implementing carefully chosen additional set of cuts.}
  \label{fig:BDT2lep}
\end{figure}

\begin{figure}[!ht]
  \centering
  \includegraphics[width=0.8\linewidth]{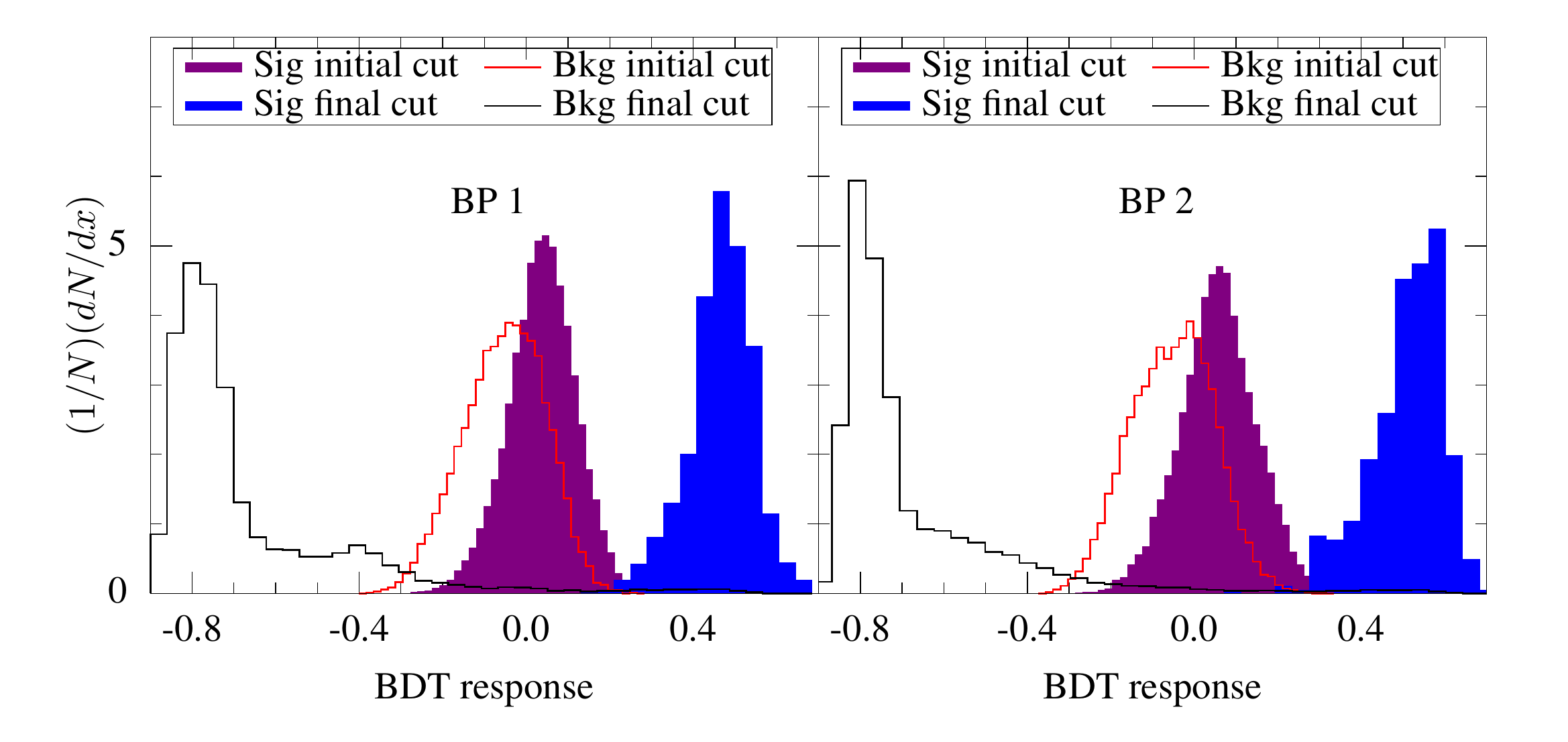}
  \caption{BDT response curves for the di-$\tau$-jet channel. The solid purple and the hollow red ones are before any additional cuts with only taking into account the preliminary selection cuts, whereas the solid blue and the hollow black ones are after implementing a carefully chosen additional set of cuts.}
  \label{fig:BDT2tau}
\end{figure}

\cref{fig:BDT2lep} (\cref{fig:BDT2tau}) shows the BDT response curves (solid filled histograms are for the signal, and the hollow ones are for the background) for the light dilepton (di-$\tau$-jet) channel for each benchmark points. Here we show two sets of BDT responses: (i) the solid purple and the hollow red ones are before any additional cuts with only taking into account the preliminary selection cuts, whereas (ii) the solid blue and the hollow black ones are after implementing a carefully chosen additional set of cuts to improve the distinguishability of the signal from the backgrounds. We will elaborate on the cuts chosen later on.

\begin{figure}[!ht]
  \centering
  \includegraphics[width=0.8\linewidth]{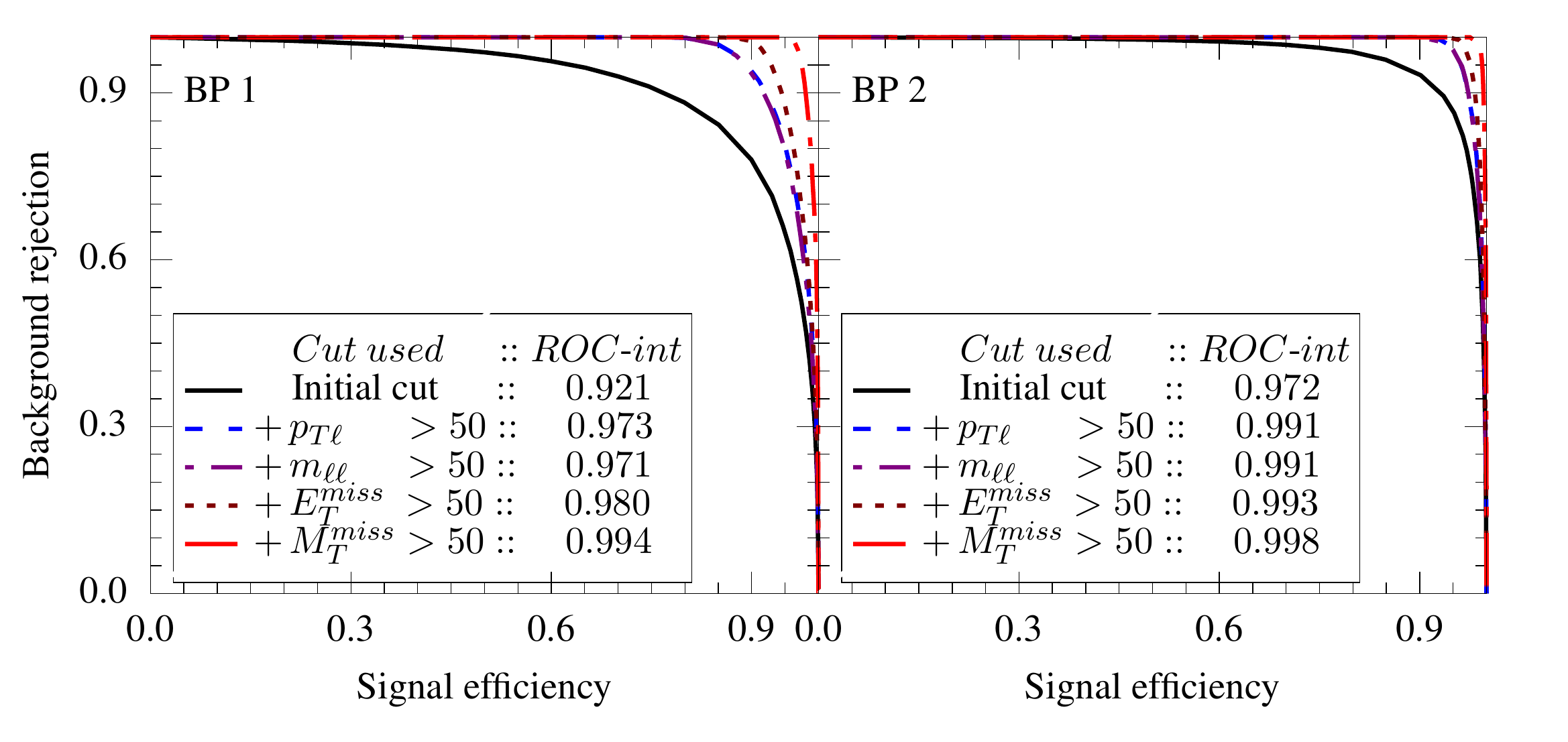}
  \caption{ROC curves for the dilepton channel. The area under the ROC curve for each cut (in GeV) is given in the inset. Each value is in addition to all the preceding cuts.}
  \label{fig:ROC2lep}
\end{figure}

\begin{figure}[!ht]
  \centering
  \includegraphics[width=0.8\linewidth]{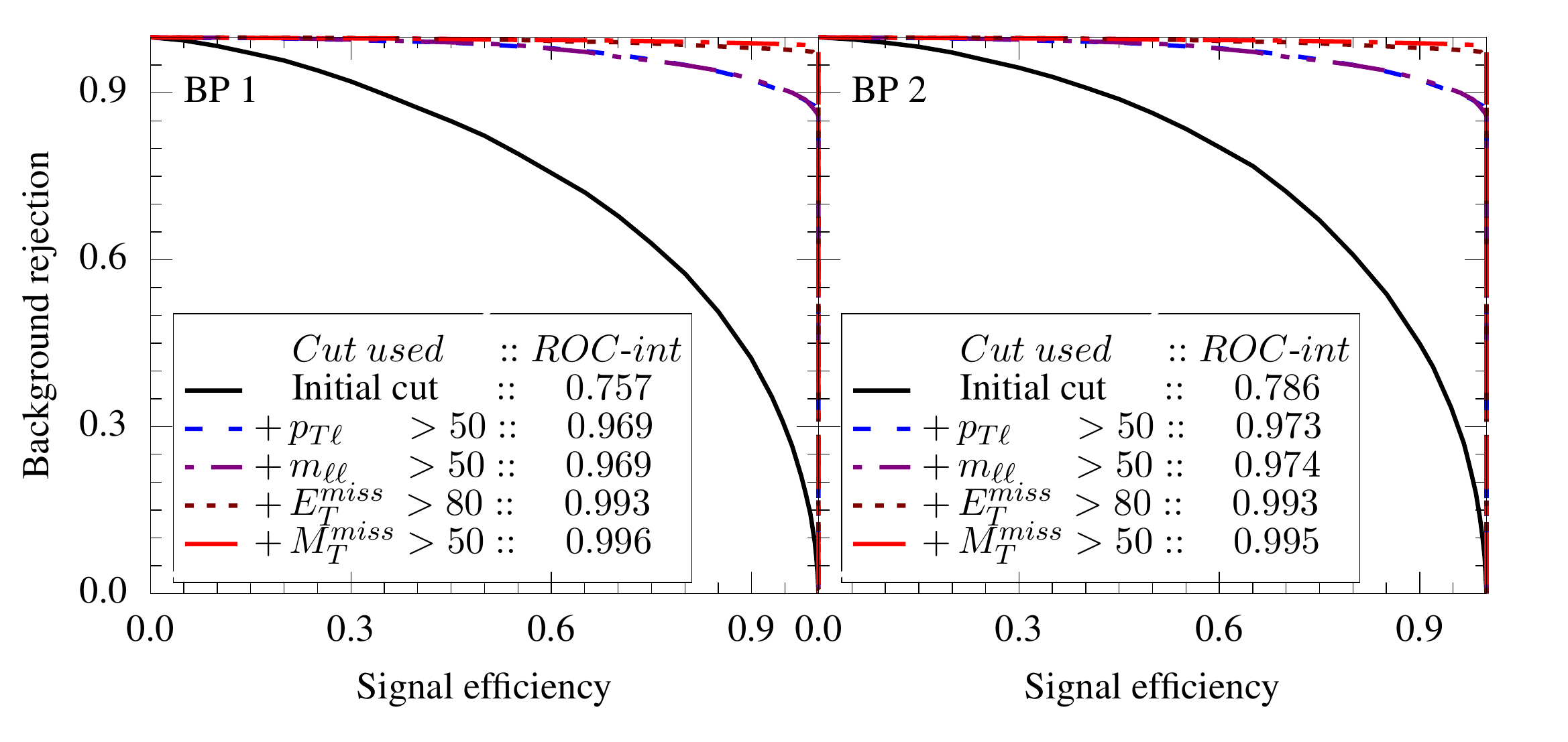}
  \caption{ROC curves for the di-$\tau$-jet channel. The area under the ROC curve for each cut (in GeV) is given in the inset. Each value is in addition to all the preceding cuts.}
  \label{fig:ROC2tau}
\end{figure}

We observe that the signal is separable from the background from the BDT response curves of \cref{fig:BDT2lep,fig:BDT2tau} after the use of additional cuts. However, we have not yet quantified the improvement. For this, we draw the \emph{receiver operating characteristic} (ROC) curves for each benchmark point using the gradual use of additional cuts. \cref{fig:ROC2lep} (\cref{fig:ROC2tau}) shows the resulting curves for the signal efficiency vs the rejection of the background for the light dilepton (di-$\tau$-jet) channel for each benchmark points. The area under each curve gives the quantitative estimate of the goodness of the separation of the signal from the backgrounds. In \cref{tab:cutflowROC}, we show the area under the ROC curve for each cut. The value of the cut is in addition to all the preceding cuts. We can see the improvement in the separation of the signal from the backgrounds from these numbers. The important point to be noted here is that, for the light dilepton case, the cuts were used only on the background events, leaving the signal events untouched, whereas for the di-$\tau$-jet case, the cuts were used only on the signal events. The reason for this can be understood as the population of events in the distribution plots of \cref{fig:2lep,fig:2tau}, respectively.

\begin{table}[!ht]
    \centering
    \begin{minipage}{0.48\linewidth}
        \centering
          {\setlength{\tabcolsep}{1em}
          \begin{tabular}{ r | c | c }
            \hline
            \multicolumn{3}{c}{dilepton channel}          \\ \hline
            \multicolumn{1}{c|}{Cuts}    &  BP 1  & BP 2   \\ \hline\hline
            \multicolumn{1}{c|}{Initial} & 0.921  & 0.972  \\ \hline
            $p_T^\ell > 50$~GeV          & 0.973  & 0.991  \\ \hline
            $m_{\ell\ell} > 50$~GeV      & 0.971  & 0.991  \\ \hline
            $E^{miss}_T > 50$~GeV        & 0.980  & 0.993  \\ \hline
            $M^{miss}_T > 50$~GeV        & 0.994  & 0.998  \\ \hline
          \end{tabular}
          }
    \end{minipage}%
    \hfill
    \begin{minipage}{0.48\linewidth}
        \centering
          {\setlength{\tabcolsep}{1em}
          \begin{tabular}{ r | c | c }
            \hline
            \multicolumn{3}{c}{di-$\tau$-jet channel}      \\ \hline
            \multicolumn{1}{c|}{Cuts}    &  BP 1  & BP 2   \\ \hline\hline
            \multicolumn{1}{c|}{Initial} & 0.757  & 0.786  \\ \hline
            $p_T^\tau > 50$~GeV          & 0.969  & 0.973  \\ \hline
            $m_{\tau\tau} > 50$~GeV      & 0.969  & 0.974  \\ \hline
            $E^{miss}_T > 80$~GeV        & 0.993  & 0.993  \\ \hline
            $M^{miss}_T > 50$~GeV        & 0.996  & 0.995  \\ \hline
          \end{tabular}
          }
    \end{minipage}
    \caption{The area under the ROC curves in \cref{fig:ROC2lep,fig:ROC2tau} for each cut. Each value is in addition to all the cuts above it. The efficiency of the cuts can be seen from the increasing values for each subsequent entry.}
    \label{tab:cutflowROC}
\end{table}

From the above discussions, plots, and numbers, we can clearly see that, for our set of benchmark points, the DM signal for a leptophilic model with coannihilating dark lepton partners may not be separable from the SM backgrounds in a collider environment by the ordinary cut-flow analysis. However, it can be easily done with the help of a set of carefully chosen cuts in BDT analysis.

Apart from the prompt decay of dark partners, this model can also accommodate delayed decays in colliders leading long-lived particle (LLP) signatures which we discuss in the following. As mentioned previously, the gauge invariance mandates the degeneracy between the dark leptons. Only the loop effect generates a small mass splitting between them. Even after the introduction of a scalar triplet in our model, the mass splitting remains sufficient but small as can be seen from \cref{eq:deltam}. The resulting phase space will be automatically suppressed in the decays of such particles because of this small mass splitting. Such a scenario leads either of the dark leptons to be long-lived depending on their mass hierarchy. If the decay length $c\tau$ be greater than the detector radius, it is obvious that the neutral partner will give a missing energy signal, whereas the charged one will leave a stable ionisation track (HSCP)~\cite{Aaboud:2018hdl,Chatrchyan:2013oca,Heisig:2012zq} in the detector.

Things become much more interesting once these particles decay inside the detector. For the charged particle, it will either give a disappearing (DT)~\cite{Gunion:1999jr,Feng:1999fu,Aad:2013yna} or a kinked track~\cite{Asai:2011wy,Jung:2015boa}.
 
\begin{figure}[!ht]
  \centering
  \subfloat[Decay of the charged dark lepton.]{\includegraphics[height=0.25\linewidth]{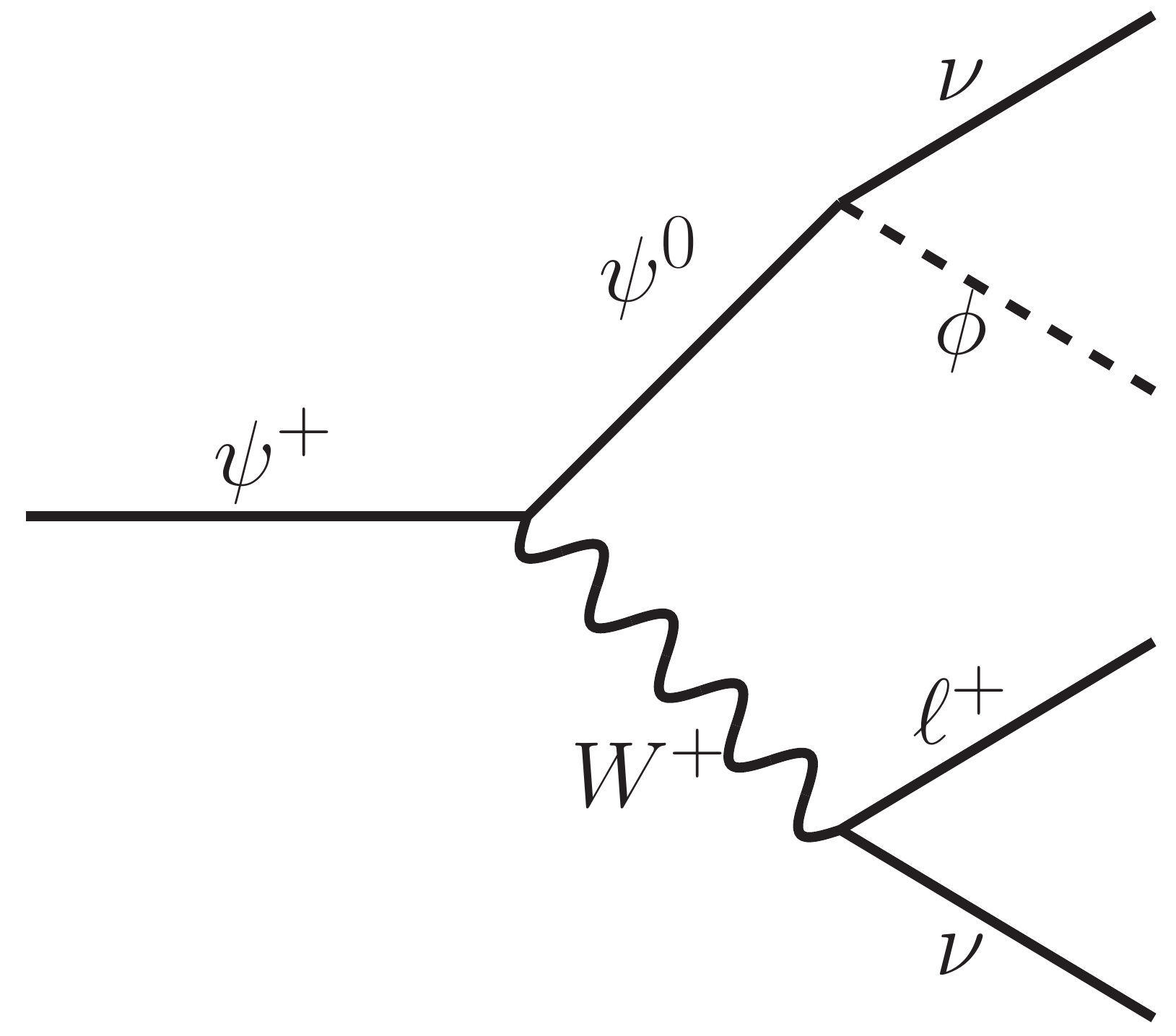}\label{fig:llp_psip_to_psi0}}
  \hskip1cm
  \subfloat[Decay of the neutral dark lepton.]{\includegraphics[height=0.25\linewidth]{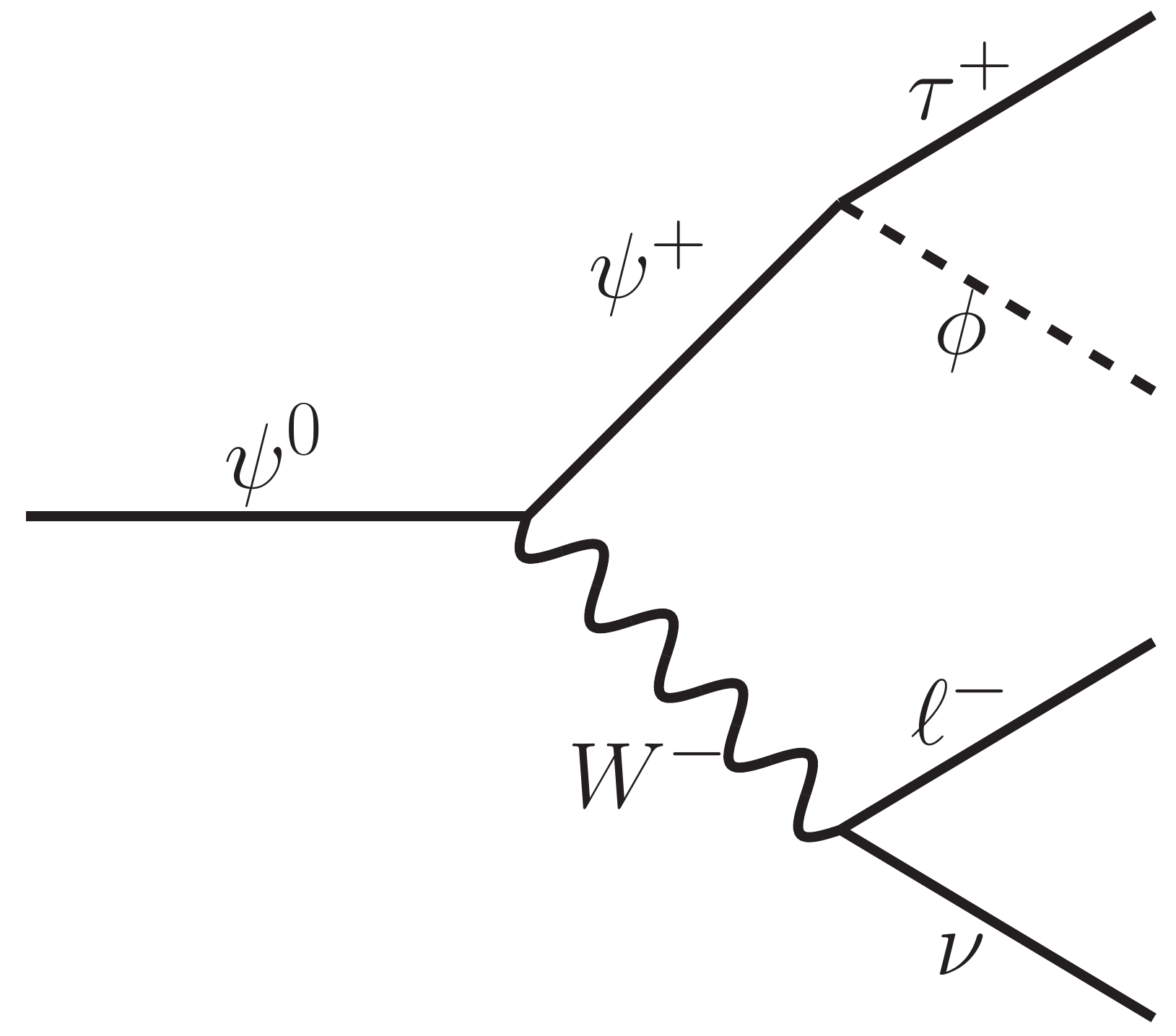}\label{fig:llp_psi0_to_psip}}
  \caption{Possible channels for LLP decays.}
   \label{fig:llp}
\end{figure}

Depending on the mass hierarchy, there can be several LLP possibilities. \cref{fig:llp} shows such possibilities in our model. In \cref{fig:llp_psip_to_psi0}, $\psi^+$ decaying into a charged lepton and MET may give disappearing tracks or kinks in the tracker of the detector. The signature is similar to the charged Higgsino decaying into neutral a gravitino and SM leptons through off-shell decay of a $W$ boson~\cite{Jung:2015boa}. In our study, we are producing an on-shell $\psi^0$ in $\psi^+$ decay and that, in turn, decays into $\phi$ and $\nu_{\tau}$ with 100\%  branching ratio. It is obvious that both the decay products of $\psi^0$ are charge neutral, so the entire $\psi^0$ decay chain will give a missing energy signal in the colliders.
 
\begin{figure}[!ht]
  \centering
  \includegraphics[width=0.48\linewidth]{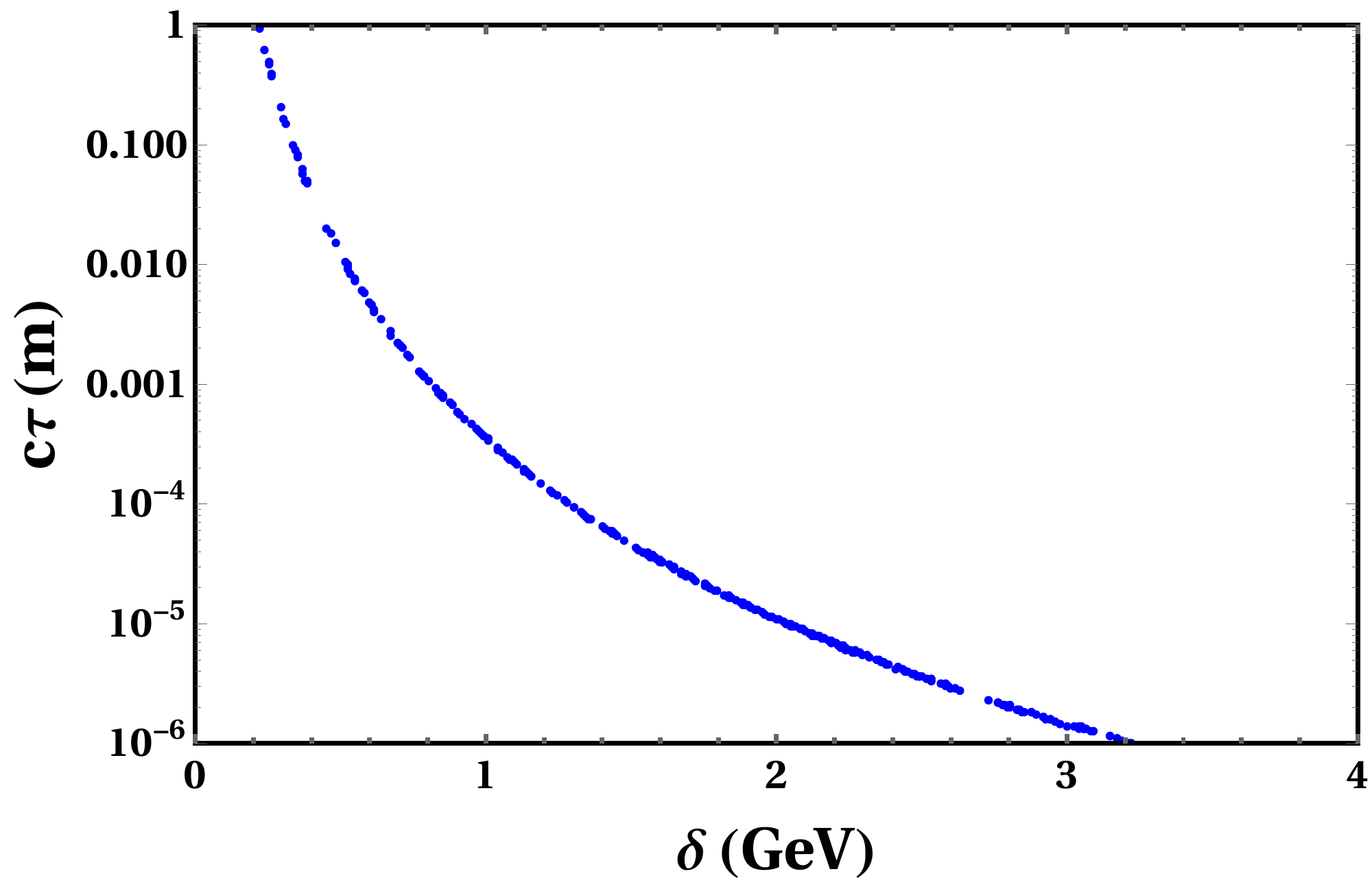}
  \caption{Variation of $c\tau$ as a function of the mass splitting between the dark leptons.}
   \label{fig:ctau}
\end{figure}

For nearly degenerate $\psi^+$ and $\psi^0$, the emitted lepton will be soft, and the track of the mother particle can be identified as disappearing tracks. As we have already pointed out, radiative correction allows $\delta\sim$ 300 MeV, and, from \cref{fig:ctau}, it is clear that for this the in-flight decay of $\psi^+$ gives only DT of decay length around a few centimetres. The kink, however, may be observed in cases where the emitted lepton carries significant energy. One obvious way to address this issue is to increase the mass splitting ($\delta$) between $\psi^+$ and $\psi^0$. The introduction of the triplet in the particle spectrum becomes important in this context because $\delta$ can be varied up to 10 GeV. \cref{fig:ctau} shows the typical decay length ($c\tau$) vs $\delta$ variation for a typical value of $m_{\psi^0}$. The decay length varies around the sensitivity region of ATLAS for $\delta \lesssim$ 2 GeV. For larger splittings, however, the decay becomes too prompt to be detected in the trackers. The kink angle, i.e., the difference of azimuthal angles of the charged parent and the daughter particle, can be measured as a function of the mass and 3-momentum of these particles and the polar angle of the parent particle with the beam axis. The expression, although model dependent, should be similar to Eq. (1) of Ref.~\cite{Asai:2011wy}. However, the detailed kinematics specific to this model along with a robust collider simulation and proper background estimation is beyond the scope of this work, and we postpone it for future study.

If the mass hierarchy between $\psi^0$ and $\psi^+$ is reversed, then the neutral lepton decays late into two emerging visible particles and MET, the final states being a displaced jet coming from $\tau$ decay and a displaced lepton obtained via $W$ decay [\cref{fig:llp_psi0_to_psip}].

%
\section{Conclusion}
\label{sec:conc}
In this work, we have proposed a singlet scalar DM with a vectorlike fermionic doublet having the same dark symmetry. The minuscule Higgs-portal coupling with scalar DM keeps the direct detection cross section below the experimental bound, which is an important handle in reviving the scenario of scalar singlet DM models. The new Yukawa coupling, on the other hand, which is irrelevant to direct search prospects, plays a vital role in dictating the relic density. We have shown that the model can provide a viable DM candidate through pair annihilation, coannihilation, and mediator annihilation channels over a wide parameter space ranging from GeV up to TeV scale. The transition from the pair annihilation to coannihilation regime is demonstrated, and the relevant limits of parameters are discussed. We have observed that coannihilation processes have a substantial contribution to relic density for a comparatively larger mass splitting between DM and the dark sector particles than what is usually discussed in the literature both in SUSY~\cite{Ellis:2015vaa,Ellis:2001nx,Arnowitt:2006jq} and non-SUSY~\cite{Khoze:2017ixx,Chakraborti:2018lso,Lu:2019lok,Dasgupta:2014hha} context. This may be attributed to the gauge couplings involved in these channels, which is a substantial contribution thanks to the dark fermion being a doublet. This is an artefact of the unconventional beyond the SM Yukawa structure considered in the proposed model. This arrangement, involving SM and dark sector lepton $SU(2)_L$ doublets and a scalar singlet, appropriately highlights the important features in the work.

Apart from the DM context, the gauge production of the fermionic doublet followed by decay to DM through the Yukawa coupling results in a substantially increased DM production at the colliders compared to scalar singlet scenarios. Using suitable kinematic observables in a BDT classifier, we separate the signal events from the backgrounds in an effective manner. We have shown that, with the use of proper cuts, we can achieve good results for both the light as well as $\tau$ leptonic channels. 

This model can also provide potential search prospects for long-lived particles because of the nearly degenerate or small mass splittings between the dark leptons. This can lead to suppressed phase space, and the delayed decay of these leptons can facilitate long-lived signatures (LLP) in the colliders, which is recently being given wide attention in the literature. In our study, we have indicated different possibilities of LLP signatures that may arise by tuning the relevant parameters.

One can interpret a limitation of the proposed model in the sense that, from the observed results in both dark matter and the collider analysis, there is no way to distinguish between the two dark leptons, although, one might assume on the contrary from \cref{fig:llp} that it is possible from one-prong (for charged dark lepton) and two-prong (for neutral dark lepton) decays. But from \cref{fig:coll_sig_2lep} and the subsequent discussion, we see that we have to reconstruct a dark lepton from the visible final state SM particles; here, these are leptons. Since these two dark leptons have almost equal masses, it will be difficult to separate them, whereas, a larger splitting between the dark leptons can give more interesting signatures in the colliders. We are pursuing a possible solution to address these issues in an ongoing work.

\appendix
\section{Appendix}
The differential cross section of the pair annihilation process $\phi\phi \to \tau^+\tau^-$ is
\begin{align}
  \frac{d\sigma}{dc_\theta}
  &= \frac{1}{32\pi}\,\frac{y_\tau^4}{4}\,\sqrt{s(s - 4m^2_\phi)}\,(1 - c_\theta^2)
  \Big[
    \frac{1}{(t - m^2_{\psi^+})^2}
  \nl
  &
    + \frac{1}{(u - m^2_{\psi^+})^2}
    - \frac{2}{(t - m^2_{\psi^+}) (u - m^2_{\psi^+})}
  \Big]\,.
  \label{eq:eq1}
\end{align}
In the above, we neglected the $\tau$ lepton mass. We will get the same expression for the process $\phi\phi \to \nu_\tau \bar\nu_\tau$ with $m_{\psi^+} \to m_{\psi^0}$ and $\Gamma_{\psi^+} \to \Gamma_{\psi^0}$. The differential cross section of the pair annihilation process $\phi\,\psi^0 \to \nu_\tau Z$ is
\begin{widetext}
\begin{align}
  \frac{d\sigma}{dc_\theta}
  &=
  \frac{1}{32 s^{3/2}}\,\frac{s - m^2_Z}{p\,m^2_Z}\,\frac{y_\tau^2\,\alpha_{EM}}{\sin^22\theta_W}
  \nl
  &\times
  \Big[
    \frac{1}{s^2}
    \Big\{
    p\sqrt{s} c_\theta (s-2 m_Z^2) (s-m_Z^2)
    + \frac{1}{2} \big[(m_\phi^2-m_{\psi^0}^2)
    (6 m_Z^4-3 m_Z^2 s-s^2)
    +s (2m_Z^4-m_Z^2 s+s^2)\big]
    \Big\}
  \nl
  &
    + \frac{1}{(u - m^2_{\psi^0})^2}
    \Big\{
    \frac{p^3 c_\theta^3}{s^{3/2}} (s-m_Z^2)^3
    -\frac{p^2 c_\theta^2}{2 s^2} (m_Z^2-s)^2 [(m_\phi^2-m_{\psi^0}^2)
    (3 m_Z^2-s) +s (5m_Z^2+s)]
  \nl
  &
    -\frac{p c_\theta}{4 s^{5/2}} (m_Z^2-s) [ (3 m_Z^4-2 m_Z^2 s-s^2)
  (m_\phi^4-2m_\phi^2m_{\psi^0}^2 + m_{\psi^0}^4)
    +2 m_\phi^2 s (5 m_Z^4-2 m_Z^2 s+s^2)
  \nl
  &
    -2 m_{\psi^0}^2 s (5 m_Z^4+6 m_Z^2 s+s^2)
    +s^2 (7 m_Z^4-2 m_Z^2 s-s^2)]
    - \frac{m_\phi^6}{8 s^3} (m_Z^2-s)^2 (m_Z^2+s)
  \nl
  &
    - m_\phi^4 (m_Z^2-s) (-3 m_{\psi^0}^2(m_Z^4-s^2)+5 m_Z^4 s+3 s^3)+m_\phi^2 (-3 m_{\psi^0}^4 (m_Z^2-s)^2 (m_Z^2+s)
  \nl
  &
    +2
   m_{\psi^0}^2 s (5 m_Z^6+3 m_Z^4 s+11 m_Z^2 s^2-3 s^3)-s^2 (7 m_Z^6-23 m_Z^4 s+5 m_Z^2 s^2+3s^3))
  \nl
  &
   +m_{\psi^0}^6 (m_Z^2-s)^2 (m_Z^2+s)+m_{\psi^0}^4 s (-5 m_Z^6-11 m_Z^4 s-19 m_Z^2 s^2+3s^3)
  \nl
  &
   +m_{\psi^0}^2 s^2 (7 m_Z^6-7 m_Z^4 s+21 m_Z^2 s^2+3 s^3)+s^3 (-3 m_Z^6+3 m_Z^4 s+7 m_Z^2 s^2+s^3)
    \Big\}
  \nl
  &
    - \frac{2}{(u - m^2_{\psi^0})}
    \Big\{
    p^2 s c_\theta^2 (m_Z^2-s)^2
    + \frac{3p}{\sqrt{s}} m_Z^2 c_\theta (m_\phi^2-m_{\psi^0}^2) (m_Z^2-s)
  \nl
  &
    + \frac{1}{4 s} ( (5 m_Z^2-s) (m_Z^2+s) (m_\phi^4-2m_\phi^2m_{\psi^0}^2 + m_{\psi^0}^4) + 2 m_\phi^2 s (2m_Z^4-m_Z^2 s+s^2)
  \nl
  &
    -2 m_{\psi^0}^2 s (-2 m_Z^4+3 m_Z^2s+s^2)-s^2 (m_Z^2+s)^2)
    \Big\}
  \Big]\,.
  \label{eq:eq2}
\end{align}
\end{widetext}
In the above, $p$ is the 3-momentum of $\phi$ in the CM frame. With the substitution $m_{\psi^+} \to m_{\psi^0}, m_Z  \to m_W$ and $\Gamma_{\psi^0} \to \Gamma_{\psi^+}$, we can arrive at the analytical expressions for other channels of coannihilation.

\section*{Acknowledgements}
RI thanks the SERB-DST, India for the research grant No. EMR/2015/000333. SC would like to thank MHRD, Government of India for research fellowship. The authors acknowledge useful discussions with Poulose Poulose, Sunando Patra and Pratishruti Saha.

\bibliography{biblio}
\bibliographystyle{apsrev4-1}

\end{document}